\documentclass[12pt]{article}
\usepackage{amssymb,amsmath}
\usepackage{graphicx}
\usepackage{epsfig}
\usepackage{epstopdf}
\usepackage{latexsym}
\usepackage{psfrag}
\usepackage{booktabs}
\usepackage{bbm}
\usepackage[toc]{appendix}
\usepackage{color}
\usepackage{datetime}
\usepackage[
      colorlinks=true,
      linkcolor=darkblue,  
      urlcolor=blue,    
      filecolor=blue,     
      citecolor=red,
      linktocpage=true,
      pdfstartview=FitV,
      bookmarksopen=true    
      ]{hyperref}

\definecolor{darkblue}{rgb}{0.2, 0, 0.8}

\numberwithin{equation}{section}

\newcommand{\ds}{\displaystyle}
\newcommand{\m}{\mu}
\newcommand{\n}{\nu}
\renewcommand{\r}{\rho}
\newcommand{\s}{\sigma}
\newcommand{\D}{\nabla}

\newcommand{\HH}{\mathbb{H}}

\newcommand{\abs}[1]{\left\lvert #1 \right\rvert}

\newcommand{\Z}{\mathbb{Z}}

\newcommand{\R}{\mathbb{R}}

\newcommand{\es}[2] {\begin{equation} \label{#1} \begin{split} #2 \end{split} \end{equation}}

\usepackage{cite,dsfont}
\usepackage{amsfonts}
\usepackage{amssymb}
\usepackage{amsmath}
\usepackage{graphicx}
\usepackage{slashed}
\usepackage{setspace}

\usepackage{color}

\newcommand{\cn}{{\cal N}}

\newcommand{\Kt}{\widehat{K}}

\def \t{\tilde t}

\newcommand{\reef}[1]{(\ref{#1})}

\newcommand{\cl}{{\cal L}}

\newcommand{\be}{\begin{equation}}
\newcommand{\ee}{\end{equation}}

\def\be{\begin{equation}}
\def\ee{\end{equation}}
\def\bea{\begin{eqnarray}}
\def\eea{\end{eqnarray}}
\def\ba{\begin{array}}
\def\ea{\end{array}}
\def\bd{\begin{displaymath}}
\def\ed{\end{displaymath}}

\def\Tr{{\rm Tr}}

\def\a{\alpha}
\def\b{\beta}

\def\d{\delta}
\def\e{\epsilon}           
  
\def\g{\gamma}
\def\h{\eta}

\def\l{\lambda}
\def\m{\mu}
\def\n{\nu}
  
\def\r{\rho}                                     
\def\s{\sigma}                                   
\def\t{\tau}

\def\D{\Delta}

\def\L{\Lambda}

\def\pa{\partial}                              
\def\>{\rangle} 
\def\<{\langle} 
\def\Dsl{D \hskip-.6em \raise1pt\hbox{$ / $ } }
\def\to{\rightarrow}
\def\pa{\partial}

\def\lab{\label}

\newcommand{\eps}{\epsilon}

\def\tchi{\tilde{\chi}}
\def\bs{\bar{\s}}
\def\tz{\tilde{z}}

\def\te{\tilde{\e}}

\begin{document}  


\begin{titlepage}

 \begin{flushright}
{\tt MCTP-13-36} \\
{\tt MIT-CTP-4513}
\end{flushright}

\vspace*{2.3cm}

\begin{center}
{\LARGE \bf Holography for $\mathcal{N}=2^*$ on $S^4$} \\

\vspace*{1.2cm}

{\bf Nikolay Bobev$^{1}$, Henriette Elvang$^{2}$, Daniel Z. Freedman$^{3,4,5}$ and Silviu S. Pufu$^{3}$}
\medskip

$^{1}$Perimeter Institute for Theoretical Physics \\
31 Caroline Street North, ON N2L 2Y5, Canada 
\bigskip

$^{2}$Randall Laboratory of Physics, Department of Physics,\\
University of Michigan, Ann Arbor, MI 48109, USA
\bigskip

$^{3}$Center for Theoretical Physics, Massachusetts Institute of Technology,\\
Cambridge, MA 02139, USA
\bigskip

$^{4}$Department of Mathematics, Massachusetts Institute of Technology,\\
Cambridge, MA 02139, USA
\bigskip

$^{5}$Stanford Institute for Theoretical Physics, Department of Physics,\\
Stanford University, Stanford, CA 94305, USA

\bigskip
nbobev@perimeterinstitute.ca, elvang@umich.edu, dzf@math.mit.edu, spufu@mit.edu  \\
\end{center}

\vspace*{0.1cm}

\begin{abstract}  
We find the gravity dual of $\mathcal{N}=2^*$ super-Yang-Mills theory on $S^4$ and use
holography to calculate the universal contribution to the corresponding $S^4$ free energy at large $N$ and large 't Hooft coupling. Our result matches the expression previously computed using supersymmetric localization in the field theory. This match represents a non-trivial precision test of
holography in a non-conformal, Euclidean signature setting.
\end{abstract}

\end{titlepage}

\setcounter{tocdepth}{2}
{\small
\setlength\parskip{-0.5mm}
\tableofcontents
}

\newpage

\section{Introduction}
\label{sec:Introduction}

The recent interest in rigid supersymmetric field theories on curved manifolds was sparked by the use of supersymmetric localization \cite{Witten:1988ze, Pestun:2007rz} to obtain exact results in many weakly and strongly-coupled supersymmetric quantum field theories.  Pestun \cite{Pestun:2007rz} was the first to use this technique to reduce the partition functions of various ${\cal N} = 2$ theories on $S^4$ to finite-dimensional matrix integrals.  The same technique was later generalized to supersymmetric theories in other dimensions \cite{Kapustin:2009kz, Jafferis:2010un, Hama:2010av, Benini:2012ui,Doroud:2012xw,Kallen:2012cs, Kallen:2012va, Hosomichi:2012ek}.  The three-dimensional generalizations \cite{Kapustin:2009kz, Jafferis:2010un, Hama:2010av} stand out in that they provide impressive checks of the AdS/CFT duality when applied to field theories with holographic duals.  In particular, the matrix models corresponding to various superconformal field theories dual to $AdS_4 \times X$ backgrounds of eleven-dimensional supergravity provide a field theory understanding of the $N^{3/2}$ scaling \cite{Klebanov:1996un} of the number of degrees of freedom on $N$ coincident M2-branes \cite{Drukker:2010nc}.

The goal of this paper is to construct the holographic dual of a supersymmetric, but non-conformal, field theory on $S^4$.  The theory we are interested in is commonly referred to as ${\cal N} =2^*$ supersymmetric Yang-Mills (SYM) theory. It is a mass deformation of the maximally supersymmetric ${\cal N} = 4$ Yang-Mills theory which preserves ${\cal N} = 2$ supersymmetry. For simplicity, we take the gauge group to be $SU(N)$. In general, one can use a Weyl rescaling to uniquely define a conformal field theory on conformally flat manifolds such as $S^d$.  There is no such luxury, in general, for non-conformal field theories, where there are many curved-space generalizations of a given theory on $\R^d$ that differ precisely by couplings proportional to various powers of the space-time curvature.  In the case of supersymmetric field theories, however, supersymmetry suffices to fix the curvature couplings \cite{Festuccia:2011ws}.   In particular, there is a unique supersymmetric Lagrangian for the ${\cal N} = 2^*$ SYM theory on $S^4$;  this Lagrangian was constructed in \cite{Pestun:2007rz} and will be described shortly. 

That the ${\cal N} = 2^*$ SYM theory is not conformal means that we should think of it as a renormalization group (RG) flow on $S^4$, and not as an RG fixed point.  At large $N$ and large 't Hooft coupling, the supergravity dual of this theory is a ``holographic RG flow'' in five Euclidean dimensions that can be foliated using $S^4$ slices.  While supersymmetric holographic RG flows with $\R^d$ slicing have been studied extensively (see, for example, \cite{Girardello:1998pd, Freedman:1999gp, Freedman:1999gk}), there is only 
a relatively small amount of literature on holographic RG flows where the dual field theory lives on a curved manifold.  Refs.~\cite{Martelli:2011fu, Martelli:2011fw, Martelli:2012sz} constructed four-dimensional holographic duals of supersymmetric field theories on certain deformations of $S^3$.  The bosonic supergravity fields that participated in these constructions were the metric and the graviphoton $U(1)$ gauge field.  Ref.~\cite{Freedman:2013oja} considered more complicated four-dimensional holographic RG flows that correspond to ${\cal N} =2$-preserving mass deformations of the ${\cal N} = 8$ superconformal ABJM theory \cite{Aharony:2008ug} on $S^3$.  The supergravity fields with non-trivial profiles in these constructions were the metric and six scalar fields. 

The present work can be thought of as a generalization of the construction in \cite{Freedman:2013oja} to one higher dimension, as both here and in \cite{Freedman:2013oja} we are studying mass deformations of a maximally-supersymmetric CFT on $S^d$, with $d = 4$ and $d=3$, respectively.  To understand which supergravity fields are needed in our $S^4$ example, let us first describe more precisely the ${\cal N} =2^*$ SYM theory starting with ${\cal N} = 4$ SYM\@.  In ${\cal N} = 2$ notation, the field content of ${\cal N} = 4$ SYM is given by a hypermultiplet consisting of two complex scalars $Z_1$ and $Z_2$ and two Weyl fermions $\chi_1$ and $\chi_2$, as well as by a vector multiplet consisting of a gauge field $A_\mu$, two Weyl fermions $\psi_1$ and $\psi_2$, and a complex scalar $Z_3$.  All these fields transform in the adjoint representation of the $SU(N)$ gauge group.   The $\cn=4$ SYM 
Lagrangian on $S^4$ can be obtained using conformal symmetry from the one on $\R^4$.  The two differ only in that on $S^4$ the scalars acquire a conformal coupling to curvature:
 \es{ConformalMass}{
  {\cal L}^{S^4}_{{\cal N} = 4} 
  ~=~ 
  {\cal L}^{\R^4}_{{\cal N} = 4}  \Big|_{\eta_{\mu\nu} \to g_{\mu\nu}}
  +~~ \frac{2}{a^2}\, \text{tr} \big( \abs{Z_1}^2 + \abs{Z_2}^2 + \abs{Z_3}^2 \big) \,.
 }
Here, $g_{\mu\nu}$ denotes the metric on a round $S^4$ whose radius is $a$, and by ``$\eta_{\mu\nu} \to g_{\mu\nu}$'' we mean that when considering the theory on $S^4$, we should introduce a minimal coupling to curvature.
The mass deformation that gives the ${\cal N}=2^*$ SYM theory is a mass term for the hypermultiplet.  In flat space, this mass term would take the form
 \es{MassHyper}{
  {\cal L}_m^{\R^4} = 
  m^2 \, \text{tr} \left( \abs{Z_1}^2 + \abs{Z_2}^2 \right) 
  + m \, \text{tr} \left(\chi_1 \chi_1 + \chi_2 \chi_2 + \text{h.c.} \right) \,.
 }
On $S^4$, the ${\cal N} = 2$ supersymmetry algebra receives curvature corrections, and the mass deformation \eqref{MassHyper} does not preserve supersymmetry by itself. Up to a discrete choice,\footnote{The discrete choice corresponds to which $OSp(2|4)$ sub-algebra of the $SU(2|2, 4)$ superconformal algebra we wish to preserve.  The other choice is obtained by formally sending $a \to -a$ in \eqref{MassS4} and subsequent formulas.} the correct supersymmetric expression on $S^4$ is \cite{Pestun:2007rz}  
 \es{MassS4}{
  {\cal L}_m^{S^4} 
  ~=~ 
  {\cal L}_m^{\R^4} 
  + ~~\frac{im}{2a}\, \text{tr} \left(Z_1^2 + Z_2^2 + \text{h.c.} \right) \,.
 }
The extra term in \eqref{MassS4} will play a crucial role in our work.  It is important to stress that, as mentioned above, the curvature couplings in \eqref{ConformalMass} and \eqref{MassS4} are uniquely fixed by requiring invariance under ${\cal N}=2$ supersymmetry.

According to the AdS/CFT dictionary, there exists a correspondence between certain gauge-invariant operators in the field theory and type IIB supergravity fields.  To describe the holographic dual of ${\cal N} =2^*$ theory, we expect that at least four bosonic bulk fields should acquire non-trivial profiles:  the bulk metric $g_{\mu\nu}$, a scalar field $\phi$ dual to the bosonic mass term 
${\cal O}_\phi$ in \reef{MassHyper}, a scalar field $\psi$ dual to the fermionic mass term ${\cal O}_\psi$ in \reef{MassHyper}, and another scalar field $\chi$ dual to the operator ${\cal O}_\chi$ in \reef{MassS4}.  Explicitly, the operators are 
\be
   {\cal O}_\phi = \text{tr} \big( \abs{Z_1}^2 + \abs{Z_2}^2 \big)\,,
   ~~~~
   {\cal O}_\psi = \text{tr} \big( \chi_1 \chi_1 + \chi_2 \chi_2 + \text{h.c.} \big)\,,
   ~~~~
   {\cal O}_\chi = \text{tr} \big( Z_1^2 + Z_2^2 + \text{h.c.} \big)\,.~~
   \label{3Os}
\ee
In ${\cal N} = 4$ SYM, or in other words at the UV fixed point of the RG flow we want to consider, the operators ${\cal O}_\phi$ and ${\cal O}_\chi$ have scaling dimension two and are part of the ${\bf 20'}$ irrep of the $SU(4)$ R-symmetry group, while ${\cal O}_\psi$ has scaling dimension three and is part of the ${\bf 10}\oplus\overline{{\bf 10}}$ of $SU(4)$, as can be seen from their expressions in terms of the matter fields of the ${\cal N} = 4$ theory.  At the linearized level, the dual fields $\phi$, $\chi$, and $\psi$ are not only part of the excitations of type IIB supergravity around $AdS_5 \times S^5$, but also part of five-dimensional ${\cal N}=8$ $SO(6)$ gauged supergravity \cite{Gunaydin:1984qu, Pernici:1985ju, Gunaydin:1985cu}, as can be seen from \cite{Kim:1985ez}.  It is therefore reasonable to assume that the whole ${\cal N} = 2^*$ flow can be described within ${\cal N} = 8$ gauged supergravity.

Using symmetry properties, we find that there exists a consistent truncation of ${\cal N} = 8$ gauged supergravity containing only the bosonic fields mentioned in the previous paragraph, namely $g_{\mu\nu}$, $\phi$, $\psi$, and $\chi$.  This consistent truncation appears to be new;  the 5D Lagrangian of the truncated theory is in Euclidean signature
 \es{LagTrunc}{
  {\cal L}_\text{5D} &=\frac{1}{4 \pi G_5} \left[  -\frac{R}{4} + \frac{3 \partial_\mu \eta \partial^\mu \eta }{\eta^2} 
   + \frac{\partial_\mu z \partial^\mu \tilde z}{\left(1 - z \tilde z\right)^2} + V \right] \,, \\
  V &\equiv -\frac{1}{L^2} \left(\frac{1}{\eta^4} + 2 \eta^2 \frac{1 + z \tilde z}{1 - z \tilde z} 
   + \frac{\eta^8}4 \frac{(z - \tilde z)^2}{(1 - z \tilde z)^2} \right)  \,,
 }
where we denoted $z = (\chi + i \psi) / \sqrt{2}$, $\tilde z = (\chi - i \psi) / \sqrt{2}$, and $\eta = e^{\phi / \sqrt{6}}$, $G_5$ is the five-dimensional Newton constant, and $L$ is a length scale equal to the radius of curvature of the (Euclidean) $AdS_5$ extremum of \eqref{LagTrunc} that has $\phi = \chi = \psi = 0$.

The equations of motion following from \eqref{LagTrunc} are second order in derivatives.  From the vanishing of the supersymmetry variations of the spin-$3/2$ and spin-$1/2$ fields of the full ${\cal N} = 8$ gauged supergravity theory, one can also find a set of BPS equations that are first order in derivatives and that imply the second-order equations.  We find that the BPS equations have a one-parameter family of smooth solutions with $S^4$ slicing, as expected from the one-parameter family of field theory deformations parameterized by $m$. These solutions are the holographic duals of the ${\cal N} = 2^*$ on $S^4$.

As a check that our supergravity solutions indeed correspond to the ${\cal N} = 2^*$ theory, we use holographic renormalization \cite{Henningson:1998gx, Skenderis:2002wp, Bianchi:2001kw, Bianchi:2001de} to compute the $S^4$ free energy and match that with known field theory results.  In the field theory, the $S^4$ free energy of the ${\cal N} = 2^*$ theory was computed by first using supersymmetric localization to reduce the path integral on $S^4$ to a finite dimensional matrix integral \cite{Pestun:2007rz} and then evaluating this matrix integral in the limit of large $N$ and large 't Hooft coupling $\lambda = g_\text{YM}^2 N$ \cite{Russo:2012kj,Russo:2012ay,Buchel:2013id, Russo:2013qaa,Russo:2013kea}. The result is
 \es{FFieldTheory}{
  F_{S^4} = -\frac{N^2}{2} (1 + m^2 a^2) \log \frac{\lambda (1 + m^2 a^2) e^{2 \gamma + \frac 12}}{16 \pi^2} \,,
 }
where $\gamma$ is the Euler-Mascheroni constant.  The appearance of the Euler-Mascheroni constant suggests that the result \eqref{FFieldTheory} was derived in a particular regularization scheme.  Indeed, the expression \eqref{FFieldTheory} is found after subtracting certain non-universal UV divergences, and this subtraction introduces ambiguities in $F_{S^4}$. However, the third derivative of \eqref{FFieldTheory} with respect to $ma$,
 \es{d3FFieldTheory}{
  \frac{d^3 F_{S^4}}{d(ma)^3} = -2 N^2 \frac{ma (m^2 a^2 + 3)}{(m^2 a^2 + 1)^2} \,,
 } 
is non-ambiguous, and it is this quantity that we will calculate from our supergravity background and match to \eqref{d3FFieldTheory}.

Note that if we take $z = - \tilde z$, or equivalently $\chi = 0$, the Lagrangian \eqref{LagTrunc} represents the bosonic part of a simpler  truncation of ${\cal N} = 8$ gauged supergravity studied in \cite{Pilch:2000ue}.  In that theory there is a flat-sliced domain wall solution of the BPS equations that is dual to the mass-deformed ${\cal N} =2^*$ SYM on $\R^4$. Indeed, in the flat space limit $a \to \infty$, the dual operator ${\cal O}_\chi$ does not appear as a deformation of the Lagrangian, and consequently the supergravity field $\chi$ is not needed.  The two scalar fields $\phi$ and $\psi$ that remain in this truncation are dual to the scalar and fermion mass operators in \eqref{MassHyper}. The truncation with $\chi = 0$ was also used in \cite{Buchel:2013fpa} as a step towards the construction of the holographic RG flow with $S^4$ slicing by solving the second order equations of motion that follow from the action.  As shown in \cite{Buchel:2013fpa}, the free energy of these solutions does not match the supersymmetric localization result \eqref{FFieldTheory}.  The discrepancy comes from the fact that the supersymmetric localization result relies on a Lagrangian that includes the extra mass term in \eqref{MassS4}, whereas the solutions constructed in \cite{Buchel:2013fpa} do not include the third bulk scalar $\chi$ that is dual to
this mass term.

The rest of this paper is organized as follows.  In Section~\ref{FIELDTHEORY} we begin with a more extensive discussion of the ${\cal N} = 2^*$ theory on $S^4$.  In Section~\ref{sec:supergravity} we present our supergravity truncation and BPS equations.  In Section~\ref{SOLUTION} we solve the system of BPS equations numerically and find a one-parameter family of regular solutions. In Section~\ref{HOLOREN} we use holographic renormalization to compute the $S^4$ free energy of the dual field theory.  We end with a discussion of our results in Section~\ref{sec:Conclusions}. In the appendices we present various technical details of the calculations summarized in the main text.

\section{The $\cn =2^* $ SYM theory on $S^4$}
\label{FIELDTHEORY}

The $\cn =2^* $ SYM theory is obtained by  mass deformation of the superconformal invariant $\cn =4 $ theory. The deformation breaks the $SU(2,2|4)$ superconformal algebra to the superalgebra  $OSp(2|4)$ of $\cn=2$ Poincar\'e supersymmetry.  Pestun \cite{Pestun:2007rz} studied the theory on the Euclidean  signature four-sphere and applied the method of localization to calculate the partition function $Z$, or equivalently the free energy $F = -\log\abs{Z}$, and the expectation value of a supersymmetric Wilson loop. He obtained the action and transformation rules by time-like dimensional reduction of the $\cn =1$ SYM theory in ten dimensions.
We have determined an equivalent form of $\cn=2^*$ on $S^4$ by ``Wick rotation" in four dimensions.  Our general approach to Euclidean supersymmetry is described in the appendices of \cite{Freedman:2013oja}.  
In this section, we present the results and discuss the global symmetries that must be matched in the construction of the gravity dual.  Additional details are given in Appendix \ref{app:S4susy}.

\subsection{$\cn=4$ SYM on $S^4$ in $\cn=2$ formulation}

The fields of  $\cn=4$ SYM theory are 
\begin{equation} \lab{n4mult}
A_{\mu}\,, \qquad X_{1,2,3,4,5,6}\,, \qquad \lambda_{1,2,3,4}\,.
\end{equation}
The $X_i$ are six real scalars in the $\bf{6}$ of the $R$-symmetry group $SO(6)_R\cong SU(4)_R$ and the $\lambda_\a$ are four Weyl fermions in the $\bf{4}$.  All fields are in the adjoint of the $SU(N)$ gauge group.
In the $\cn = 2^*$ theory the multiplet in (\ref{n4mult}) decomposes into one $\cn = 2$ vector multiplet consisting of 
\be  \lab{vmult}
A_\mu  \,,~~~~~~~
\psi_1 \,=\, \l_4 \,,~~~~~~~
\psi_2  \,=\, \l_3\,,~~~~~~~
\Phi  \,=\, Z_3=\displaystyle \frac{1}{\sqrt2}\big(X_3 + i X_6\big)\,,
\ee
and one massive hypermultiplet which contains 
\be \lab{hmult}
\chi_i = \l_i \,,~~~~~~~
Z_i =   \frac{1}{\sqrt2}\big(X_i+i X_{i+3}\big),
~~~~~~~\quad i=1,2 \,.  
\ee
We have introduced complex scalars $Z_i$ because we will describe the theory largely using $\cn=1$ language.  As discussed in \cite{Freedman:2013oja}, \emph{fields, both fermions and bosons, that are complex conjugate in Lorentzian signature are algebraically independent in Euclidean signature 
supersymmetry.} To emphasize their independence we use the notation $\tchi,~ \tilde{Z}_i$, etc.~to denote the ``formal conjugates" of $\chi,~Z_i$.
All adjoint fields have gauge covariant derivatives  (with gauge coupling $g_{\text{YM}}=1$), e.g. 
\be
 D_\m Z_i^a \equiv \pa_\m Z_i^a + \,f^{abc}A^b_\m Z_i^c\,,~~~~~~~
 D_\m \chi_i^a \equiv \nabla_\m \chi_i^a + \,f^{abc}A^b_\m \chi_i^c\,,
\ee 
in which $\nabla_\mu$ indicates the spinor covariant derivative on $S^4$ and $a,b,c$ are gauge group indices.

The action of the massless $\cn=4$ theory on $S^4$ can be written as 
\be \lab{massless}
S_{m=0} = \int d^4x \,\sqrt{g} \Big[\cl_{\rm kin} + \cl_{\rm Yukawa} +\cl_{4}\Big]\,.
\ee
The kinetic Lagrangian is\footnote{In Euclidean signature, the ``Weyl" matrices are $\s^\m = (\vec{\sigma},- iI)$ and $\bar{\s}^\m = (\vec{\sigma},+ iI)$.
}
\be \lab{Lkin}
\begin{split}
  \cl_{\rm kin} ~=~& 
  \frac{1}{4} F_{\mu\nu} ^aF^{\mu\nu a} 
  - \tilde \psi_\a^{aT} \sigma_2 \bar \sigma^  \mu {D_\mu} \psi_\a^a  
  - \tilde \chi_i^{aT} \sigma_2 \bar \sigma^\mu {D_\mu} \chi_i^a  
  \\[1mm]
  & +D^\mu \Phi^a D_\mu \Phi^a 
  +  D^\mu \tilde Z_i^a D_\mu Z_i^a  
  +\frac{2}{a^2}\Big(\tilde\Phi^a\Phi^a+ \tilde{Z}_i^a Z_i^a\Big)\,.
\end{split}
 \ee
The last bracket contains the conformal coupling of the scalars to the curvature scalar $R$ of the four-sphere with radius $a$,   i.e.~$2/a^2 = R/6$.

The Yukawa term is simply a rearrangement of the Yukawa term of $\cn =4$ SYM as written in terms of $4\times 4$ 't Hooft matrices (see Appendix \ref{app:S4susy}):
\bea \lab{lyuk}
\begin{split}
{\cal L}_{\rm Yukawa} 
&\,=\,\sqrt2 f^{abc} \bigg( 
   \frac12 \e_{\a\b}\big(\psi_\a^{aT} \s_2\psi_\b^b\big) \tilde{\Phi}^c 
   - \frac12 \e_{ij}\big(\chi_i^{aT} \s_2\chi_j^b\big) {\Phi}^c\\
&~~
  \hspace{1.3cm}
  +\begin{pmatrix} 
   \psi_1^{aT} \s_2 &  \psi_2^{aT} \s_2
   \end{pmatrix}
   \begin{pmatrix}
    0 & 1 \\
    -1 & 0 
   \end{pmatrix}
  \bigg[
   \chi_1^b
   \bigg(\!\!
   \begin{array}{c}
     Z_2^c \\ \tilde Z_1^c
   \end{array}\!\!
   \bigg)
  -\chi_2^b
     \bigg(\!\!
   \begin{array}{c}
     Z_1^c \\ -\tilde Z_2^c
   \end{array}\!\!
   \bigg)
   \bigg]\bigg)  \,+ \text{h.c.}\, \\
&\,=\,\sqrt2 f^{abc} \bigg( 
   \frac12 \e_{\a\b}\big(\psi_\a^{aT} \s_2\psi_\b^b\big) \tilde{\Phi}^c 
   - \frac12 \e_{ij}\big(\chi_i^{aT} \s_2\chi_j^b\big) {\Phi}^c+ (\psi_1^{aT}\sigma_2 \chi_i^b)\tilde{Z}_i^c - (\epsilon_{ij}\psi_2^{aT}\sigma_2 \chi_i^b)Z_j^c \bigg)
    + \text{h.c.}
\end{split}
\eea
Here, and in the following, ``h.c." stands for  the formal Hermitian conjugate, i.e.~terms that in the Lorentzian theory are obtained by Hermitian conjugation and are converted to  Euclidean signature via the analytic continuation detailed in \cite{Freedman:2013oja}. 
In the first form \reef{lyuk}, the global symmetries of ${\cal L}_{\rm Yukawa}$ are manifest, as we will discuss shortly.  The second form is neater.

The quartic term is also obtained directly from that of the $\cn=4$ theory, viz.
\bea \lab{Lfour}
\cl_4&=&\frac12 f^{abc}f^{ab'c'}
\sum_{i,j=1}^3\Big( -\,\tilde Z^b_i Z^c_i \tilde Z^{b'}_j Z^{c'}_j 
    + 2\tilde Z^b_j\tilde Z^c_i Z^{b'}_j Z^{c'}_i \Big)\,.
\eea
In the $\cn=1$ formulation with three adjoint scalars $Z_i$, the first quartic term in \reef{Lfour} is simply the D-term potential $V_D = \frac{1}{2}D^a D^a$ and the second term is the F-term potential 
$V_F = \tilde{F}^a F^a = \sum_{i=1}^3 \big| \frac{\pa W}{\pa Z_i^a}\big|^2$ for the cubic superpotential, 
\be
  W = -\sqrt{2} f^{abc}  Z_1^a Z_2^b Z_3^c\,.
\ee 
For the $\cn=2^*$ formulation, we replace  $Z_3 \to \Phi$ in the bilinear sums of \reef{Lfour}, for example
$\sum_{i=1}^3 \tilde Z^b_i Z^c_i = \tilde\Phi^b\Phi^c + \sum_{i=1}^2 \tilde Z^b_i Z^c_i$. 

This massless theory is invariant under transformation rules in which the spinor parameters are Killing spinors on $S^4$.  They are Weyl spinors that satisfy the equations
\be \lab{Kspin}
\nabla_\m \e_{\pm} = \pm \frac{i}{2a} \s_\m \tilde{\e}_{\pm}\,,\qquad\qquad 
\nabla_\m \tilde{\e}_\pm= \pm \frac{i}{2a} \bs_\m \e_{\pm}\,.
\ee
For each sign $\pm$ there are two linearly independent solutions. The explicit form of these solutions is known \cite{Lu:1998nu} but is not needed for our purposes. The massless $\cn=4$ theory is superconformal, and thus invariant under transformations involving both signs.  There is a further doubling of the number of spinors because of  $\cn =2$ supersymmetry. It is incorporated by adding the subscript $I=1,2$, i.e.~$\e_\pm \to \e_{\pm,I}.$ 
We will not exhibit the complete transformation rules because they are not needed, but the subset used to determine the mass deformation of the action is discussed in Appendix \ref{app:S4susy}.

So far we have just rewritten the $\cn=4$ SYM theory on  $S^4$ in  a notation which incorporates the split into vector and hypermultiplet.  The subgroup of the $R$-symmetry group $SU(4)$ that preserves this split may be denoted by
$SU(2)_V\times SU(2)_H\times U(1)_R$. 
The specific implementation of these symmetries is discussed in Appendix \ref{app:S4susy};  see also \cite{AlvarezGaume:1996mv} and \cite{Okuda:2010ke}. 
The results are summarized in the following table: 
\begin{displaymath}
 \begin{array}{c|c|c|c}
  & SU(2)_V & SU(2)_H& U(1)_R\\
\hline  A_\mu & 0 & 0& 0 \\
\Phi & 0 & 0 & +2\\
\psi_{1,2} & {1}/{2} & 0 & + {1} \\
\tilde{\psi}_{1,2} & {1}/{2} & 0 & - {1} \\
\hline  Z_{1,2} & 1/2^\dagger & 1/2 & 0 \\
\chi_{1,2} & 0 & 1/2 & - 1 \\
\tchi_{1,2} & 0& 1/2 & +1 \\
\hline
 \end{array}
 \end{displaymath}
The action of $SU(2)_V$ on the scalars $z_{1,2}$ is flagged to indicate its special form: the basic doublets are 
$\bigg(\!\!
   \begin{array}{c}
     Z_2^c \\ \tilde Z_1^c
   \end{array}\!\!
   \bigg)$
and    
$\bigg(\!\!
   \begin{array}{c}
     Z_1^c \\ -\tilde Z_2^c
   \end{array}\!\!
   \bigg)$. All fields of the vector multiplet are $SU(2)_H$ singlets and all fields of the hypermultiplet belong to the $s=1/2$ representation. With this information we can now understand the structure of $\cl_{\rm Yukawa}$ in  \reef{lyuk}.  
The quantities $\chi_1 Z_2 - \chi_2 Z_1$ and $\chi_1 \tilde Z_1 + \chi_2\tilde Z_2$ are both $SU(2)_H$ invariants. Thus $\cl_{\rm Yukawa}$ is invariant under all global symmetries.

\subsection{The mass deformation}

In flat space one can introduce the hypermultiplet mass term via the $\cn =1$ superpotential 
(see (3.1) of \cite{Buchel:2000cn}) 
\be \lab{N=2supot}
W_{2^*} =  -\sqrt2 f^{abc} Z^a_1Z^b_2Z^c_3+  \frac{1}{2}m(Z_1^aZ_1^a+Z_2^aZ^a_2)\,.
\ee
This produces two new terms in the Lagrangian, a cubic coupling of the scalars (recall that $\Phi =Z_3$)
\be \lab{cubic}
\cl_{3}  = -\sqrt2 m [f^{abc}(\tilde Z^a_1Z^b_2 -\tilde Z^a_2Z^b_1)\Phi^c + \text{h.c.}]\,,
\ee
 and the mass term proper
 \be \lab{Lmass}
\cl_{\rm mass} = -\frac12 m (\chi_i^{aT}\s_2\chi_i^a +\tilde{\chi}_i^{aT}\s_2\tilde{\chi}_i^a) + m^2 \tilde Z^a_iZ^a_i\,.
\ee

The hypermultiplet mass breaks the symmetry group $SU(2)_V \times SU(2)_H \times U(1)_R$ of the Euclidean theory on $\R^4$ and we find that there is further breaking on $S^4$. 
One can  see that $U(1)_R$  is broken by the fermion mass term and that $SU(2)_H$ is broken to the $U(1)_H$ subgroup generated by the Pauli matrix $\t_2$.\footnote{Pauli matrices are denoted by $\sigma_i$ when they act on spacetime spinors, and otherwise by $\tau_i$.}\,\, $SU(2)_V$ is preserved by $\cl_{\rm mass}$.  $\cl_3$ also preserves $SU(2)_V\times U(1)_H$, and $U(1)_R$ is broken.  The parameter $m$ in \reef{cubic}--\reef{Lmass} may be real or complex.
The mass term obviously breaks conformal symmetry in flat space, but $\cn=2$ Poincar\'e supersymmetry is unbroken.  

On $S^4$,  the $\cn=2$ transformation rules permit one choice of the sign in the equations  \reef{Kspin} obeyed by Killing spinors.
One can choose either sign. For the upper sign in \reef{Kspin},  supersymmetry requires one additional term in the action,  given by\footnote{For the lower sign one can simply change $a \to -a$ throughout.}
\be\lab{LS4}
\cl_{S^4} = \frac{im}{2a} (Z^a_iZ^a_i + \tilde Z^a_i\tilde Z^a_i)\,.
\ee
In this  term $SU(2)_V$ is broken to the $U(1)_V$ subgroup generated by $\t_2$.  Therefore the global symmetry of the complete $\cn=2^*$ theory on $S^4$ is the Abelian product group $U(1)_V\times U(1)_H$.

A minor generalization of \reef{Lmass} is possible.   $U(1)_R$ is not a symmetry, and we may use it to make a change of variables in the presentation.  Specifically, we define $\chi_i = e^{-i\theta}\chi'_i,~ \psi_i =e^{i\theta}\psi'_i,~~\Phi = e^{2i\theta}\Phi'.$   Since $U(1)_R$ is a symmetry when $m=0$,  this change affects only $\cl_3$ and $\cl_{\rm mass}$.  The latter becomes
\be \lab{Lmassprime}
\cl_{\rm mass} = -\frac12 (m' \chi_i^T\s_2\chi_i +\tilde{m}'\tilde{\chi}_i^{'T}\s_2\tilde{\chi}'_i) + m^2 \tilde Z^a_iZ^a_i\,,
\ee
where $m' = m e^{-2i\theta}$ and $\tilde{m}'= m e^{2i\theta}$.  $\cn =2$ supersymmetry is maintained if we make the same $U(1)_R$ phase change in the supersymmetry transformation rules.   The role of the parameter $\theta$ will be made clear in the holographic setup after \eqref{LagTruncMainText2}.

\subsection{Summary}

In this section we have presented the action and discussed the symmetries of the $\cn=2^*$ SYM theory on the Euclidean manifold $S^4$.  The massless theory  is superconformal and indeed just  a rewrite of the well-known 
$\cn=4$ theory.
When so written the Lagrangian consists of the three terms in \eqref{Lkin}--\eqref{Lfour}. It is invariant under  $\cn=2$ transformation rules  (not given above)  
with  Killing spinor parameters that satisfy either of  
the $\pm$ equations in \eqref{Kspin}. The presentation is invariant under the $R$-symmetry group $SU(2)_V\times SU(2)_H\times U(1)_R$.   The mass deformation requires the two additional Lagrangian terms in \eqref{Lmass}--\eqref{LS4}  or more generally \eqref{LS4}--\eqref{Lmassprime}.   The transformation rules of its reduced supersymmetry involve half the number of Killing spinors, which comprise eight real supercharges.  The global symmetry group is reduced to  $U(1)_V\times U(1)_H$.

In the next section, we construct the gravity dual of the mass-deformed theory.  The dual of the undeformed theory is type IIB superstring theory on $AdS_5\times S^5$. Its low-energy limit is type IIB supergravity, which is believed to contain the maximal gauged $\cn=8$ supergravity theory in five spacetime dimensions as a consistent truncation.  We will assume that the dual of the mass-deformed theory can be described within a further consistent truncation of $\cn=8$.  In its general form it should contain three scalar fields  that source the three gauge invariant operators of the deformation in \eqref{LS4}--\eqref{Lmassprime} (at fixed $\theta$), namely $\tilde Z^a_iZ^a_i$,  $e^{-2i \theta} \chi_i^T\s_2\chi_i + e^{2 i \theta} \tilde{\chi}_i^{'T}\s_2\tilde{\chi}'_i$, and $Z^a_iZ^a_i + \tilde Z^a_i\tilde Z^a_i$.  The dual description is valid in the limit $N\to\infty$ and $g_{\text{YM}}^2N  \gg 1$.  In that limit the undeformed $\cn=4$ field theory has an additional   $U(1)_Y$ symmetry whose presence was inferred \cite{Intriligator:1998ig} from the gravity dual and which is also expected in the $\cn =2^*$ deformation \cite{Buchel:2000cn}.  This $U(1)_Y$ symmetry is a diagonal subgroup of the $U(1)_R$ subgroup of $SO(6)_R$ and an $SO(2)$ subgroup of the S-duality group, which in the limit mentioned above is enhanced from $SL(2, \Z)$ to $SL(2, \R)$.  Our gravity dual therefore should have Euclidean signature and a gauged $U(1)^3$ symmetry.

\section{Supergravity}
\label{sec:supergravity}

We now describe how to construct the five-dimensional holographic dual of the ${\cal N} =2^*$ SYM theory on $S^4$ presented in the previous section.  As mentioned above, this holographic dual must contain (at the very least) a non-trivial bulk profile for three bulk scalar fields that we identify based on their transformation properties under the various symmetries. The strategy  followed in this section is to study first the relevant sector of ${\cal N} = 8$ gauged supergravity in Lorentzian signature, and  then continue this analysis to Euclidean signature.

\subsection{Lorentzian truncation}

The {\em ungauged} ${\cal N} = 8$ supergravity theory in five-dimensions has an $E_{6(6)}$ global duality group that acts as a symmetry of the equations of motion, but not of the Lagrangian.  
This theory can also be written in a description with $USp(8)$ composite local symmetry, where the target space for the $42$ scalars can be identified with the coset $E_{6(6)} / USp(8)$.  If the local $USp(8)$ acts on the $E_{6(6)}$ group element $g$ by multiplication on the right, the global $E_{6(6)}$ transformations act on the left:
 \es{Multip}{
  g \to h g k^{-1} \,, \qquad g, h \in E_{6(6)} \,, \qquad k \in USp(8) \,.
 }
It is simplest, however, to describe this theory (and its subsequent gauging) in a particular ``symmetric'' $USp(8)$ gauge defined such that $g$ stays invariant under \eqref{Multip} whenever $h = k \in USp(8)$.  In this gauge, all the supergravity fields transform in totally anti-symmetric symplectic traceless representations of the diagonal $USp(8)$.  These fields are: the metric $g_{\mu\nu}$, 8 gravitini $\psi^a_\mu$, 27 vector fields $A^{ab}_\mu$, 56 spin-$1/2$ fields $\chi^{abc}$, and 42 real scalars $\phi^{abcd}$, where the upper indices are fundamental $USp(8)$ indices that run from $1$ to $8$.\footnote{We hope that no confusion will arise from the fact that we used $a,b,c$ to denote $SU(N)$ gauge indices in the previous section. The latter will not appear in the supergravity discussion.}

There are several inequivalent gaugings that one can perform by promoting part of the global $E_{6(6)}$ symmetry to a local symmetry.  The desired subsector of type IIB string theory on $S^5$ involves a gauging of  the $SO(6)$ subgroup in 
 \es{Subgroups}{
  SO(6) \times SO(2) \subset USp(8) \subset E_{6(6)}\,.
 }
Within $E_{6(6)}$, this $SO(6)$ commutes with an $SL(2, \R)$ of which only an $SO(2)$ subgroup is contained in $USp(8)$.   The ${\cal N} = 8$ gauged supergravity theory therefore has $SO(6) \times USp(8)$ local and $SL(2, \R)$ global invariance.  Fixing the $USp(8)$ gauge as above, it is then straightforward to determine the $SO(6) \times SO(2)$ charges of the various supergravity fields by decomposing the $USp(8)$ irreps listed in the previous paragraph with respect to the $SO(6) \times SO(2)$ subgroup.  One effect of the gauging is that $15$ of the $27$ vector fields become the $SO(6)$ gauge fields while the other $12$ must be represented as rank-two antisymmetric tensor fields that are charged under $SO(6)$.

To characterize the embedding of $SO(6) \times SO(2)$ into $USp(8)$, let $H_i$, with $i = 1, \ldots , 4$, 
 be the Cartan elements of $USp(8)$ defined such that the fundamental eight-component vectors $v_{\pm i}$ satisfy $H_j v_{\pm i} = \pm \delta^i_j v_{\pm i}$.  Choosing the Cartan of $SO(6)$ to be generated by rotations in the $13$, $24$, and $56$ planes, it is straightforward to work out that, up to equivalence, one must have:
\es{Identifications}{
  U(1)_{13}: \quad &-H_1 + H_2 + H_3 - H_4 \,, \\
  U(1)_{24}: \quad &H_1 - H_2 + H_3 - H_4 \,, \\
  U(1)_{56}: \quad &H_1 + H_2 - H_3 - H_4 \,, \\
  SO(2): \quad &H_1 + H_2 + H_3 + H_4 \,.
 }
where $U(1)_{ij} \in SO(6)$ corresponds to rotations in the $ij$ plane.  Another characterization of the embedding of $SO(6) \times SO(2)$ into $USp(8)$ is that the fundamental irrep ${\bf 8}$ of the latter group decomposes as ${\bf 4}_1 + \overline{\bf 4}_{-1}$ under the former.  The $SO(6)$ gauge group in supergravity corresponds to the $SO(6)_R \cong SU(4)_R$ symmetry group of the ${\cal N} = 4$ SYM theory described in the previous section.  Similarly, the $SO(2)$ invariance of the supergravity theory corresponds to the $SO(2)$ ``symmetry'' of ${\cal N} = 4$ SYM that emerges at large $N$ and 't Hooft coupling \cite{Intriligator:1998ig} .
 
As discussed in the previous section,  the mass deformed ${\cal N} = 2^*$ theory on $S^4$ is invariant under  a $U(1)_V \times U(1)_H \times U(1)_Y$ subgroup of $SO(6) \times SO(2)$  in the large $N$ limit and at large 't Hooft coupling.  
The supersymmetries  of the field theory transform under $U(1)_V$, which is the only $R$-symmetry in the product group.   The holographic dual we seek must reflect these symmetries.  This means that the bulk scalar fields should be $U(1)^3$ invariant, while the gravitini\footnote{None of the eight gravitini of the ${\cal N} = 8$ theory, which transform in the ${\bf 8}$ of $USp(8)$, are invariant under \eqref{U1VHa}.} are  $U(1)_H \times U(1)_Y$ invariant but  charged under  $U(1)_V$.
From the previous  analysis we  identify
 \es{U1VHa}{
  U(1)_V: \quad &H_3 - H_4 \,, \\
  U(1)_H: \quad &H_1 - H_2 \,, \\
  U(1)_Y: \quad &H_1 + H_2 \,.
 }

One may therefore consider the sector of ${\cal N} = 8$ gauged supergravity that is invariant only under $U(1)_H$ and $U(1)_Y$.  The fields that are invariant are:  the metric, 4 gravitini, 5 vector fields (corresponding to the $SU(2)_V \times U(1)_H \times U(1)_R$ subgroup of $SO(6)$), 2 anti-symmetric tensors, 8 spin-$1/2$ fields, and 6 real scalars.  This theory is an ${\cal N} = 4$ gauged supergravity theory with a gravity multiplet, one vector multiplet, and gauge group $SU(2) \times U(1)^2$.  The Lagrangian of this supergravity theory is rather constrained.  For instance, the six real scalars parameterize a $\R \times \HH^5$ target space.  Vector and antisymmetric  tensor fields are omitted because they have vanishing profiles in the $S^4$-sliced solutions we need.  The bosonic Lagrangian is then\footnote{We use the conventions of \cite{Gunaydin:1984qu}, in particular a  mostly minus Lorentzian signature metric.}
\es{LagTruncMainText}{
  {\cal L} &= \frac{1}{2 \kappa^2} \left[-R +  12 \frac{\partial_\mu \eta \partial^\mu \eta}{\eta^2} +
    \frac{4\, \partial_\mu \vec{X} \cdot \partial^\mu \vec{X} }{\left(1 - \vec{X}^2 \right)^2} - V \right] \,, \\
  V &= -\frac{4}{L^2} \left[\frac{1}{\eta^4} + 2 \eta^2 \frac{1 + \vec{X}^2}{1 - \vec{X}^2} - \eta^8 \frac{(X_1)^2 + (X_2)^2}{\left(1 - \vec{X}^2 \right)^2}\right]   \,,
 }
where $\vec{X} = (X_1, X_2, X_3, X_4, X_5)$ are five of the scalars and $\eta$ is the sixth.  The scalars $(X_1, X_2)$ form a doublet under the $U(1)_R$ part of the gauge group, while $(X_3, X_4, X_5)$ form a triplet under $SU(2)_V$ and $\eta$ is neutral.  The overall normalization of the potential was chosen such that the $AdS_5$ extremum of \eqref{LagTruncMainText}, which is obtained with $\vec{X} = 0$ and $\eta =1$, has curvature radius $L$. See Appendix~\ref{app:truncation} for details on how to derive \eqref{LagTruncMainText} from ${\cal N} = 8$ gauged supergravity. 

In the ${\cal N} = 4$ supergravity theory we can use the $SU(2)_V \times U(1)_R$ gauge transformations to set, say, $X_2 = X_4 = X_5 = 0$.  The resulting action is
 \es{LagTruncMainText2}{
  {\cal L} &=\frac{1}{2 \kappa^2} \left[  R - \frac{12 \partial_\mu \eta \partial^\mu \eta }{\eta^2} 
   - \frac{4\, \partial_\mu z \partial^\mu z^*}{\left(1 - \abs{z}^2 \right)^2} - V \right] \,, \\
  V &\equiv -\frac{4}{L^2} \left(\frac{1}{\eta^4} + 2 \eta^2 \frac{1 + \abs{z}^2 }{1 - \abs{z}^2} 
   + \frac{\eta^8}4 \frac{(z - z^*)^2}{(1 - \abs{z}^2 )^2} \right)  \,,
 }
where $z = X_3 + i X_1$ and $z^* = X_3 - i X_1$.  The fields $\eta$, $z$, and $z^*$ are invariant under $U(1)_V \times U(1)_H \times U(1)_Y$, and they correspond to the three independent operators in \eqref{LS4}--\eqref{Lmassprime} after $U(1)_R$ was used to fix the value of $\theta$.  Changing the value of the field theory parameter $\theta$ corresponds to a constant rotation in the $(X_1, X_2)$ plane and does not yield new physics.

As in any five-dimensional 
${\cal N} = 4$ theory, the supersymmetries can be written as two pairs of symplectic Majorana spinors $(\epsilon_1, \epsilon_3)$ and $(\epsilon_2, \epsilon_4)$.  Following~\cite{Gunaydin:1984qu}, we use a basis of five-dimensional gamma matrices $\gamma_m$, where $m = 0, \ldots, 4$, that satisfy the Clifford algebra $\{\gamma_m, \gamma_n\} = 2 \eta_{mn} = 2\text{diag}\{1, -1, -1, -1, -1\}$, where $\gamma_m$, with $m = 0, \ldots, 3$ are pure imaginary and $\gamma_4$ is pure real.  In this basis, the symplectic Majorana condition is
\es{SympMajorana}{
  \epsilon_3 = \gamma_5  \epsilon_1^* \,, \qquad \epsilon_4 = \gamma_5 \epsilon_2^* \,,
 }
where $\gamma_5$ is defined as $\gamma_5 \equiv -i \gamma_4$.  Because $\gamma_5$ is pure imaginary, the conditions \eqref{SympMajorana} imply $\epsilon_1 = -\gamma_5 \epsilon_3^*$ and $\epsilon_2 = -\gamma_5 \epsilon_4^*$.  Instead of writing the supersymmetry variations in terms of all four spinors $\epsilon_i$, we will use \eqref{SympMajorana} to write the supersymmetry variations only  in terms of $\epsilon_i$ and $\epsilon_i^*$ with $i = 1, 2$.

In Lorentzian signature the vanishing of the supersymmetry variations of the spin-$1/2$ fields in the Majorana basis takes the form 
\es{Spin12}{
  \frac {3 \gamma^\mu \partial_\mu \eta}{2 \eta} \gamma_5 \epsilon_i^* - \frac{1}{2L} \frac{1 + (z^*)^2 + \left((z^*)^2 - 1\right) \eta^6}{(1 - \abs{z}^2) \eta^2}\epsilon_i &=0 \,, \\
  \frac{\gamma^\mu \partial_\mu z}{1 - \abs{z}^2} \gamma_5 \epsilon_i^* + \frac{1}{2L}  \frac{2 (z + z^*) + (z - z^*) \eta^6}{(1 - \abs{z}^2) \eta^2}\epsilon_i &=0 \,,
 }
with $i = 1,2 $.  The vanishing of the gravitino variation takes the form
 \es{Spin32}{
  \nabla_\mu \epsilon_i + \frac{z^* \partial_\mu z - z \partial_\mu z^*}{2 ( 1- \abs{z}^2)} \epsilon_i
   + \frac{1}{6 L} \frac{2 (1 + z^2) + \eta^2 (z^2 - 1)}{(1 - \abs{z}^2) \eta^2} \gamma_\mu \gamma_5 \epsilon_i^* = 0 \,,
 }
where $\nabla_\mu$ is the usual covariant derivative acting on a spinor, and again $i = 1,2$.  That the vanishing of the supersymmetry transformations parameterized by $\epsilon_1$ and $\epsilon_2$ leads to identical equations is a consequence of the fact that all three scalars $\eta$, $z$, and $z^*$ are invariant under $U(1)_V \times U(1)_H \times U(1)_Y$, while the fermions are invariant only under $U(1)_H \times U(1)_Y$ and transform under $SU(2)_V$.  The $U(1)_V$ subgroup of $SU(2)_V$ acts on the fermions by rotating $\epsilon_1$ and $\epsilon_2$ as an $SO(2)$ doublet, so if the supersymmetry variations with parameter $\epsilon_1$ vanish, then so do those corresponding to $\epsilon_2$.

\subsection{Euclidean continuation}

In Euclidean signature the fields that in Lorentzian signature were related by complex conjugation are now independent.  As in Section \ref{FIELDTHEORY} we emphasize this fact by replacing the complex conjugation symbol by a tilde, and write $\tilde z$ instead of $z^*$, $\tilde \epsilon$ instead of $\epsilon^*$, and so on.  The Euclidean continuation of the Lagrangian \eqref{LagTruncMainText2} is then
\es{LagEuclidean}{
  {\cal L} &=\frac{1}{2 \kappa^2} \left[  -R + \frac{12 \partial_\mu \eta \partial^\mu \eta }{\eta^2} 
   + \frac{4\, \partial_\mu z \partial^\mu  \tilde z}{\left(1 - z \tilde z \right)^2} +V \right] \,, \\
  V &\equiv -\frac{4}{L^2} \left(\frac{1}{\eta^4} + 2 \eta^2 \frac{1 + z \tilde z }{1 - z \tilde z} 
   + \frac{\eta^8}4 \frac{(z - \tilde z)^2}{(1 - z \tilde z )^2} \right)  \,.
 }
The Euclidean continuation of the supersymmetry variations \eqref{Spin12}--\eqref{Spin32} requires more care.  It can be done in two steps.  The first step is to stay in Lorentzian signature and go from mostly minus to mostly plus signature.  This change requires replacing $\gamma_\mu \to i \gamma_\mu$ and $\gamma^\mu \to -i \gamma^\mu$ everywhere in \eqref{Spin12}--\eqref{Spin32}.  Note, however, that $\gamma_5$ should not be replaced by $i \gamma_5$, because the symplectic Majorana condition \eqref{SympMajorana} remains unchanged.  The second step is to rotate the time direction to Euclidean signature, which amounts to multiplying the gamma matrix corresponding to the time direction by a factor of $i$, as well as relaxing the complex conjugation condition on all the fields, as discussed above.  The Euclidean continuation of the spin-1/2 equations \eqref{Spin12} is
 \es{Spin12Euclidean}{
  - \frac {3 i \gamma^\mu \gamma_5 \partial_\mu \eta}{2 \eta}  \tilde \epsilon_i - \frac{1}{2L} \frac{1 + \tilde z^2 + \left(\tilde z^2 - 1\right) \eta^6}{(1 - z \tilde z) \eta^2}\epsilon_i &=0 \,, \\
  -\frac{i \gamma^\mu  \gamma_5 \partial_\mu z}{1 - z \tilde z} \tilde \epsilon_i + \frac{1}{2L}  \frac{2 (z + \tilde z) + (z - \tilde z) \eta^6}{(1 - z \tilde z) \eta^2}\epsilon_i &=0 \,.
 }
In Lorentzian signature, the equations \eqref{Spin12} are equivalent to their complex conjugates.  When continuing to Euclidean signature, however, we should also continue the complex conjugates of \eqref{Spin12}, and obtain
 \es{Spin12ConjEuclidean}{
  \frac {3 i  \gamma_5 \gamma^\mu \partial_\mu \eta}{2 \eta} \epsilon_i - \frac{1}{2L} \frac{1 + z^2 + \left(z^2 - 1\right) \eta^6}{(1 - z \tilde z) \eta^2} \tilde \epsilon_i &=0 \,, \\
  \frac{i  \gamma_5 \gamma^\mu \partial_\mu \tilde z}{1 - z \tilde z} \epsilon_i + \frac{1}{2L}  \frac{2 (z + \tilde z) - (z - \tilde z) \eta^6}{(1 - z \tilde z) \eta^2} \tilde \epsilon_i &=0 \,.
 }
The equations \eqref{Spin12Euclidean} and \eqref{Spin12ConjEuclidean} are now independent, and should be satisfied simultaneously if there is unbroken supersymmetry.  Similarly, the Euclidean continuation of the spin-$3/2$ equation \eqref{Spin32} is 
 \es{Spin32Euclidean}{
   \nabla_\mu \epsilon_i + \frac{\tilde z \partial_\mu z - z \partial_\mu \tilde z}{2 ( 1- z \tilde z)} \epsilon_i
   + \frac{i}{6 L} \frac{2 (1 + z^2) + \eta^2 (1 - z^2 )}{(1 - z \tilde z) \eta^2} \gamma_\mu \gamma_5 \tilde \epsilon_i = 0 \,.
 }
The Euclidean continuation of its complex conjugate is 
 \es{Spin32ConjEuclidean}{
  \gamma_5  \nabla_\mu\gamma_5 \tilde \epsilon_i - \frac{\tilde z \partial_\mu z - z \partial_\mu \tilde z}{2 ( 1- z \tilde z)} \tilde \epsilon_i
   - \frac{i}{6 L} \frac{2 (1 + \tilde z^2) + \eta^2 (1 - \tilde z^2 )}{(1 - z \tilde z) \eta^2} \gamma_5 \gamma_\mu  \epsilon_i = 0 \,.
 }
In order to have backgrounds with ${\cal N} = 2$ supersymmetry, equations \eqref{Spin12Euclidean}--\eqref{Spin32ConjEuclidean} must have simultaneous solutions where the four independent four-component complex spinors $\epsilon_i$ and $\tilde \epsilon_i$ depend on eight free complex parameters.

\subsection{Solution Ansatz and equations of motion}

We are looking for Euclidean backgrounds that are invariant under the isometries of $S^4$.  The metric and the scalars should therefore take the form
 \es{MetricAnsatz}{
  ds^2 = L^2 e^{2 A(r)} ds_{S^4}^2 + dr^2 \,, \qquad
   \eta = \eta(r) \,, \qquad z = z(r) \,, \qquad \tilde z = \tilde z(r) \,,
 } 
for some function $A(r)$.  A convenient frame is
 \es{Frame}{
   e^i = Le^{A} \hat{e}^i\;, \qquad\qquad e^5 = dr \,, 
 }
where the $\hat{e}^i$, $i = 1, \ldots 4$, form a frame on the $S^4$ of unit radius. The non-zero components of the spin connection are
 \es{SpinConn}{
   \omega^{ij} = \hat{\omega}^{ij} \;, \qquad\qquad \omega^{i5} = - \omega^{5i} = LA'e^{A}\hat{e}^i\;,
  }
where $\hat \omega^{ij} $ is the spin connection on the unit radius $S^4$.

The equations of motion following from the Lagrangian \eqref{LagEuclidean} are 
\es{SecondOrder}{
&6A'' + 12A'^2 + \dfrac{4z'\tilde{z}'}{(1-z\tilde{z})^2} + \dfrac{12\eta'^2}{\eta^2} +V - \dfrac{6}{L^2}e^{-2A} =0\;,\\
& \eta'' + 4A'\eta'-\dfrac{\eta'^2}{\eta^2} - \dfrac{\eta^2}{24} \partial_{\eta} V=0\;, \\
& 4z'' + 16A'z' + \dfrac{8\tilde{z}}{1-z\tilde{z}}z'^2 - (1-z\tilde{z})^2\partial_{\tilde{z}} V=0\;,\\
& 4\tilde{z}'' + 16 A'\tilde{z}' + \dfrac{8 z}{1-z\tilde{z}}\tilde{z}'^2 - (1-z\tilde{z})^2\partial_{z} V=0\;,\\
& 12A'^2 - \dfrac{12\eta'^2}{\eta^2} - \dfrac{4z'\tilde{z}'}{(1-z\tilde{z})^2}+ V -\dfrac{12}{L^2}e^{-2A}=0\;.
 }
%

\subsection{The BPS equations}
\label{subsec:BPSeqns}

With the Ansatz \eqref{MetricAnsatz}--\eqref{Frame}, the spin-$1/2$ variations \eqref{Spin12Euclidean}--\eqref{Spin12ConjEuclidean} take the form
 \es{Spin12Simp}{
  \begin{pmatrix}
     1 + \tilde z^2 + \left(\tilde z^2 - 1\right) \eta^6 
      &  3 i L \eta \eta' (1 - z \tilde z)  \\
     -3 i L \eta \eta' (1 - z \tilde z)
      & 1 + z^2 + \left(z^2 - 1\right) \eta^6   \\
   2 (z + \tilde z) + (z - \tilde z) \eta^6 & -2 i L z' \eta^2  \\
  2 i L \tilde z' \eta^2   &   2 (z + \tilde z) - (z - \tilde z) \eta^6  
  \end{pmatrix} \begin{pmatrix}
    \epsilon_i \\
    \tilde \epsilon_i 
  \end{pmatrix} = 0 \,.
 }
This system of equations has non-trivial solutions only if the $4 \times 2$ matrix in \eqref{Spin12Simp} has rank $1$, or in other words only if all its $2 \times 2$ minors vanish.  This condition requires
 \es{BPS}{
   z' &= \frac{3 \eta' (z \tilde z -1) \left[2(z+\tilde z) + \eta^6 (z- \tilde z)  \right]}{2 \eta \left[\eta ^6\left(\tilde z^2-1\right)+\tilde z^2+1\right]} \,, \\
   \tilde z' &= \frac{3 \eta' (z \tilde z -1) \left[2 (z+\tilde z) - \eta^6 (z - \tilde z)\right]}{2 \eta \left[\eta^6 \left( z^2-1\right)+z^2+1\right]}   \,, \\
   (\eta')^2 &= \frac{ \left[\eta^6 \left(z^2-1\right)+z^2+1\right] \left[\eta^6 \left(\tilde z^2-1\right)+\tilde z^2+1\right]}{9 L^2 \eta^2 (z \tilde z-1)^2}  \,.
 }
The first equation comes from the minor constructed from the first and third row of \eqref{Spin12Simp};  the second equation comes from the second and fourth rows;  and the last equation comes from the top two rows.

Next we should consider the spin-$3/2$ variations \eqref{Spin32Euclidean}--\eqref{Spin32ConjEuclidean} in the case where the index $\mu$ points along the $S^4$ directions.  The equations take the form of the generalized eigenvalue problem
 \es{Spin32Simp}{
  \hat  \nabla_\mu \begin{pmatrix} \epsilon_i \\
  \tilde \epsilon_i 
  \end{pmatrix} = \begin{pmatrix} \frac12 L A' e^A &
    \displaystyle{\frac{i e^A}{6 } \frac{2 (1 + z^2) + \eta^6 (1-z^2 )}{(1 - z \tilde z) \eta^2} } \\
     \displaystyle{\frac{i e^A}{6} \frac{2 (1 + \tilde z^2) + \eta^6 (1- \tilde z^2)}{(1 - z \tilde z) \eta^2} }
    & 
    -\frac12 L A' e^A 
   \end{pmatrix}
   \gamma_5 \hat \gamma_\mu 
   \begin{pmatrix}
    \epsilon_i \\
  \tilde \epsilon_i
   \end{pmatrix} \;,
 }
where $\hat \gamma_\mu \equiv \hat e^m_\mu \gamma_m$, and $\hat  \nabla_\mu$ is the covariant derivative on $S^4$. 

We expect that $\epsilon_i$ and $\tilde \epsilon_i$ should be linear combinations of the Killing spinors on $S^4$ with $r$-dependent coefficients.  One way to write the Killing spinor equation is 
 \es{Killing}{
  \hat  \nabla_\mu \zeta_{\pm} = \pm \frac 12 \gamma_5 \hat \gamma_\mu \zeta_{\pm} \,.
 }
This equation has four linearly independent complex solutions for each sign.  In fact, $\zeta_+$ and $\zeta_-$ can be related through $\zeta_- = \gamma_5 \zeta_+$.  Since the equations \eqref{Spin12Simp} and \eqref{Spin32Simp} do not mix $\zeta_+$ and $\zeta_-$, let us take
 \es{epsExpansion}{
  \begin{pmatrix} \epsilon_i \\
  \tilde \epsilon_i 
  \end{pmatrix}
   = \begin{pmatrix} a_i(r) \\
  \tilde a_i(r) 
  \end{pmatrix} \zeta_\pm  \,.
 }
Then \eqref{Spin32Simp} becomes
 \es{Spin32Again}{
   \begin{pmatrix}  L A' e^A \mp 1 &
    \displaystyle{\frac{i e^A}{3 } \frac{2 (1 + z^2) + \eta^6 (1-z^2 )}{(1 - z \tilde z) \eta^2} } \\
     \displaystyle{\frac{i e^A}{3} \frac{2 (1 + \tilde z^2) + \eta^6 (1- \tilde z^2)}{(1 - z \tilde z) \eta^2} }
    & 
    - L A' e^A  \mp 1
   \end{pmatrix}
   \begin{pmatrix}
    \epsilon_i \\
  \tilde \epsilon_i
   \end{pmatrix}   = 0 \,.
 }
This equation needs to hold together with \eqref{Spin12Simp}.  Thus, to have non-trivial solutions for $\epsilon_i$ and $\tilde \epsilon_i$, one should impose two more equations in addition to \eqref{BPS}.  Constructing $2\times 2$ matrices from the first / second row of \eqref{Spin32Again} and the second / first row of \eqref{Spin12Simp}, and requiring that the determinants of those matrices vanish, we obtain
 \es{etapEqs}{
  L \frac{\eta'}{\eta} &= \frac{\left[1 + z^2 + \eta^6(z^2 - 1) \right] (L A' \mp e^{-A} )}{2 ( 1 + z^2) + \eta^6 (1 -z^2)} \,, \\
  L \frac{\eta'}{\eta} &= \frac{\left[ 1 + \tilde z^2 + \eta^6(\tilde z^2 - 1) \right] (L A' \pm e^{-A} )}{2 ( 1 + \tilde z^2) + \eta^6 (1 - \tilde z^2 )} \,.
 }

We have therefore derived five first order equations (three in \eqref{BPS} and two in \eqref{etapEqs}) for four functions ($A$, $\eta$, $z$, and $\tilde z$).  Quite remarkably, these equations are consistent with each other and with the second order equations \eqref{SecondOrder}!  Moreover, one can obtain an algebraic equation
 for $A$ by solving \eqref{etapEqs} for $\eta'$ and plugging the result into the last equation in \eqref{BPS}.  The algebraic equation is 
 \es{Algebraic}{
  e^{2 A}  =   \frac{(z \tilde z -1)^2\left[ \eta^6  \left(z^2-1\right)+z^2+1\right]  \left[\eta^6 \left(\tilde z^2-1\right)+\tilde z^2+1\right]}{\eta^8 \left(z^2-\tilde z^2\right)^2} \,,
 }
which holds regardless of the sign choice in \eqref{etapEqs}.

Putting things together, our independent BPS equations are \eqref{BPS} and \eqref{Algebraic}.  We will solve these equations numerically in the next section.

\section{Solution to the BPS equations}
\label{SOLUTION}
The BPS equations \eqref{BPS} and \eqref{Algebraic} can be solved systematically in the UV and IR asymptotic regions. In the UV, we find a  two-parameter family of solutions, parameterized by a mass (or source) parameter $\m$ and a vev-parameter $v$. Requiring smoothness of the IR solution allows for a one-parameter family of solutions. Interpolation from the IR to the UV allow us to fix $v$ in terms of $\mu$ numerically: and from the numerics, we extract an analytic formula for $v=v(\mu)$. This result is an important ingredient 
for matching the $S^4$ free energy, identified as the on-shell action in the bulk, to the same quantity as computed from the field theory.

\subsection{UV asymptotics}

In the coordinates used in the metric \eqref{MetricAnsatz}, the UV region is at large $r$, where at leading order the metric should approach $\HH^5$ (Euclidean $AdS_5$),
\be
ds^2_5 = dr^2 + L^2\sinh^2\Big(\frac{r}{L}\Big)\,ds^2_{S^4}\;.
\label{pureH5}
\ee
This means that we have $e^{2A} = \frac{1}{4}e^{2r/L} + {\cal O}(1)$ as $r \to \infty$.   We set the $AdS_5$ scale $L=1$ for simplicity; it is easily restored by sending $r \to r/L$ in all the formulas presented below.  The scalar $\eta$ approaches $1$ while $z$ and $\tilde{z}$ vanish at a rate that can be found by linearizing their BPS equations.  
Solving the BPS equations \eqref{BPS} and \eqref{Algebraic} iteratively, order by order in the asymptotic expansion as $r\to \infty$, we find 
\be
\begin{split}  
  e^{2A} 
  &=
  \dfrac{e^{2r}}{4}
  + \frac{1}{6}(\mu^2-3) 
      + \mathcal{O}\big( r^2 \,e^{-2r}\big) 
  \;, \\[2mm]
  \eta
  &=
  1  + e^{-2r} 
  \bigg[
     \dfrac{2\mu^2}{3}r+ \dfrac{\mu(\mu+v)}{3}
  \bigg] 
    + \mathcal{O}\big(r^2 \,e^{-4r}\big) 
  \;, \\
  \frac{1}{2}(z+\tilde{z})  
  &=
   e^{-2r} \Big[ 2\mu \,r +v \Big]
    + \mathcal{O}\big(r^2 \,e^{-4r}\big) 
  \;, \\
  \frac{1}{2}(z-\tilde{z})  
  &=
  \mp \mu\, e^{-r} 
  \mp  e^{-3r} \bigg[
  \frac{4}{3}\mu\big(\mu^2-3\big)\,r 
  +\frac{1}{3}\Big( 2v (\mu^2 -3) + \mu (4\mu^2 -3)\Big)
  \bigg]
      + \mathcal{O}\big(r^2 \,e^{-5r/L}\big) 
  \,.
\end{split}  
\label{UVsol}
\ee
Here $\mu$ and $v$ are integration constants, and the choice of sign in the last equation corresponds to a choice of sign in \eqref{etapEqs}. 
We emphasize that $z$ and $\tilde{z}$ are \emph{not} each other's conjugates because the model is Euclidean. 

\subsection{IR asymptotics}

One expects that at some value $r = r_*$ of the radial coordinate, the $S^4$ shrinks to zero size.   We can also solve the BPS equations approximately close to $r=r_*$, where we require that the solution is smooth.  Specifically, the warp factor $e^{2A}$ starts out as $(r-r_*)^2$ for small $r-r_*$, while the scalars approach constant values. Taking $\eta = \eta_0$ at $r=r_*$ for some constant $\eta_0$, the BPS equations imply that both $z$ and $\tilde{z}$ approach constant values determined by $\eta_0$. The BPS equations can be solved successively for higher powers in small $r-r_*$; since the BPS equations are invariant under flipping the sign of $r-r_*$, the expansion only depends on even powers of $r-r_*$.  We find
\begin{equation}
\begin{split}
e^{2A} &= (r-r_*)^2 + \dfrac{7\eta_0^{12}+20}{81 \eta_0^4}\,(r-r_*)^4 
+  \mathcal{O}\big((r-r_*)^6\big)\;, \\[3mm]
\eta &= \eta_0 -
\left(\dfrac{\eta_0^{12}-1}{27\eta_0^3}\right) (r-r_*)^2
\bigg[ 1-
\left(\dfrac{85+131\eta_0^{12}}{810 \eta_0^4}\right)\, (r-r_*)^2
+  \mathcal{O}\big((r-r_*)^4\big)
\bigg]
\;, \\[3mm]
\frac{1}{2}(z+\tilde{z})  
  &=\ds\sqrt{\dfrac{\eta_0^6-1}{\eta_0^6+1}} 
  \Bigg[
  \frac{\eta_0^6}{\eta_0^6+2} \,
  - \frac{2 \eta_0^8 (4 \eta_0^6 +5)}{15(\eta_0^6 +2)^2} \,(r-r_*)^2
   + \mathcal{O}\big((r-r_*)^4\big)
  \Bigg]\,,\\[3mm]
\frac{1}{2}(z-\tilde{z})  
  &=\mp \ds\sqrt{\dfrac{\eta_0^6-1}{\eta_0^6+1}} \,
  \Bigg[
  \frac{2}{\eta_0^6+2} 
  +\frac{ \eta_0^2 (3 \eta_0^{12}- 10 \eta_0^6 -20)}{15(\eta_0^6 +2)^2}\, (r-r_*)^2
   + \mathcal{O}\big((r-r_*)^4\big)
  \Bigg]  \,.
\end{split}
\label{IRsol}
\end{equation}
Here, $\eta_0 $ and $r_*$ are the only free parameters, and the sign in the last equation is correlated with the choice of sign in \eqref{etapEqs}. We have determined the IR expansion up to $O((r-r_*)^{14})$, but we only display the first few terms here.

\subsection{Matching UV onto IR}

\begin{figure}[t]
\begin{center}
\includegraphics[width=\textwidth]{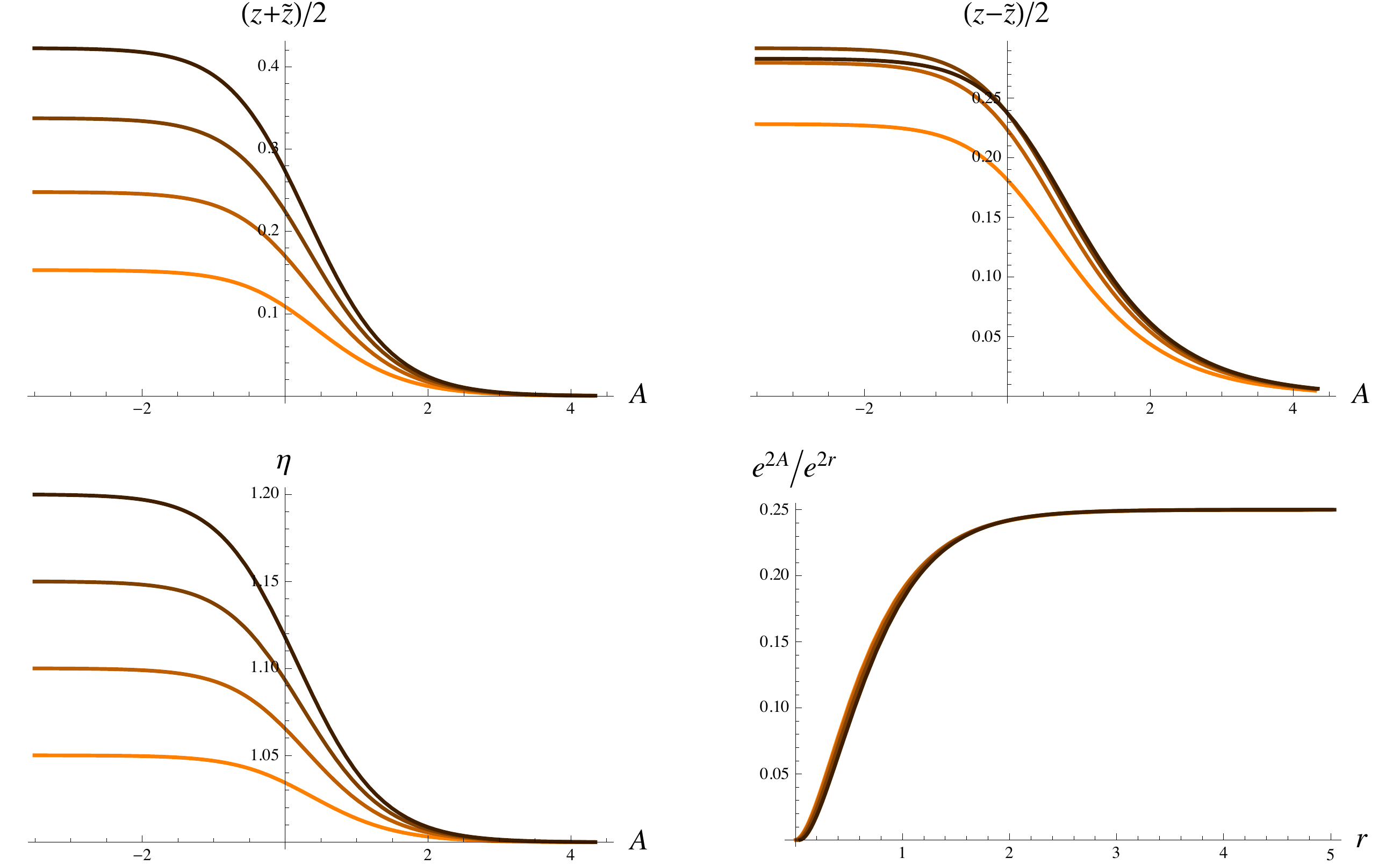}

\caption{Plots of the numerical solutions for $A(r)$, $\eta(r)$, and $\frac{1}{2}\big(z(r)\pm\tilde{z}(r)\big)$ for $\eta_0=\{1.05,1.10,1.15,1.20\}$ (orange to black).  The functions $z$ and $\tilde z$ are real in this case.  Note that the scalar fields are plotted as a function of $A$ as defined in \eqref{MetricAnsatz} and not as a function of the radial coordinate $r$.   \label{numsolnRe}}
\end{center}
\end{figure}

\begin{figure}[t]
\begin{center}
\includegraphics[width=\textwidth]{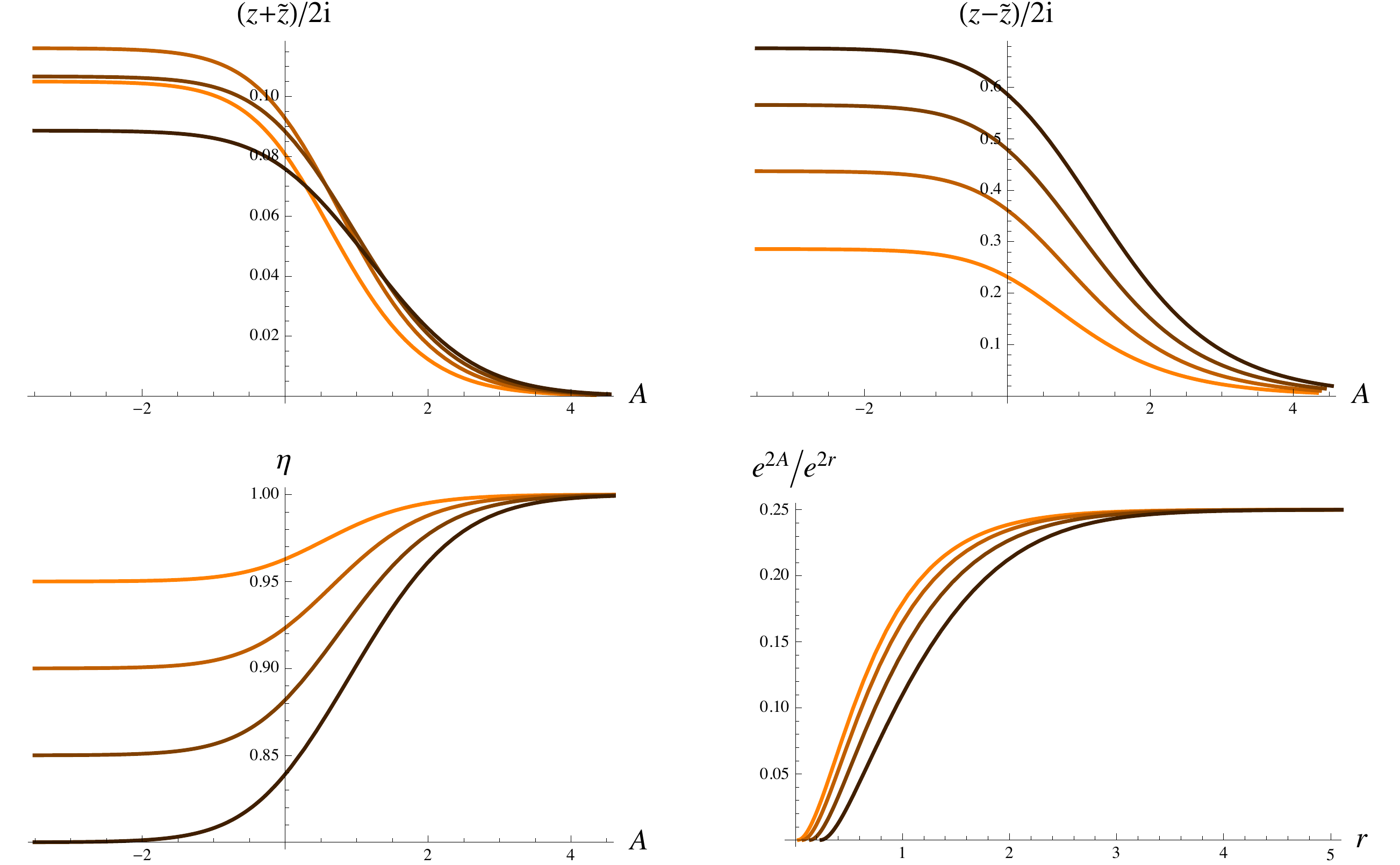}

\caption{Plots of the numerical solutions for $A(r)$, $\eta(r)$, and $\frac{1}{2i}\big(z(r)\pm\tilde{z}(r)\big)$ for $\eta_0=\{0.95, 0.90, 0.85, 0.80\}$ (orange to black).     The functions $z$ and $\tilde z$ are pure imaginary in this case. Again the scalar fields are plotted as a function of $A$ as defined in \eqref{MetricAnsatz}.  \label{numsolnIm}}
\end{center}
\end{figure}

From now on we will focus on solving the BPS equations corresponding to the lower choice of signs in \eqref{Killing}--\eqref{IRsol}.  One can obtain the solutions corresponding to the upper choice of signs by simply interchanging $z$ with $\tilde z$.

The BPS equations can be solved numerically over the whole range of $r$. In doing so, it is convenient to use the fact that these equations are invariant under shifting $r$ by a constant, and set $r_* = 0$.  The IR solution \eqref{IRsol} then has only one free parameter $\eta_0$.  One can integrate  the BPS equations numerically by shooting from near $r=0$ with input parameter $\eta_0$ towards the UV at $r\to \infty$.  After obtaining this solution, one can shift back $r \to r + r_*$ and compare the numerical solution to the UV asymptotics \eqref{UVsol}, from which one can extract the functions $r_*(\eta_0)$, $\mu(\eta_0)$, and $v(\eta_0)$.  

As can be seen from the IR asymptotics \eqref{IRsol}, when $\eta_0>1$ the functions $z(r)$ and $\tilde z(r)$ are both real, while for $\eta_0<1$, $z(r)$ and $\tilde z(r)$ are pure imaginary.  In both cases, $A(r)$ and $\eta(r)$ are real.  See Figure~\ref{numsolnRe} for a few examples of numerical solutions in the case $\eta_0>1$ and Figure~\ref{numsolnIm} for a few examples in the case $\eta_0<1$.  Note that $e^{2A}$ approaches $e^{2r}/4$ at large $r$ and that it vanishes at some radial coordinate $r_*(\eta_0)$.

\begin{figure}[t]
\begin{center}
\includegraphics[width=\textwidth]{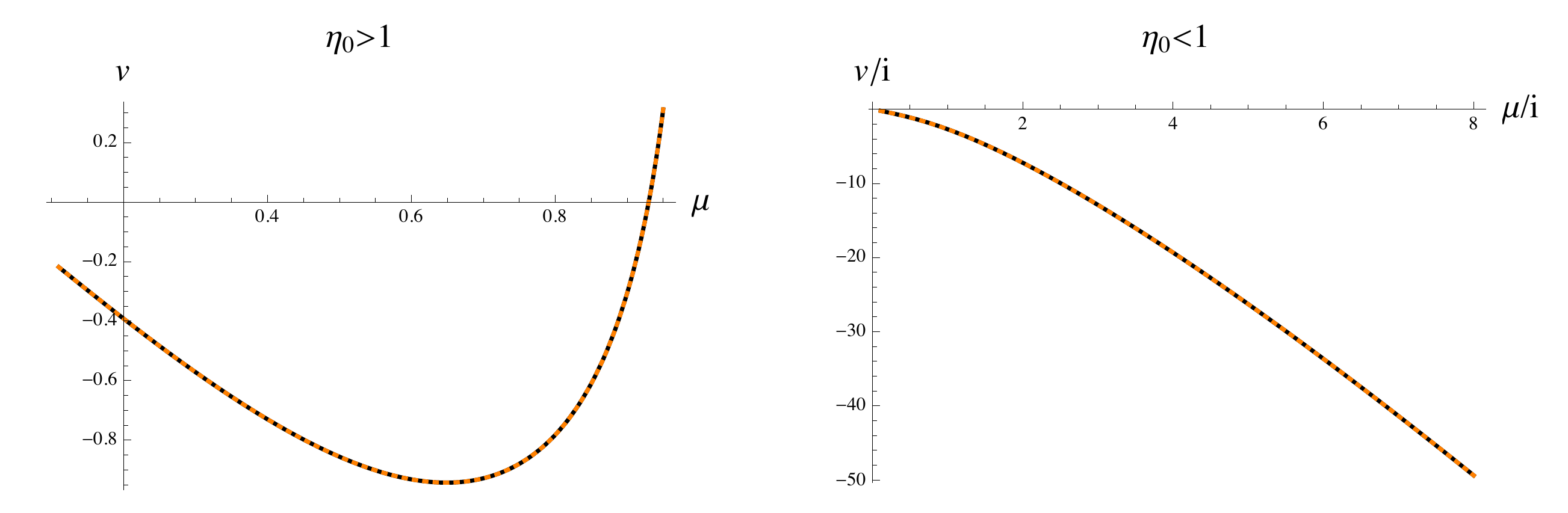}

\caption{$v(\mu)$ as a function of $\mu$ for both $\eta_0>1$ (left) and $\eta_0<1$ (right).  The orange curve is obtained numerically, while the black curve is a plot of the analytical relation \eqref{vofmu}.  Note that for $\eta_0<1$ both $\mu$ and $v$ are pure imaginary. \label{vPlot}}
\end{center}
\end{figure}

From the numerics, we were able to extract the following relation between $v$ and $\mu$:
 \es{vofmu}{
   v(\mu) = -2 \mu - \mu \,\log(1-\mu^2) \,.
 }
See Figure~\ref{vPlot}.  In the next section we will use this relation to show that the $S^4$ free energy of our solutions matches the corresponding quantity as computed from field theory.

\section{Calculation of the free energy}   
\label{HOLOREN}

In the AdS/CFT correspondence, the free energy of the field theory is encoded in the on-shell action. However, the action integral evaluated on a classical solution diverges at large values of the radial coordinate. The method of holographic renormalization is a systematic  technique to determine the \emph{infinite} counterterms needed to extract finite predictions for field theory observables. These counterterms are universal. They must cancel divergences for all solutions of the equations of motion of a given bulk theory, not just the BPS solutions.  However, the procedure leaves open the possibility of \emph{finite} counterterms, which can be important because the radial cutoff used is not necessarily compatible with supersymmetry. Incompatibility can be detected within the gravity dual if the vacuum energy of a Lorentz invariant, BPS state fails to vanish.  This situation was first encountered in \cite{Bianchi:2001de} and more recently in \cite{Freedman:2013oja}. The second case is a close analogue of the present study; it involved four-dimensional Euclidean domain wall solutions of extended supergravity, dual to deformations of the ABJM theory on $S^3$. The extra finite counterterm found in \cite{Freedman:2013oja} was essential to the agreement between the supergravity results and the dual field theory. Thus holographic renormalization is a necessary preliminary to the extraction of the free energy.  In this section we summarize the procedure of holographic renormalization for the action of interest. A more detailed and systematic treatment is presented in Appendix \ref{app:holoren}. The calculation of the infinite counterterms closely follows \cite{Bianchi:2001kw,Bianchi:2001de}, but the derivation of the finite counterterm is considerably more subtle than in previous cases.

The starting point is the 5D bulk action\footnote{We will temporarily set $4\pi G_5=1$ to reduce clutter in the formulas below. We will restore this important normalization factor later in this section.}
\be
  S= S_\text{5D}+S_\text{GH} 
  =
   \int d^5 x \,\sqrt{G} \, \mathcal{L}_\text{5D}
   -\frac{1}{2} \int_{\pa M} \sqrt{\gamma}\, \mathcal{K}
 \,,
 \label{Sbulk}
\ee
where ${\cal L}_\text{5D}$ was given in \eqref{LagTrunc}, $S_\text{GH}$ is the Gibbons-Hawking term and $\mathcal{K}$ is the trace of the second fundamental form. 
We rewrite the Lagrangian in \reef{LagTrunc} in terms of canonically normalized fields by writing
\be
  \eta = e^{\phi/\sqrt{6}}\,,~~~~~~
  z = \frac{1}{\sqrt{2}} \big( \chi + i \psi \big)\,,~~~~~~
  \tilde{z} = \frac{1}{\sqrt{2}} \big( \chi - i \psi \big)\,,
  \label{fieldredef}
\ee
 and obtaining
\be
  S_\text{5D}=\int_M d^5x\,\sqrt{G} \,
  \bigg\{ 
     -\frac{1}{4} R 
     + \frac{1}{2} (\pa \phi)^2
     + K  \Big(  \frac{1}{2} (\pa \chi)^2 +  \frac{1}{2} (\pa \psi)^2  \Big)
     + V
  \bigg\} 
  \,,
  \label{5Daction2-MT}
\ee
with $K = \big(1 - \frac{1}{2}(\chi^2 + \psi^2)\big)^{-2}$. The contraction of the 5D Einstein equation gives an expression for the Ricci scalar $R$, 
\be
  R = 2 K ((\pa \chi)^2 +  (\pa \psi)^2) + 2 (\pa \phi)^2 + \frac{16}{3} V\,,
  \label{trR}
\ee
and using this expression in the action \reef{5Daction2-MT},  we find 
$
S_\text{5D} \to \int_M d^5x\,\sqrt{G} \big( -\frac{1}{3} V \big)\,.
$
This simple result conceals considerable detail, as we will see.

To facilitate the near-boundary analysis, the 5D metric is written in Fefferman-Graham form as  
\be\lab{5dmetricMain}
  ds^2 ~=~ G_{\mu\nu} dx^\m dx^\n
  ~=~
  \frac{d\rho^2}{4\rho^2} + \frac{1}{\rho}\,g_{ij}(x,\rho)\, dx^i dx^j\,.
\ee
In terms of the radial variable $\r$, related by $\r= e^{-2r}$ to the radial coordinate $r$ used in Sections~\ref{sec:supergravity} and~\ref{SOLUTION}, the $AdS_5$ boundary is at $\r=0$. Note that we have fixed the scale of $AdS_5$ (or $\mathbb{H}^5$) 
 by setting $L=1$ in \eqref{pureH5}.  The fields of a general solution of equations of motion behave near the boundary  as
\be
   \begin{split}
  g_{ij} &=~ g_{0\,ij} + \rho\, g_{2\,ij}
    + \rho^2 \big[
    g_{4\, ij}
    + h_{1\,ij}\,\log\rho
    + h_{2\,ij}\,(\log\rho)^2
    \big] + \dots\,,\\[2mm]
   \phi &=~ 
   \rho\log\rho \big( \phi_0 + \phi_2\, \rho + \phi_4 \,\rho \log\rho \big) 
   + \rho \big( \tilde{\phi}_0 + \tilde{\phi}_2 \, \rho \big) + \dots\,,\\[2mm]
   \chi &=~ 
   \rho\log\rho \big( \chi_0 + \chi_2\, \rho + \chi_4 \,\rho \log\rho \big) 
   + \rho \big( \tilde{\chi}_0 + \tilde{\chi}_2 \, \rho \big) + \dots\,,\\[2mm]
   \psi &=~
   \psi_0\, \rho^{1/2} + \psi_2 \,\rho^{3/2} \log\rho 
   + \tilde{\psi}_0\,\rho^{3/2} + \dots \,.
 \end{split}
 \label{fieldexp-MT}
\ee
The independent data for the  scalar fields are the non-normalizable modes (or ``sources")  
$\phi_0$,  $\chi_0$, $\psi_0$ and the normalizable modes (or ``vevs") 
$\tilde{\phi}_0$, $\tilde{\chi}_0$, $\tilde{\psi}_0$.\footnote{The words ``sources" and ``vevs" are used rather imprecisely here. As we will show later, the renormalized one-point functions involve both the ``source" and ``vev" coefficients.}  Since we are interested in  $S^4$-invariant solutions,  we choose the  boundary metric $g_{0\,ij}$ to describe a round four-sphere with radius $1/2$,  
\be   \lab{gzero}
g_{0ij} = \frac{1}{4} g_{\text{unit,}ij}\,.  
\ee
This value of the radius is compatible with the asymptotic normalization chosen in \eqref{UVsol}. The asymptotic equations of motion determine the non-leading coefficients in \eqref{fieldexp-MT} in terms of the independent data. The results for the first few subleading 
coefficients are given in \reef{g2}, \reef{psi2}, \reef{trh2}, and \reef{trg4}.   Non-asymptotic information  on the solution, such as a regularity condition in the interior, is needed to relate ``source" and ``vev" coefficients.

The BPS equations are first order.  The asymptotic data of a BPS solution are of course compatible with (\ref{fieldexp-MT}), but contain fewer independent coefficients.
From the   UV expansion \reef{UVsol}  we see that the parameter $\mu$ determines  the ``sources"
\be
  \psi_0 = - i \sqrt{2} \m\,, \qquad
  \phi_0 =- \sqrt{\frac{2}{3}} \mu^2\,, \qquad
  \chi_0 = -\sqrt{2} \mu\,,
  \label{sub0s}
\ee
and the second parameter $v$ determines the ``vevs".  In the full solution in Section \ref{SOLUTION},  
we required regularity in the interior,  so that $v$ becomes the function of $\mu$ in (\ref{vofmu}).

The next steps in the holographic renormalization procedure are
\begin{enumerate}
\item Insert the general form of an asymptotic solution \eqref{fieldexp-MT} in the action \eqref{Sbulk} with the radial integral  $\int d^5 x  \to \int d^4 x \, \int_\eps d\rho$ cutoff near the boundary, $\rho=\epsilon \to 0$, and $S_\text{GH}$  evaluated at $\e$.  After integration one finds  a set of $1/\eps^2$, $1/\eps$, and $\log\eps$ divergences whose coefficients are given solely in terms of the ``sources", $g_0$, $\psi_0$, $\chi_0$, and $\phi_0$. 
\item Invert the field expansion  (\ref{fieldexp-MT})  and express the divergences in the action \eqref{Sbulk} in terms of the bulk fields evaluated at the cutoff surface $\rho = \epsilon$ and in terms of the induced metric $\g_{ij} =g_{ij}/\e$ (as opposed to expressing these divergences in terms of the asymptotic coefficients that appear in \eqref{fieldexp-MT}).
\item
The  counterterm action  that should be added to the action \eqref{Sbulk} is simply the negative of the divergences found in step 2. We find that the result is
\bea
  \label{CTeom-MT}
 S_\text{ct}
 \!&\!=\!&\!
  \int_{\pa M_\eps} d^4x \, \sqrt{\gamma}\,
  \bigg[
  \frac{3}{2}
  +  \frac{1}{8}R[\gamma] 
  +\frac{1}{2} \psi^2 
   +\Big( 1 + \frac{1}{\log \eps}\Big) \big( \phi^2 + \chi^2 \big)
     \\[1mm]
   && \hspace{1.1cm}
   -\log\eps \, 
    \bigg\{
      \frac{1}{32}
      \Big[ R[\gamma]^{ij} R[\gamma]_{ij} - \frac{1}{3} R[\gamma]^2 \Big]
      + \frac{1}{4} \psi \Box_\gamma \psi
      -\frac{1}{24} R[\gamma]\,\psi^2
      -\frac{1}{6} \psi^4\bigg\}
  \bigg] \,,
   \nonumber
\eea
where $R[\gamma]_{ij}$ and $R[\gamma]$ are the Ricci tensor and Ricci scalar, respectively, of the induced metric $\gamma_{ij}$.

\end{enumerate} 

There are other five-dimensional holographic flows in the literature that involve supergravity scalars dual to dimension 2 and 3 operators in the dual field theory,  such as the GPPZ \cite{Girardello:1998pd}, FGPW \cite{Freedman:1999gp}, Coulomb branch \cite{Freedman:1999gk}, and Pilch-Warner \cite{Pilch:2000ue} flows.
It is interesting to note that, when expressed using canonically normalized scalars, all terms in the counterterm action \eqref{CTeom-MT}, except the final $ \psi^4 \log\eps$, appear in the same form with the same coefficients in these models.  Only the last term in \eqref{CTeom-MT} is model-dependent in the sense that its coefficient (here $1/6$) is sensitive to details of the scalar potential.

Let us now consider finite counterterms. 
If supersymmetry is preserved in the vacuum state of a supersymmetric field theory in flat space, the vacuum energy  must vanish.  This means that the renormalized on-shell action of the dual gravity theory must vanish when the boundary metric is Lorentz invariant and operator sources are constant on the boundary.  This criterion may be tested when the dual supergravity theory has flat-sliced BPS domain walls, i.e.~solutions with metric  $ds^2 = dr^2 + e^{2A(r)}\d_{ij} dx^idx^j$, that are controlled by a superpotential.  In five-dimensional supergravity, the superpotential is a real function of the fields. For a theory with several real scalars
$\phi^i$ and target space metric $K_{ij}(\phi) $, the superpotential $W(\phi)$ is related to the potential $V(\phi)$
by
\be \lab{wvcanon}
V= \frac12  K^{ij}\pa_i W\, \pa_j W-\frac43 W^2 
\,.
\ee
In this case the BPS equations of flat sliced walls take the form  (see \cite{D'Hoker:2002aw} or Ch.~23 of \cite{Freedman:2012zz}) of simple gradient flow equations that are compatible with the Lagrangian equations of motion. Further,  the action integral for flat sliced solutions can be rearranged by the Bogomolnyi  maneuver into the form 
\be  \lab{actbogo}
S = \int d^4x\int^{r_0}\!\bigg( e^{4A}\bigg[-3 \Big(A' -\frac23 W\Big)^2 + \frac12 K_{ij}\big(\phi^{i}{}'- K^{il}\pa_l W\big)\big(\phi^{j}{}'-K^{jm} \pa_m W\big) \bigg]
-\frac{d}{dr}\big(e^{4A} W\big)\bigg)\,,
\ee
where $r_0$ is a UV cutoff.  When the flow equations  (e.g.~$A' =\frac23 W$ and  $\phi^{i}{}' = K^{il}\pa_l W$) 
are satisfied, i.e.~for a BPS solution,  the on-shell action vanishes, except for the surface term evaluated at the cutoff $r_0$. Supersymmetry requires this term to be cancelled, so one must add to the action a supersymmetry counterterm, 
\be \lab{ssusygen}
S_W = \int d^4x \,e^{4A(r_0)} \,W\big(\phi_i(r_0)\big) \,.
\ee
This surface term contains the infinite counterterms of (\ref{CTeom-MT}) (evaluated for the $r$-dependent fields of flat-sliced domain walls) plus  finite counterterms needed for supersymmetry (if any) plus terms which vanish as $r_0\to \infty$. 

There is one problem with this scenario for our model; there is no superpotential $W$ that obeys \eqref{wvcanon} for our potential \reef{LagTrunc} in its complete form with three scalars $\phi$, $\chi$, $\psi$. 
The reason is that the integrability condition needed to convert the BPS equations (\ref{BPS})  into gradient flow form is not satisfied.  (We show this in Appendix \ref{app:noW}.)
Alternatively, one can show that flat-sliced solutions of the BPS equations with all three scalars turned on do not satisfy the equations of motion. 

We now show how to overcome the problem of not having an exact superpotential. The strategy is first to study two consistent truncations of our model which do have planar domain walls and 
superpotentials.\footnote{In Appendix \ref{app:analytic} we give the analytic solution of the BPS flow equations 
with $\mathbb{R}^4$ slicing for both truncations.} Second, we show that an approximate superpotential is sufficient for the analysis. 
Let us begin with the two truncated models:%
\begin{itemize}
\item Set $\chi(r) \equiv 0$ and retain $\eta(r),~\psi(r)$.  In this truncation our model reduces to the truncation of $\cn=8$ supergravity studied by Pilch and Warner \cite{Pilch:2000ue}.  The superpotential, expanded to the order needed to include all infinite and finite terms as $r_0\to\infty$, is
\be \lab{supota}
 W_a =       \frac{3}{2}
        + \phi^2 + \frac{1}{2}\psi^2
        + \sqrt{\frac{2}{3}} \phi\, \psi^2
        + \frac{1}{4} \psi^4 \;.
\ee          
The first three terms contribute divergent counterterms when $W_a$ is inserted   in  (\ref{ssusygen}).
 In Appendix \ref{s:bogo}, we determine the UV behavior of BPS domain wall solutions and show that the infinite terms of (\ref{ssusygen})   agree with  (\ref{CTeom-MT}).  The last term of (\ref{supota}) gives  the extra finite counterterm required by supersymmetry.  

\item Set $\psi(r)\equiv 0$ and retain $\eta(r),~\chi(r)$.  This truncation of our model appears to be new. 
The exact superpotential is expanded as 
\be \lab{supotb}
W_b = \frac{3}{2}
        + \phi^2 
        +    
        \frac{1}{2}     
        \chi^2\;,
        \ee
and contributes divergent terms in (\ref{ssusygen}).  In Appendix \ref{s:bogo2}, we show that these are in agreement with \reef{CTeom-MT}.
In this truncation there is no residual finite counterterm. 
  \end{itemize} 
  
The results in the two truncations are relevant to our complete model because the planar domain wall solutions for each of the two truncations are also solutions of the equations of  motion of the complete model.  In this spirit, we note that the union of 
$W_a$ and $W_b$, namely 
\be \lab{aunionb}
W_{a\cup b} =    \frac{3}{2}
        + \phi^2 + \frac{1}{2}\psi^2 + \frac12 \chi^2 
        + \sqrt{\frac{2}{3}} \phi\, \psi^2
        + \frac{1}{4} \psi^4\,,
\ee          
provides an \emph{approximate}  superpotential\footnote{The $\Z_2$ reflection symmetries in $\chi,~\psi$ forbid ``mixed" terms such as $\chi\psi^2$. Other terms, such as $\chi^2\psi^2$, are negligible at the boundary.} for the potential of the complete model when expanded to the order necessary to produce all the divergent counterterms of (\ref{CTeom-MT}).  Specifically,  $W_{a\cup b}$ is related to 
\be \lab{vtrunc}
V_{a\cup b} =
   -3 - 2 \phi^2 - 2 \chi^2 -\frac{3}{2} \psi^2 - \frac{1}{2} \psi^4 \;,
\ee  
by (\ref{wvcanon}) (with the target space metric  $K$ in \eqref{5Daction2-MT}) if we drop terms that are higher order in the fields and therefore vanish as $r_0\to \infty$.  $V_{a\cup b}$ is the expansion of the exact  potential of (\ref{LagTruncMainText2}) with asymptotically negligible terms dropped.  It has already been shown that 
(\ref{ssusygen}) with $W_{a\cup b}$ inserted reproduces the correct infinite 
counterterms of planar  BPS domain walls of the two truncations, but the additional finite term 
\be \lab{sfinite}
S_\text{finite} =  \int d^4x\, \sqrt{\gamma} \,\frac{1}{4} \psi^4 
= \int d^4x\, \sqrt{g_0} \,\frac14\psi_0^4 \;,
\ee
is required by supersymmetry.  

Universality then implies that this term must be included for all solutions of the equations of motion of the full theory,  and therefore for the $S^4$-sliced domain walls of interest here. The major conclusion of this argument is that the renormalized action
\be\lab{satlast}
  S_\text{ren}
  ~=~ 
  S_\text{5D}+S_\text{GH}+ S_{\text{ct}}+ S_\text{finite} 
  \;,
\ee
with the actions as in \eqref{Sbulk}, \eqref{5Daction2-MT}, \eqref{CTeom-MT}, and \eqref{sfinite}, should be used to derive the renormalized free energy of our $S^4$-sliced BPS solutions. Alternatively, we can write 
\be\lab{satlast2}
  S_\text{ren}
  ~=~
  S_\text{5D}+S_\text{GH}+ S_\text{susy}\,,
  ~~~~\text{with}  ~~~~
  S_\text{susy}
  =
   \int d^4x\, \sqrt{\gamma} \,W_{a\cup b}\,.
\ee
The two forms of $S_\text{ren}$ in \reef{satlast} and \reef{satlast2} are equivalent up to terms that vanish as $r_0 \to \infty$.

We now return to our main task, namely the calculation of the free energy $F$. We show in Appendix \ref{s:freeE} that the derivative of $F$ with respect to the common source parameter $\mu$ of  the asymptotic fields is\footnote{Here we have restored the factor of $1/4\pi G_{5}$ in the normalization of the five-dimensional supergravity  action. When this factor is expressed in terms of the ten-dimensional Newton constant in type IIB supergravity compactified on $S^5$ one finds $1/4\pi G_{5}=N^2/2\pi^2$, where $N$ is the number of units of D3-brane flux, or equivalently the rank of the gauge group in the dual $\mathcal{N}=4$ SYM theory \cite{Bianchi:2001de}.} 
\be   
   \frac{dF}{d\mu}
   =
   \frac{N^2}{2\pi^2}
  \int d^4 x\, \sqrt{g_0}
  \bigg( 
  \< \mathcal{O}_\psi \> \frac{\pa \psi_0}{\pa\mu}
  + \< \mathcal{O}_\phi \> \frac{\pa \phi_0}{\pa\mu}
  + \< \mathcal{O}_\chi \> \frac{\pa \chi_0}{\pa\mu}
  \bigg)\,.~
  \label{dFdmu-MT}
\ee
The one-point functions $\< \mathcal{O}\>$  in the dual field theory are computed by taking derivatives of the action  
$S_\text{ren}$ in \reef{satlast2} with respect to the sources. Holographic renormalization ensures that these one-point functions are finite.
For example, the one-point function of the dimension-three  operator is\footnote{Without the finite counterterm $S_\text{finite}$, the one-point function $\<\mathcal{O}_\psi\>$ would have included an additional term $-\psi_0^3$.}
\be
   \<\mathcal{O}_\psi\> = \lim_{\eps\to 0}\frac{1}{\eps^{3/2}} \frac{1}{\sqrt{\gamma}}
   \frac{\delta S_\text{ren}}{\delta \psi}
   =   
   - 2 \psi_2 -2\tilde{\psi}_0 \,.
   \label{O3-MT}
\ee
The holographic calculation of the one-point function for dimension-two operators requires an extra $\log\eps$ factor, so one finds
\be
   \<\mathcal{O}_\phi\> = \lim_{\eps\to 0}\frac{\log\eps}{\eps} \frac{1}{\sqrt{\gamma}}
   \frac{\delta S_\text{ren}}{\delta \phi}
   = 2 \tilde{\phi}_0\,.
\label{Ophi-MT}
\ee
Similarly, $\<\mathcal{O}_\chi\>= 2 \tilde{\chi}_0$.

Now we use the asymptotic data \reef{sub0s} and 
\reef{psi0} for our solution to
express the one-point functions as
\begin{eqnarray} 
\< O_\psi \> \frac{\pa \psi_0}{\pa\mu} &=& 
 \frac{N^2}{2\pi^2}
 \Big(
4 \mu + \frac{8}{3}\mu^3 - 8v(\mu) + \frac{8}{3} \mu^2 \, v(\mu)
\Big)\,,
\notag\\   
\< O_\phi \> \frac{\pa \phi_0}{\pa\mu} &=& 
 \frac{N^2}{2\pi^2}
 \Big(
-\frac{8}{3} \mu^3 - \frac{8}{3} \mu^2 \, v(\mu) \Big) \,,\\
\< O_\chi \> \frac{\pa \chi_0}{\pa\mu} &=& 
 \frac{N^2}{2\pi^2}
 \Big( -4v(\mu) \Big)\,. \notag
\label{new1point}
\end{eqnarray}
Adding these three expressions  to obtain the free energy \reef{dFdmu-MT}, we note that the 
$\mu^3$ and $\mu^2\,v(\mu)$ terms cancel, so that 
\be
 \frac{dF}{d\mu}
 ~=~
 \frac{N^2}{2\pi^2}\,
 \text{vol}_0(S^4)
 \Big(
 4 \mu -12v(\mu)
 \Big)
   ~=~ 
  N^2 \Big(\,\frac{1}{3} \mu -v(\mu)\Big)  
 \,.\label{resdFdmu-MT}
\ee
The volume factor $\text{vol}_0(S^4)$ is produced by the integral in \reef{dFdmu-MT}. In the last step we used that this is the volume of the round four-sphere described by the metric $g_0$. It has  radius $1/2$, so 
\be
  \text{vol}_0(S^4) = \frac{1}{2^4} \times \frac{8\pi^2}{3} = \frac{\pi^2}{6}\,.
\ee

As we discussed in the Introduction, we must take three derivatives of the free energy \reef{FFieldTheory} to obtain an unambiguous result in the field theory. Taking two more $\mu$-derivatives of $dF/d\mu$ in \reef{resdFdmu-MT}, the linear term in $\mu$ is eliminated and we find
\be
  \label{d3F}
  \frac{d^3F}{d\mu^3} 
  ~=~ 
  - N^2 \, v''(\mu) 
  ~=~
  -2 N^2 \,\frac{\mu\, (3 -\mu^2)}{(1-\mu^2)^2}
  \,.
\ee
In the second step, we used \reef{vofmu} to evaluate $v''(\mu)$. 
The result  \reef{d3F} for $d^3F/d\mu^3$ 
exactly matches the field theory result \reef{d3FFieldTheory} after identifying  $\mu = \pm i ma$.

We end this section with a few of comments on the match with the field theory. First, note that without the finite counterterm $\frac{1}{4} \psi^4$ provided by the supersymmetric 
counterterm \eqref{sfinite}, the coefficient of the $\mu^3$ term in $\< O_\psi \>$ would have been $-\frac{4}{3}\mu^3$ and thus the cubic terms in $\mu$ would not have cancelled in $dF/d\mu$. This term would then survive in $d^3F/d\mu^3$ and create a mismatch with the field theory result \eqref{d3FFieldTheory}. Second, we argue in Appendix \ref{s:freeE} that  finite counterterms cannot contribute any $v$-dependence to $dF/d\mu$; they can only contribute to the $\mu$ or $\mu^3$ terms. Thus even without computing the finite counterterm $\frac{1}{4} \psi^4$ required by  supersymmetry, we have a perfect match of $d^5F/d\mu^5$ with the field theory result.

\subsection{Further comments}

The match between the field theory expression for $d^3 F / d\mu^3$ and our holographic computation is related to the fact that on general grounds in a supersymmetric theory $d^3 F / d\mu^3$ is independent of the regularization scheme, as long as this scheme does not itself break supersymmetry.  (If the renormalization scheme breaks supersymmetry, then $d^3 F / d\mu^3$ would be scheme-dependent, but $d^5 F / d\mu^5$ would still be universal.)  That $d^3 F / d\mu^3$ is free of renormalization-scheme ambiguities can be shown through the following argument.  If one studies a superconformal field theory on $S^4$ in the presence of a small distance cutoff $\epsilon$, the free energy takes the form
 \es{FS4SCFT}{
  F = \alpha_{2} \frac{a^2}{\epsilon^2} + \alpha_0 - a_\text{anom} \log \frac{a}{\epsilon} + {\cal O}(\epsilon/a) \,,
 }
where the coefficients $\alpha_{2}$ and $\alpha_0$ multiply non-universal UV divergences,\footnote{In a non-supersymmetric theory, $F_{S^4}$ would also contain a more singular non-universal UV divergent contribution $\alpha_4 a^4 / \epsilon^4$.  In a supersymmetric theory, however, the coefficient $\alpha_4$ vanishes provided that one employs a supersymmetric regularization scheme.} and $a_\text{anom}$ is the $a$-anomaly coefficient, which is universal.  For instance, in the case of ${\cal N} = 4$ SYM, a free field computation shows that $a_\text{anom} = N^2-1$.  For our ${\cal N} = 2^*$ deformation of the ${\cal N} = 4$ theory, the $S^4$ free energy is not only a function of the radius of the sphere $a$ and the UV cutoff $\epsilon$, but also of the mass parameter $m$.  The coefficients $\alpha_2$ and $\alpha_0$ in \eqref{FS4SCFT} can now depend on the dimensionless combination $m^2 \epsilon^2$.  At small $\epsilon$, we can expand $\alpha_2 = \tilde \alpha_2 + m^2 \epsilon^2 \beta_2 + O(m^4 \epsilon^4)$ and $\alpha_0 = \tilde \alpha_0 + O(m^2 \epsilon^2)$, for some constants $\tilde \alpha_2$, $\tilde \alpha_0$, and $\beta_2$ that are renormalization scheme-dependent.   The non-universal contributions to $F_{S^4}$ then take the form 
 \es{FNonConf}{
  \tilde \alpha_2 \frac{a^2}{\epsilon^2} + \tilde \alpha_0 + \beta_2 m^2 a^2 \,.
 }
It follows that the quantity 
 \es{d3FAgain}{
  \frac{d^3 F}{d (ma)^3} \;,
 }
is non-ambiguous, because after taking three derivatives with respect to $ma$, the non-universal contribution \eqref{FNonConf} vanishes.  Consequently, if we identify $\mu = \pm i m a$, we conclude that $d^3 F / d\mu^3$ is non-ambiguous in a supersymmetric theory.

Notice that the free energy displayed in \eqref{FFieldTheory} has a branch cut singularity when $m^2a^2 =-1$.\footnote{It was noticed in \cite{Pestun:2007rz,Okuda:2010ke} that precisely at this mass value there are cancellations in the supersymmetric localization computation. In the large $N$ limit and at large 't Hooft coupling it can be seen that the free energy vanishes. We thank J.~Russo and K.~Zarembo for comments on this issue.}  Restricting to pure imaginary values of $ma$, one can understand this singularity as the onset of a tachyonic instability, where the field theory path integral diverges.  To get a feel for how this singularity arises, one can consider the theory of a free complex scalar $Z = (A + i B) / \sqrt{2}$ with the same mass as the complex scalars $Z_1$ and $Z_2$ in our ${\cal N} = 2^*$ SYM theory as given in \eqref{ConformalMass}--\eqref{MassS4}.  In other words, the mass term in the $S^4$ Lagrangian for the complex scalar $Z$ is:\footnote{Different squared masses for $A$ and $B$ are to be expected for a supersymmetric field theory on $S^4$. A similar situation occurs for the chiral multiplet on $AdS_4$, see \cite{Breitenlohner:1982jf}.}
\begin{equation} \lab{Lscalar}
\begin{split}
\cl_{\rm scalar} &=\frac12\left[\left(\frac{2}{a^2}  + i \frac ma + m^2\right)A^2 +\left(\frac 2{a^2} -i \frac ma + m^2\right)B^2\right]\\
&= \frac{1}{2a^2}\left[(1+i ma)(2-i ma)A^2 + (1-i ma)(2+i ma)B^2\right]\;.
\end{split}
\end{equation}
If we restrict to pure imaginary values of $ma$, it is not hard to see that the squared mass of $A$ is positive for $-1 < i ma < 2$ and of $B$ when
$-2 <i ma<1$.  Thus there are tachyon thresholds at $ ma=\pm i$, which is precisely where the free energy has branch points!

\section{Discussion}
\label{sec:Conclusions}

In this paper we have performed a precision test of holography in a non-conformal setup. We first found the five-dimensional supergravity solution dual to the $\mathcal{N}=2^*$ theory on $S^4$ and then calculated the on-shell supergravity action after carefully implementing holographic renormalization to cancel all divergent terms. The result for the third derivative of the free energy $F$ with respect to the mass is in perfect agreement with the field theory calculation in \cite{Russo:2012ay,Buchel:2013id,Russo:2013qaa}, which used the matrix integral arising from the path integral localization formula of Pestun \cite{Pestun:2007rz} to compute the partition function of the theory. In the matrix model calculations in the dual field theory \cite{Russo:2012kj,Russo:2012ay,Buchel:2013id,Russo:2013qaa,Russo:2013kea} it was assumed that the instantons do not contribute to the partition function at large $N$ and large $\lambda$. The fact that our supergravity result for the partition function matches the one in field theory should serve as strong evidence for this assumption. More generally it would be interesting to understand when instantons are important in the 't Hooft limit, both from field theory and holography (see \cite{Azeyanagi:2013fla} for a recent discussion in the current context).

One of the lessons from our analysis is that constructing the gravity dual of a non-conformal theory on a curved manifold is a nontrivial task. Even if such a curved manifold is conformal to $\mathbb{R}^4$ (as is $S^4$), the field theory action may contain new couplings that in the five-dimensional holographic description correspond to additional bulk fields developing nontrivial space-time dependence. Thus even if the gravity dual of a given supersymmetric theory on $\mathbb{R}^4$ is known, finding the gravity dual of the same theory defined in a supersymmetric way on $S^4$ requires ``starting from scratch''.

There is a simple generalization of the construction we presented here. One can consider $\mathcal{N}=2$ quiver gauge theories which are orbifold generalizations of $\mathcal{N}=2^*$ SYM\@. One way to obtain these theories is to first take a $\mathbb{Z}_k$ orbifold\footnote{To describe the orbifold action consider the $SO(6)$ R-symmetry of $\mathcal{N}=4$ SYM as acting on $\mathbb{R}^6$ with coordinates $x_i$ with $i=1,\ldots, 6$. Then the orbifold acts as simultaneous rotations by angle $2 \pi/ k$ in the $(x_1, x_2)$ and $(x_3, x_4)$ planes, while leaving $x_5$ and $x_6$ unchanged.} of $\mathcal{N}=4$ SYM preserving $\mathcal{N}=2$ supersymmetry as described in \cite{Kachru:1998ys,Klebanov:1998hh} and then deform the resulting superconformal quiver gauge theory by equal mass terms for all the hypermultiplets. One can study this class of orbifold theories on $S^4$ in much the same way as $\mathcal{N}=2^*$ and compute their free energy in the large $N$ and large 't Hooft coupling limit \cite{Azeyanagi:2013fla}. The result is that the free energy of the $\mathbb{Z}_k$ orbifold theory with gauge groups $U(N)$ is given by
\begin{equation}\label{FZkorbifold}
F_{\mathbb{Z}_k} (U(N)) = k F_{\mathcal{N}=2^*} (U(N))\;.
\end{equation}
Here $F_{\mathcal{N}=2^*} (U(N))$ is the free energy on $S^4$ of $\mathcal{N}=2^*$ SYM with gauge group $U(N)$ as written in \eqref{FFieldTheory}. It is not hard to reproduce \eqref{FZkorbifold} from our supergravity solution. First one should uplift our solution of five-dimensional gauged supergravity to a solution of the ten-dimensional type IIB supergravity.  While this is not an easy task (and is beyond the scope of this paper), we will not need the details of the full ten-dimensional solution to extract the relevant information concerning the $\Z_k$ orbifold. The only relevant fact about the (unorbifolded) ten-dimensional background is that it has a $\Z_k$ symmetry (which is a subgroup of $U(1)_H$) that acts within the internal directions.  The $\Z_k$ orbifold will decrease the volume of the internal space by a factor of $k$. Upon compactification of the resulting orbifolded solution to five dimensions, one finds that the five-dimensional Newton constant, $G_5$, is proportional to the volume of the (orbifolded) internal space \cite{Gubser:1998vd}. Since the five-dimensional gravitational action is proportional to $1/G_5$ and the holographic calculation of the free energy reduces to evaluating the renormalized gravitational on-shell action, we find that the holographic calculation yields the same result for the free energy as in \eqref{FZkorbifold}.

As mentioned above, an interesting problem that we have left unsolved is the uplift of our solution to type IIB supergravity. Having the explicit form of this solution at hand would allow for a holographic calculation of expectation values of Wilson, 't Hooft, and dyonic loop operators, as well as of expectation values of supersymmetric surface operators. One could also probe the solution with branes as was done in \cite{Buchel:2000cn,Evans:2000ct}. The uplift of the holographic dual of $\mathcal{N}=2^*$ on $\mathbb{R}^4$ was found in \cite{Pilch:2000ue}. The Pilch-Warner solution has an internal manifold with the same topology as $S^5$ and an $SU(2)\times U(1)$ isometry reflecting the global symmetry of the dual field theory. Most importantly it has a nontrivial profile for the ten-dimensional axion-dilaton as a function of the spacetime radial variable. The uplift of our solution to ten dimensions will also have such a nontrivial axion-dilaton profile.  However, the internal manifold will have only a $U(1)\times U(1)$ isometry due to the reduced symmetry of the $\mathcal{N}=2^*$ theory on $S^4$. One can use the uplift formulae of \cite{Pilch:2000ue} to find the ten-dimensional metric and axion-dilation for this solution. Finding the ten-dimensional metric is in principle straightforward once the five-dimensional solutions is known. The nontrivial problem is to find the R-R and NS-NS fluxes along the directions of the topological $S^5$. We postpone this problem for future work.

The Pilch-Warner solution in type IIB was later generalized in \cite{Pilch:2003jg} to include more general distributions of D3-branes. It will be interesting to study similar generalizations to $\mathcal{N}=2$ solutions with an $S^4$ boundary. A supergravity solution dual to pure $\mathcal{N}=2$ SYM on $\mathbb{R}^4$ was found in \cite{Gauntlett:2001ps,Bigazzi:2001aj}. It will be most interesting to find the corresponding BPS supergravity solution with an $S^4$ boundary and calculate the free energy of the dual field theory. The result should then be compared with the field theory calculation in \cite{Russo:2012ay} performed using path integral localization.

It was found in \cite{Russo:2013qaa} that the $\mathcal{N}=2^*$ theory on $S^4$ undergoes an infinite number of phase transitions at large $N$ as one varies the 't Hooft coupling $\lambda=g_{\text{YM}}^2N$. Our supergravity solution is dual to $\mathcal{N}=2^*$ on $S^4$ with both $N$ and $\lambda$ large. It will be very interesting to understand the nature of the phase transitions observed in field theory from the dual type IIB string theory. For that purpose one will probably need to find the $\alpha'$ corrections to the type IIB uplift of our five-dimensional supergravity solution. 

Another interesting avenue for further explorations is to study gravitational dual solutions to $\mathcal{N}=1$ supersymmetric field theories on $S^4$ and other curved manifolds. It is known how to put such field theories on various curved manifolds while preserving $\mathcal{N}=1$ supersymmetry \cite{Festuccia:2011ws}. To the best of our knowledge there are no exact results known from path integral localization for $\mathcal{N}=1$ theories and thus holography may provide some valuable insights into their structure.  A particularly interesting example which will be amenable to analysis using the techniques we employed in the current work is the gravity dual of $\mathcal{N}=1^*$ on $S^4$. The five-dimensional supergravity solution should be a generalization of the GPPZ flow \cite{Girardello:1999bd}, and its type IIB uplift should be similar to the Polchinski-Strassler solution \cite{Polchinski:2000uf}.

\section*{Acknowledgements}
We would like to thank Dionysios Anninos, Francesco Benini, Nadav Drukker, Krzysztof Pilch, Jorge Russo, Kostas Skenderis, Balt van Rees, Nick Warner, and Kostya Zarembo for useful discussions. Most of this work was done while NB was a postdoc at the Simons Center for Geometry and Physics and he would like to thank this institution for its support and great working atmosphere. The work of NB is supported by Perimeter Institute for
Theoretical Physics. Research at Perimeter Institute is supported by the Government of Canada through Industry Canada and by the Province of Ontario through the Ministry of Research and Innovation. HE is supported by NSF CAREER Grant PHY-0953232 and by a Cottrell Scholar Award from the Research Corporation for Science Advancement, and in part by the US Department of Energy under DoE Grant 
\#DE-SC0007859.   The research of DZF is supported in part by NSF grant PHY-0967299.  The work of SSP is supported in part by a Pappalardo Fellowship in Physics at MIT\@. Both DZF and SSP are supported in part by the U.S. Department of Energy under cooperative research agreement DE-FG02-05ER41360. 

\appendix

\section{Supersymmetry on $S^4$}
\label{app:S4susy}
\subsection{$\cn=4$ SYM  expressed in $\cn =1$ component fields}

The Lagrangian of $\cn=4$ SYM theory on Euclidean $\R^4$ can be expressed in terms of the fields $A_\mu, \lambda_\a, \tilde \lambda_\a, X_i$ with $\a =1,\ldots,4,~i=1,\ldots,6$.
It has manifest $SU(4)_R$ symmetry, and is given by\footnote{ The Lorentzian action is given in (23.1) of \cite{Freedman:2012zz}, although the 't Hooft matrices of (23.2) must be modified as indicated below.}
\begin{equation} \lab{n=4lag}
\begin{split}
\cl= &\frac14 (F^a_{\m\n})^2 - \tilde{\lambda}^{aT}_\a \s_2 \bar{\s}^\m D_\m \l^a_\a + \frac12 (D_\m X^{a}_i)^2 \\
&-\frac12 (f^{abc} C^{\a\b}_i (\lambda^{aT}_\a \s_2 \lambda_\b^b )X^{c}_i+ \text{h.c.}) +\frac14 f^{abc}f^{ab'c'}X^{b}_iX^{c}_j X^{b'}_iX^{c'}_j\,.
\end{split}
\end{equation}
The $4\times 4 $ anti-symmetric matrices $C_i$ are
\begin{equation}\label{alphabeta}
\begin{array}{lll}
C_1 = 
\begin{pmatrix}
  0 & \sigma_1 \\ -\sigma_1 & 0 
\end{pmatrix}\,, &
C_2 = 
\begin{pmatrix}
   0 & -\sigma_3 \\ \sigma_3 & 0 
\end{pmatrix}\,, &
C_3 = 
\begin{pmatrix}
   i \sigma_2 & 0 \cr 0 &  i \sigma_2 
\end{pmatrix}\,, \\[4mm]
C_4 = -i
\begin{pmatrix}
   0 & i \sigma_2  \cr i \sigma_2  & 0 
\end{pmatrix}\,, &
C_5 = -i
\begin{pmatrix}
   0 & 1 \cr -1 & 0 
\end{pmatrix}\,, &
 C_6 =-i 
\begin{pmatrix}
   -i \sigma_2 & 0 \cr 0 &  i \sigma_2 
\end{pmatrix}\,,
\end{array}
\end{equation}
and $\sigma_i$ are the usual Pauli matrices
\begin{equation}
\sigma_1 = \begin{pmatrix} 0  & 1 \cr 1& 0  \end{pmatrix}\;, \qquad  \sigma_2 = \begin{pmatrix} 0  & -i \cr i & 0  \end{pmatrix}\;, \qquad \sigma_3 = \begin{pmatrix} 1  & 0 \cr 0& -1  \end{pmatrix}\;.
\end{equation}
Eqs.~\eqref{Lkin}--\eqref{Lfour} in the main text can be derived as follows.  First we rewrite the theory in terms of $\cn=1$ component fields: $A_\m,\,\l^a =\l_4^a, \, \chi_i^a = \l_i^a, \,Z_i=(X_i+i X_{i+3})/\sqrt2,\,i=1,2,3$. Using the explicit form of the $C_i$ matrices, one can
with due care transform the Yukawa term in \reef{n=4lag} to the form
\be \lab{lyukN1}
\cl_{\rm Yukawa} =\sqrt2 f^{abc}\bigg((\l^{aT}\s_2\chi^b_i) \tilde Z_i^c - \frac12\e_{ijk}(\chi_i^{aT}\s_2\chi_j^b)Z_k^c\bigg) + \text{h.c.} \,.
\ee
Equation \eqref{lyuk} of the main text can then be obtained by the substitutions
$\l \to \psi_1,\, \l_3\to\psi_2,\, Z_3\to \Phi.$

The quartic term in \reef{n=4lag} can also be rewritten in terms of chiral scalars $Z_i,\,\tilde Z_i$ with a little help from the Jacobi identity.  One finds
\be \lab{l4N1}
\cl_4 =\frac12 f^{abc}f^{ab'c'}\bigg( -\,\tilde Z^b_i Z^c_i \tilde Z^{b'}_j Z^{c'}_j + 2 \tilde Z^b_j\tilde Z^c_i Z^{b'}_jZ^{c'}_i \bigg)\,.
\ee
The two terms displayed are exactly the $D$-term potential, $V_D = D^aD^a/2$, and the $F$-term potential, $V_F =\tilde{F}^aF^a$, of an $\cn=1$ supersymmetric theory with three adjoint chiral multiplets and superpotential $W=-\sqrt2 f^{abc}Z_1^aZ_2^bZ_3^c\,.$  

\subsection{Symmetries of the $\cn=2^*$ theory on $S^4$}

As discussed in Section \ref{FIELDTHEORY}, the $SU(4)_R$ symmetry of the $\cn=2$ theory is broken to $SU(2)_V\times SU(2)_H\times U(1)_R$ by the split into $\cn=2$ vector and hypermultiplets.  To define these symmetries explicitly  it is useful to begin with the $SU(4)$ transformation properties of the matrices \reef{alphabeta}.
Suppose that $U_\a{}^{\a'}$ is a unitary matrix in the fundamental  of $SU(4)$ and $\L_{ij}$ is the corresponding orthogonal matrix in the fundamental of $SO(6).$  The group transformation of the $C_i$ matrices is
\be \lab{citrf}
 C_i^{\a\b}U_\a{}^{\a'}U_\b{}^{\b'}  = \L_{ij} C_j^{\a'\b'}.
\ee 
By definition, the hypermultiplet fermions $\chi_i$ are in the fundamental of the $SU(2)_H$ subgroup, and the vector multiplet fermions $\psi_\a$ are in the fundamental of $SU(2)_V$.  These subgroups and  $U(1)_R$ act on the fermions via the following $4\times 4$ unitary matrices:
 \begin{equation} \lab{su2v}
 \left( \begin{array}{cc}
 I & 0 \\
0 & U_v
 \end{array} \right) \;, \qquad
 \left( \begin{array}{cc}
 U_h & 0 \\
0 & I
 \end{array} \right) \;, \qquad 
 \left( \begin{array}{cc}
e^{i\theta}I  & 0 \\
0 &  e^{-i\theta}I
 \end{array} \right) \;.
 \end{equation}
 The matrices $C_i^{\a\b}$ transform in the fundamental of $SO(6)$ and the sum $C_i X_i$ is an invariant.   If $U=I + iT+\ldots$, then to first order in the Hermitian generator $T$,  \reef{citrf} reduces to
 \be \lab{infin}
 i (T^T C_i + C_i T ) = \l_{ij}C_j \,,\qquad {\rm with} \qquad \l_{ji} =-\l_{ij}\,.
 \ee
One can study this infinitesimal transformation for generators $T=\s_i$ of $SU(2)_v$ and $SU(2)_H$  for the various matrices $C_i$.  One soon
verifies the $R$-symmetry properties of the scalars stated in Section \ref{FIELDTHEORY}.

\subsection{Massive $\cn=1$ chiral multiplets on $S^4$}

As stated at the beginning of Section \ref{FIELDTHEORY},  we do not give full details on the derivation of the component form of the Euclidean $\cn=2^*$ theory, because there is considerable information on the process in the appendices of \cite{Freedman:2013oja}.  Nevertheless we now give readers a closer look at the simpler subsystem of a free massive hypermultiplet on $S^4$.  Actually we start here with the even simpler case of a pair of $\cn=1$ chiral multiplets.
Assuming that the chiral multiplets are conformally coupled to curvature, the action on $S^4$ is:
\bea \lab{chiralaction}
  S_\text{chiral}^\text{Euc} &=& \int d^4 x\,\sqrt{g}\, \bigg[ g^{\m\n} \pa_\mu \tilde Z_i \pa_\nu Z_i - \tilde \chi^{T}_i \sigma_2\, \bar \sigma^\mu \nabla_\mu \chi_i- \tilde F_iF_i+ \frac{2}{a^2} \tilde Z_iZ_i\bigg] \,.
\eea
It is invariant under the transformation rules: 
\be \lab{chimulttrfs}
\begin{array}{rclcrcl}
\delta Z_i &=& - \epsilon^T  \sigma_2\, \chi_i \,, 
&&
\delta \tilde Z_i &=&- \tilde \epsilon^T \sigma_2\,\tilde \chi_i\;,  \\[1mm]
\delta \chi_i &=& \sigma^\mu \partial_\mu Z_i \tilde \epsilon +  \big( F_i+\frac{i}{a} Z_i\big) \epsilon \,, &&
  \delta \tilde \chi_i &=& \bar \sigma^\mu \partial_\mu \tilde Z_i \epsilon + (\tilde F_i + \frac{i}{a} \tilde Z_i) \tilde \epsilon \,, \\[1mm]
 \delta F_i &=& - \tilde \epsilon^T\sigma_2\,\bar \sigma^\mu \nabla_\mu \chi_i 
 \,, 
&& 
  \delta \tilde F_i &=& 
  - \epsilon^T  \sigma_2\sigma^\mu \nabla_\mu \tilde \chi_i  \,.
\end{array}
  \ee
The spinors $\e$ and $\te$ are Killing spinors on $S^4$ that satisfy \reef{Kspin} with the upper sign.  (Note that we have dropped gauge indices on the fields and  subscripts on the Killing spinors because they are not needed.)

It is straightforward to demonstrate invariance if we organize things to focus on the corrections needed to accommodate the $S^4$ geometry rather than the more common case of flat $\mathbb{R}^4$.  These are the $1/a$ and $1/a^2$ terms  above.    Consider first the proof of supersymmetry in flat space with conventional transformation rules, but allow the spinor parameters $\e(x),~\tilde\e(x)$ to be arbitrary functions.   Of course supersymmetry holds for constant $\e, \tilde\e$, so the result must be an integral involving only $\pa_\m \e(x)$.  Indeed the result is
\be \lab{SvarR4}
\d S_{\R^4} = - \int d^4x [\tilde{\cal J}^{\m T} \pa_\m\tilde{\e} + \text{h.c.}] = -\int d^4x [\tchi^T_i\s_2 \bar{\s}^\m \s^\n\pa_\n Z_i\,\pa_\m \te(x)+ \text{h.c.}]\,,
\ee
where ${\cal J}^{\m}$ is (a chiral component of) the Noether supercurrent.  (See (6.24) of \cite{Freedman:2012zz} for this type of expression and its derivation in flat space.)  On $S^4$ this expression ``covariantizes" to
\begin{equation}  \lab{SvarS4}
\d S_{S^4}  =-\int d^4x \sqrt{g} [\tchi^T_i\s_2 \bar{\s}^\m \s^\n\pa_\n Z_i\,\nabla_\m \te(x)+ \text{h.c.}]
\to  \frac{i}{a} \int d^4x \sqrt{g} [\tchi^T_i\s_2 \bar{\s}^\n\pa_\n Z_i\e+ \text{h.c.}]\,.
\end{equation}
The last expression is valid for Killing spinors, as needed for our work on supersymmetry on $S^4$. (Note  $\bar\s^\m\s^\n\bar\s_\m = - 2 \bs^\n$.)   It must be canceled to gain invariance, and the conventional transformation $\d\chi$  must be modified by adding the $(i Z_i/a)\e$ term in \reef{chimulttrfs}.   This modification generates the new term 
\begin{equation} \lab{dSnew}
(\d S)' = -\int d^4x \sqrt{g} [\tchi^T_i\s_2 \bar\s^\m\nabla_\m(\frac{i}{a} Z_i\,\e) +\text{h.c.}]
=-\frac{i}{a} \int d^4x \sqrt{g} [\tchi^T_i\s_2 \bar\s^\m(\pa_\m Z_i\,\e + Z_i \nabla_\m{\e})+\text{h.c.}]\,.
\end{equation}
The first term cancels $(\d S)'$ above; in the second term we use the Killing spinor equation   to obtain
\be 
(\d S)''= \frac{2}{a^2} \int d^4x \sqrt{g}[\tchi^T_i\s_2\te \,Z_i + \text{h.c.}]\,.
\ee
To cancel this, the term $2\tilde Z_i Z_i/a^2$  is added to the Lagrangian. Its $\d\tilde Z_i$ and $\d Z_i$ variations cancel $(\d S)'$,  and supersymmetry on $S^4$ is established.\footnote{It may appear that we have been a little careless in our ``jump" to
the Noether form of $\d S_{\R^4}$.  It is justified if the flat space
calculations are organized to avoid second derivatives of $\e(x)$.
Avoiding second derivatives is always possible using partial
integration.}

\subsection{The superpotential sector}

We now introduce a general $\cn=1$ superpotential $W(Z_i)$.  This superpotential leads to the following action on $S^4$:
\be \lab{swtotal}
S_W = -\int d^4x \,\sqrt{g}\,\left[ F_i W_i + \frac 12 ( \chi^{T}_i \s_2\,\chi_j )W_{ij}   - \frac{i}{a}( 3W - W_i Z_i) \right]\,.
\ee
Derivatives of $W$ are denoted by subscripts.  As above, we focus on the $1/a$ terms that are $S^4$ corrections to the result for flat Euclidean space.  In flat space, the variation of the first two terms with general spinors $\e(x)$ is
\be  \lab{flatvarw}
\d S_{W;\,\R^4} = \int d^4x W_i\, (\chi_i^T\s_2\s^\m\pa_\m\te)\,,
\ee
in which the quantity contracted with $\pa_\m\e$ is the change in the supercurrent due to $W$.  Using  $S^4$ Killing spinors and adding the $1/a$ correction to $\d\chi$, \eqref{flatvarw} becomes
\begin{equation}
\begin{split}
\d S_{W}|_{\text{1st~2~ terms}} &= \int d^4x \sqrt{g}\bigg[W_i\, (\chi_i^T\s_2\s^\m\nabla_\m\te)
- W_{ij} \chi_i^T\s_2 (\frac{i}{a}Z_j \e)\bigg]\\
&= \frac{i}{a} \int d^4x \sqrt{g}\bigg[2W_i\,(\chi_i^T\s_2\e)- W_{ij}( \chi_i^T\s_2 Z_j \e)\bigg]\,. \lab{ds12}
\end{split}
\end{equation}
This undesired residuum requires further modification of the action, namely the addition of the term proportional to $i/a$ in \reef{swtotal}.  Its
 variation is 
$\d(3W -W_iZ_i) = (2W_i -W_{ij}Z_j)\d Z_i$ which neatly cancels \reef{ds12}.  Note that the order $1/a$ modification of the action vanishes for a purely cubic superpotential, due to  the superconformal invariance of this case.

A similar discussion can be given for the ``formal conjugate" superpotential $\tilde{W}(\tilde Z_i)$.  The action 
\be  \lab{Sbarw}
S_{\tilde{W}} = -\int d^4x \,\sqrt{g}\,\left[ \tilde{F}_i \tilde{W}_i + \frac 12 ( \tchi^{T}_i \s_2\,\tchi_j )\tilde{W}_{ij}   - \frac{i}{a}( 3\tilde{W} - \tilde{W}_i \tilde{Z}_i) \right]\;,
\ee
is invariant under the transformation rules (\ref{chimulttrfs}).   It is significant that invariance holds even when the functions $W(Z_i)$ and $\tilde{W}(\tilde Z_i)$ are completely unrelated.  One should also note that the $i/a$ correction terms of
(\ref{swtotal}) and (\ref{Sbarw}) are not complex conjugates of each other even when $W$ and $\tilde{W}$ are.

 The chiral multiplet on $S^4$ was also discussed in Section 2 of \cite{Festuccia:2011ws}.   The relation of the actions \reef{chiralaction} and \reef{swtotal} to those of  \cite{Festuccia:2011ws} is quite simple (for  a flat
K\"ahler target space).  One can see that they are related by redefinition of the $F$ auxiliary field;   $F'$ of that reference is related to ours by $F'=F- \frac{i}{a} Z_i$.

\subsection{The $\cn=2$ massive multiplet and its supersymmetry algebra}

In the special case of the quadratic superpotential $W=m(Z_1^2+Z_2^2)/2$, the theory discussed above possesses $\cn=2$ supersymmetry.  With  auxiliary fields eliminated, the action on $S^4$ takes the form
\bea
       \lab{hyperaction}
  S_\text{chiral}^\text{Euc} &=& \int d^4 x\,\sqrt{g}\, \bigg[ \, \pa^\mu \tilde Z_i \pa_\mu Z_i - \tilde \chi^{T}_i \sigma_2\, \bar \sigma^\mu \nabla_\mu \chi_i\\ 
  \nonumber
  &&\hspace{2.2cm}
  +\,
   \Big(\frac{2}{a^2} + m^2\Big) \tilde Z_iZ_i + \frac{i\,m}{a} \Big(Z_iZ_i +\tilde Z_i\tilde Z_i\Big)
    - \frac{m}{2}\Big( \chi_i^T\s_2\chi_i + \tchi_i^T\s_2\tilde\chi_i\Big)   \bigg  ]\,.
  \eea
If we substitute $Z_i =(A_i+iB_i)/\sqrt2$, the scalar mass term becomes 
\be \lab{vmass}
V= \frac12 \bigg[ \Big( \frac{2}{a^2} + m^2 + \frac{i\,m}{a}\Big) A_iA_i  
+   \Big( \frac{2}{a^2} + m^2 - \frac{i\,m}{a}\Big)B_iB_i\bigg]\,.
\ee
Note the distinct mass values for scalars and pseudoscalars.  The same occurs for the chiral multiplet on $AdS_4$; see \cite{Breitenlohner:1982jf}.  The parameter $m$ can be complex, and the presence of  complex scalar masses is one indication that the correlation functions of the theory on $S^4$ do not obey reflection positivity \cite{Festuccia:2011ws}.

Let us write down the transformation rules, using $\cn=2$ Killing spinors $\e_i,~i=1,2$.  The $\d_1$ set are just a rewrite of \reef{chimulttrfs} with $F_i=-m \tilde Z_i$:
\be \lab{del1trfs}
\begin{array}{rclcrcl}
\delta_1 Z_i &=& - \epsilon^T_1  \sigma_2\, \chi_i \,, 
&& \delta_1 \chi_i &=& \sigma^\mu \pa_\mu Z_i \tilde \epsilon_1 +  \big( -m\tilde Z_i+\frac{i}{a} Z_i\big) \epsilon_1 \,,
  \\[1mm]
 \delta_1 \tilde Z_i &=&- \tilde \epsilon^T_1 \sigma_2\,\tilde \chi_i\;,&&
  \delta _1\tilde \chi_i &=& \bar \sigma^\mu \pa_\mu \tilde Z_i \epsilon_1 + (-m Z_i+ \frac{i}{a} \tilde Z_i) \tilde \epsilon_1 \,.  \end{array} 
\ee
The $\d_2$ transformations are obtained by making a finite $U(1)_V$ rotation,
specifically $\exp(-i \t_2\theta/2)$ with $\theta =\pi$, on the scalars in the $\d_1$ set.  Fermions are inert under $U(1)_V$.  This prescription gives 
\begin{equation} \lab{del2trfs}
\begin{split}
\delta_2 Z_i &= - \e_{ij}\tilde\epsilon^T_2  \sigma_2\, \tchi_j\,, \qquad \delta_2 \chi_i =-\e_{ij}[ \sigma^\mu \pa_\mu \tilde Z_j \tilde \epsilon_2 +  ( -m Z_j+\frac{i}{a} \tilde Z_j) \epsilon_2 ]\,,\\
\delta_2 \tilde Z_i &=- \e_{ij}\epsilon^T_2 \sigma_2\, \chi_j\;, \qquad
  \delta _2\tilde \chi_i =-\e_{ij}[ \bar \sigma^\mu \pa_\mu  Z_j
\epsilon_2 + (-m \tilde Z_j+ \frac{i}{a} Z_j) \tilde\epsilon_2] \,.  \end{split} 
\end{equation}
Since $U(1)_V$ is a symmetry of the theory, no further calculation is needed to confirm invariance under \reef{del2trfs}.

It is of some interest to study the commutator algebra of $\cn =2$ supersymmetry to check for possible modification due to the geometry of $S^4$.   The commutator of two transformations with the same $\cn=2$ index (i.e. $k=1$ or $k=2$ with no sum) is: 
\begin{equation} \lab{susycomsame}
\begin{split}
[\d_k, \d'_k] Z_i &= (\e_k^T\s_2\s^\m \te_k' -\e_k^{T'}\s_2\s^\m \te_k)\pa _\m Z_i\;, \\
\left[\d_k,\d'_k\right]  \chi_i &= (\e_k^T\s_2\s^\m \te_k' -\e_k^{T'}\s_2\s^\m \te_k)\nabla_\m\chi_i - \frac{i}{4a}(\e_k^T\s_2\s^{[\m}\bs^{\n]}\e'_k)\,\s_{[\m}\bs_{\n]}\chi_i \, .
\end{split}
\end{equation}
Since the spinor bilinear $\e_k\s_2\s^\m \te_k'$ is a Killing vector on $S^4$, this commutator  just gives an infinitesimal isometry of the sphere, as expected.  Note that the fermion calculation requires a gentle Fierz rearrangement and holds only when the fermion equation of motion is satisfied.
The last term, proportional to $i/a$, may be interpreted as a local frame rotation.

The commutator $[\d_1,\d_2']$ is more interesting; the result is
\begin{equation} \lab{susycomdiff}
\begin{split}
\left[\d_1,\d_2'\right]Z_i &= 
  \e_{ij}(\e_1^T\s_2\e_2' + \te_1^T\s_2\te_2')(m Z_j -\frac{i}{a}\tilde Z_j)\;, \\
 \left[\d_1,\d_2' \right] \chi_i &= \e_{ij}(\e_1^T\s_2\e_2' + \te_1^T\s_2\te_2')\,m\chi_j\,.
 \end{split}
\end{equation}
The  $m$ term is just the usual central charge for a massive hypermultiplet; 
see \cite{AlvarezGaume:1996mv}.  In fact it is a transformation of  $U(1)_H$.  The $i/a$ term is an $S^4$ modification of the algebra. It is an infinitesimal $U(1)_V$ transformation which is not central, but rather a genuine $R$-symmetry.  An analogue occurs in the $\cn=2$ deformation of the ABJM theory on $S^3$ constructed in \cite{Jafferis:2010un}. 

The $\cn=2$  transformations for the free hypermultiplet extend to the interacting $\cn=2^*$ theory.  Two new features  occur. First (as expected), the derivatives on the right side of \reef{susycomsame} become gauge covariant derivatives. 
Second,  one  finds in \reef{susycomdiff} a field-dependent gauge transformation involving the scalar $\Phi^a$ of the gauge mutliplet.  The interacting version of 
\reef{susycomdiff}  is 
\begin{equation} \lab{susycomint}
\begin{split}
\left[\d_1,\d_2'\right]Z_i^a &= 
  \e_{ij}(\e_1^T\s_2\e_2' + \te_1^T\s_2\te_2')(m Z_j^a -\frac{i}{a}\tilde Z_j^a)+\sqrt2 f^{abc}(\e_1^T\s_2\e_2' \tilde\Phi^b +\text{h.c.})Z^c\;, \\
 \left[\d_1,\d_2' \right] \chi_i^a &= \e_{ij}(\e_1^T\s_2\e_2' + \te_1^T\s_2\te_2')\,m\chi_j^a+\sqrt2 f^{abc}(\e_1^T\s_2\e_2' \tilde\Phi^b +\text{h.c.})\chi^c_i\,.
 \end{split}
\end{equation}

\section{Consistent truncation}
\label{app:truncation}

\subsection{Consistent truncation with 6 scalars}

Let us explain how to obtain the scalar part of the ${\cal N} = 4$ supergravity theory\footnote{It should be possible to write down this $\mathcal{N}=4$ gauged supergravity theory in a more canonical form as in \cite{Dall'Agata:2001vb,Schon:2006kz}. Since the gauge group of the $\mathcal{N}=4$ supergravity theory is $SU(2)\times U(1) \times U(1)$ it should not be possible to describe it in the formalism of  \cite{Dall'Agata:2001vb} and one should resort to the more general treatment in \cite{Schon:2006kz}. We will not discuss the details of this canonical construction here.} as a consistent truncation of ${\cal N} = 8$ gauged supergravity, following the notation of~\cite{Gunaydin:1984qu}.  In doing so, it is more convenient to fix the local $USp(8)$ symmetry by making a different gauge choice from the symmetric $USp(8)$ gauge described in the main text, such that the $SO(6) \times SO(2)$ symmetry is made more explicit.  Let $I$, $J$, $K$, with values 1--6, denote  $SO(6)$ indices, and $\a,~\b$, with values 1,\,2 denote 
$SO(2)$ indices. The $42$ scalars of the five-dimensional theory are parameterized as follows:  
\begin{itemize}
  \item 20 scalars are represented as a real traceless symmetric tensor $\Lambda^I{}_J$.  These scalars transform in the ${\bf 20}'$ of $SO(6)$, and so do the dual bosonic bilinear operators in the ${\cal N} = 4$ SYM theory.
  \item 20 scalars are parameterized by a real tensor $\Sigma_{IJK\alpha}$, which is totally anti-symmetric in the indices $IJK$.  These scalars transform in ${\bf 10} \oplus \overline{\bf 10}$ of $SO(6)$ and the dual operators are fermionic bilinears in ${\cal N} = 4$ SYM.
  \item 2 scalars are parameterized by a real traceless symmetric tensor $\Lambda^\alpha{}_\beta$ and are dual to the complexified gauge coupling of the $\mathcal{N}=4$ theory.
 \end{itemize}
The $SO(6)$ generators are real anti-symmetric matrices $\Lambda^I{}_J$, and the $SO(2)$ generator is represented as a real anti-symmetric matrix $\Lambda^\alpha{}_\beta$.  

We are looking for the scalars that are invariant under $U(1)_H \times U(1)_Y$.  We take $U(1)_H$ to be generated by
 \es{U1H}{
  U(1)_H:  \qquad \lambda^I{}_J = \begin{pmatrix}
   0 & 0 & 1 & 0  & 0 & 0 \\
   0 & 0 & 0 & -1 & 0 & 0 \\
   -1 & 0 & 0 & 0 & 0 & 0 \\
   0 & 1 & 0 & 0 & 0 & 0 \\
   0 & 0 & 0 & 0 & 0 & 0 \\
   0 & 0 & 0 & 0 & 0 & 0
  \end{pmatrix} \,.
 }
$U(1)_Y$ is a diagonal combination between the rotations in the $56$ plane generated by
 \es{U56}{
  U(1)_{56}: \qquad \lambda^I{}_J = \begin{pmatrix}
   0 & 0 & 0 & 0 & 0 & 0 \\
   0 & 0 & 0 & 0 & 0 & 0 \\
   0 & 0 & 0 & 0 & 0 & 0 \\
   0 & 0 & 0 & 0 & 0 & 0 \\
   0 & 0 & 0 & 0 & 0 & 1 \\
   0 & 0 & 0 & 0 & -1 & 0 
  \end{pmatrix} \;,
 }
and the $SO(2)$ rotations.

There are 6 scalar fields that are invariant under $U(1)_H \times U(1)_Y$.  We have
 \es{LambdaIJ}{
  \Lambda^I{}_J = \begin{pmatrix}
     -\alpha + \beta & \gamma_1 & 0 & \gamma_2 & 0 & 0 \\
     \gamma_1 & -\alpha - \beta & \gamma_2 & 0 & 0 & 0 \\
     0 & \gamma_2 & -\alpha + \beta & -\gamma_1 & 0 & 0 \\
     \gamma_2 & 0 & -\gamma_1 & -\alpha - \beta & 0 & 0 \\
     0 & 0 & 0 & 0 & 2 \alpha & 0 \\
     0 & 0 & 0 & 0 & 0 & 2 \alpha 
  \end{pmatrix} \,,
 }
$\Lambda^\alpha{}_\beta = 0$, and 
 \es{Sigma}{
  \Sigma_{IJ52} &= - \Sigma_{IJ61} = \chi_1 \begin{pmatrix}
    0 & 0 & 1 & 0 & 0 & 0 \\
   0 & 0 & 0 & -1 & 0 & 0 \\
   -1 & 0 & 0 & 0 & 0 & 0 \\
   0 & 1 & 0 & 0 & 0 & 0 \\
   0 & 0 & 0 & 0 & 0 & 0 \\
   0 & 0 & 0 & 0 & 0 & 0 
  \end{pmatrix} \,, \\
  \Sigma_{IJ51} &= \Sigma_{IJ62} = \chi_2 \begin{pmatrix}
   0 & 0 & -1 & 0 & 0 & 0 \\
   0 & 0 & 0 & 1 & 0 & 0 \\
   1 & 0 & 0 & 0 & 0 & 0 \\
   0 & -1 & 0 & 0 & 0 & 0 \\
   0 & 0 & 0 & 0 & 0 & 0 \\
   0 & 0 & 0 & 0 & 0 & 0 
  \end{pmatrix} \,.
 } 
Defining
 \es{AnglesDef}{
   \eta = e^\alpha \,, \qquad \vec{X} &= \frac{\tanh(\sqrt{\beta^2 + \gamma_1^2 + \gamma_2^2 + \chi_1^2 + \chi_2^2}) }{\sqrt{\beta^2 + \gamma_1^2 + \gamma_2^2 + \chi_1^2 + \chi_2^2}} \begin{pmatrix} \chi_1 
    & \chi_2 & \beta & \gamma_1 & \gamma_2 \end{pmatrix} \,, \\
 }
and following \cite{Gunaydin:1984qu}, one arrives at the Lagrangian 
 \es{LagAppendix}{
  {\cal L} &= \frac{1}{2 \kappa^2} \left[-R +  12 \frac{\partial_\mu \eta \partial^\mu \eta}{\eta^2} +
    \frac{4\, \partial_\mu \vec{X} \cdot \partial^\mu \vec{X} }{\left(1 - \vec{X}^2 \right)^2} - V \right] \,, \\
  V &= -g^2 \left[\frac{1}{\eta^4} + 2 \eta^2 \frac{1 + \vec{X}^2}{1 - \vec{X}^2} - \eta^8 \frac{(X_1)^2 + (X_2)^2}{\left(1 - \vec{X}^2 \right)^2}\right]   \,.
 }
The gauge coupling $g$ is related to the radius of the $AdS_5$ extremum of \eqref{LagAppendix} through $g = 2/L$.  As described in the main text, our three-scalar truncation \eqref{LagTruncMainText} is obtained from \eqref{LagAppendix} by setting $X_2 = X_4 = X_5 = 0$ and $z = X_3 + i X_1$.

\subsection{Supersymmetry variations}

In the notation of \cite{Gunaydin:1984qu}, the supersymmetry variations of the spin-$1/2$ fields are
\begin{equation}\label{spin12GRW}
\delta\chi_{abc} = \sqrt{2}\left[ \gamma^{\mu}P_{\mu abcd}- \frac 1L A_{dabc} \right] \varepsilon^d\;,
\end{equation}
and those of the spin-$3/2$ fields are
\begin{equation}\label{spin32GRW}
\delta\psi_{\mu a} = \nabla_{\mu} \varepsilon_a + Q_{\mu a}\;^{b} \varepsilon_b - \frac{1}{3L} W_{ab} \gamma_{\mu} \varepsilon^{b}\;.
\end{equation}
Here, the indices $a$, $b$, and $c$ are fundamental $USp(8)$ indices that run from $1$ to $8$, and can be raised and lowered with a real antisymmetric matrix $\Omega_{ab} = i (\Gamma_0)^{ab}$ as in $X_a = \Omega_{ab} X^b$.  We choose the $SO(7)$ gamma matrices as in Appendix C.1 of \cite{Freedman:1999gp}, namely,   
 \es{GammaDefs}{
  \Gamma_1 &= \sigma_3 \otimes \sigma_0 \otimes \sigma_2 \,, \\
  \Gamma_2 &= -\sigma_3 \otimes \sigma_2 \otimes \sigma_3 \,, \\
  \Gamma_3 &= \sigma_3 \otimes \sigma_2 \otimes \sigma_1 \,, \\
  \Gamma_4 &= \sigma_1 \otimes \sigma_3 \otimes \sigma_2 \,, \\
  \Gamma_5 &= \sigma_1 \otimes \sigma_2 \otimes \sigma_0 \,, \\
  \Gamma_6 &= \sigma_1 \otimes \sigma_1 \otimes \sigma_2 \,, \\
  \Gamma_0 &= i \Gamma_1 \Gamma_2 \Gamma_3 \Gamma_4 \Gamma_5 \Gamma_6 \,.
 }
The supersymmetry parameters that are invariant under $U(1)_H\times U(1)_Y$ are then
\begin{equation}
\begin{split}
  \varepsilon^a &= \begin{pmatrix}
   -\epsilon_2 & \epsilon_4 & \epsilon_1 & \epsilon_3 & \epsilon_4 & \epsilon_2 & -\epsilon_3 & \epsilon_1
  \end{pmatrix} \,, \\
  \varepsilon_a &= \begin{pmatrix}
   \epsilon_4 & \epsilon_2 & -\epsilon_3 & \epsilon_1 & \epsilon_2 & -\epsilon_4 & -\epsilon_1 & -\epsilon_3
  \end{pmatrix}\,.
  \end{split}
 \end{equation}
The symplectic Majorana condition $\varepsilon = \gamma_5 (i \Gamma_0) \varepsilon^* $ implies 
 \es{SympMaj}{
  \epsilon_3 &= \gamma_5 \epsilon_1^* \,, \qquad \epsilon_4 = \gamma_5 \epsilon_2^* \,,
 }
as in \eqref{SympMajorana}.  With this at hand it is straightforward to use \eqref{spin12GRW} and \eqref{spin32GRW} to work out the supersymmetry variations presented in \eqref{Spin12}--\eqref{Spin32}.

\section{Holographic renormalization}
\label{app:holoren}
In this appendix, we provide a detailed description of the  holographic renormalization procedure outlined in Section~\ref{HOLOREN}.

\subsection{Infinite counterterms}
\label{s:infiniteCT}

We begin with a derivation of the infinite counterterms needed to obtain a finite on-shell action and finite correlation functions. Here, as in Section~\ref{HOLOREN}, we set $4\pi G_5 =1$ to simplify the expressions; the overall normalization is restored in Section \ref{s:freeE}. 

\subsubsection{Setup}
\label{s:HRsetup}

The Euclidean action for our model is $S=S_\text{5D}+S_\text{GH}$ with 
\be
  S_\text{5D}=\int_M d^5x\,\sqrt{G} \,
  \Big\{ 
     -\frac{1}{4} R 
     + \frac{1}{2}\Kt \,G^{\m\n} \pa_\mu \eta  \pa_\nu \eta
     + K  \,G^{\m\n} \pa_\mu z \pa_\nu \tilde{z}
     + V
  \Big\} 
  \,,
  \label{5Daction1}
\ee
with $K = (1- z \tilde{z})^{-2}$ and $\Kt=6/\eta^2$. The scalar potential is
\be  \lab{vs4}
  V=-\bigg(
   \eta^{-4} + 2\eta^2 \frac{1+z \tilde{z}}{1-z \tilde{z}}
   + \frac{\eta^8}{4} \frac{(z - \tilde{z})^2}{(1-z \tilde{z})^2}
  \bigg)\,.
\ee
The Gibbons-Hawking action $S_\text{GH}$ will be discussed in Section~\ref{s:onshellaction}. 

For the purpose of studying holographic renormalization (see for example \cite{Bianchi:2001de, Bianchi:2001kw}) we perform the field redefinition
\reef{fieldredef} 
to  canonical fields with definite mass. 
The action \reef{5Daction1} then takes the form \reef{5Daction2-MT}.

As described in Section \ref{HOLOREN}, it is useful for the near-boundary analysis to write the 5D metric as  
\be\lab{5dmetric}
  ds^2 ~=~ G_{\mu\nu} dx^\m dx^\n
  ~=~
  \frac{d\rho^2}{4\rho^2} + \frac{1}{\rho}\,g_{ij}(x,\rho)\, dx^i dx^j\,.
\ee
In these coordinates, the $AdS_5$ boundary is at $\r=0$.  The mass $m$ of a scalar field and the scale dimension $\D$ of its dual field theory operator are related by\footnote{We fix the scale of $AdS_5$ by setting $L=1$.}  $\Delta = 2 + \sqrt{4+m^2}$.  A field $\phi_\D$ with $\D>2$ approaches the $AdS_5$ boundary at the rate $\phi_\D \sim \phi_{\D,0}(x)\,\r^{2 - \D/2}$. For $\D=2$, there is a logarithmic term of the form $\phi_2 \sim \phi_{2,0}(x)\, \r\ln\r$.

To implement holographic renormalization we place a lower cutoff $\r=\e\to 0$ on the 
radial integral in the action \reef{5Daction2-MT}.  When a solution of the classical equations of motion is inserted,  we obtain the on-shell action.  The radial integral then diverges at the leading rate $1/\e^2$ which comes from the integral $\int_{\r=\e} d\r\, \sqrt{G} \sim  \int d\r \,\r^{-3}$. This and  subleading divergences of order $1/\e$ and $\log \e$ must be cancelled by the counterterms. The goal of this section is to construct these counterterms and use them to perform holographic renormalization of our model.
 
It simplifies the analysis of the on-shell action to exclude \emph{ab initio} all contributions to the curly bracket $\big\{\dots \big\}$ in \reef{5Daction2-MT} that vanish faster than $\r^k$ with $k>2$ (to within logarithms). 
 It is thus sufficient to expand the potential \reef{vs4} as a truncated power series
\be
  V = -3 - 2 \phi^2 - 2 \chi^2 -\frac{3}{2} \psi^2 + \frac{c}{4} \psi^4 + \dots \;,
  \label{Vexpd}
\ee  
where $c=-2$ for our potential \reef{vs4}. We choose to keep $c$ general in the analysis since this allows us to compare with other holographic models. 
It follows from the scalar potential \reef{Vexpd} that the model contains two fields $\phi$ and $\chi$ with $m^2=-4$, and thus $\D=2$, and one field $\psi$ with $m^2=-3$ and $\D=3$. 

The ``$+\dots$'' in \reef{Vexpd} denotes terms that vanish faster than ${\cal O}(\rho^2)$ asymptotically and therefore do not give divergences.  Note that the terms 
$\chi\,\psi^2$, $\chi\,\phi$, $\phi\,\psi^2$ have the same asymptotic falloff rate as $\psi^4$, but they do not appear in the series expansion for our potential \reef{Vexpd}. There is a basic reason for the absence of $\chi\,\psi^2$ and $\chi\,\phi$, namely that the symmetry $\chi \to -\chi$ of our model prohibits them. The absence of $\phi\,\psi^2$ is more interesting: its presence is inconsistent with having a source term falloff $\psi_{0}(x)\,\r^{1/2}$ for $\psi$. This can be seen from an asymptotic analysis of the $\phi$ equation of motion.

\subsubsection{Bulk EoMs}

The five-dimensional equations of motion of the scalars in \reef{5Daction2-MT} are
\bea
  \label{5dscEOMphi}
  \Box_G \phi &=& \frac{\pa V}{\pa \phi}\,,\\[1mm]
  \label{5dscEOMpsi}
  K \Box_G \psi 
     +\pa_\mu K \pa^\mu \psi  
   - \frac{1}{2} \frac{\pa K}{\pa \psi} 
       \big(  (\pa \chi)^2 +  (\pa \psi)^2\big) 
  &=& \frac{\pa V}{\pa \psi}
  \,,\\[1mm]
  \label{5dscEOMchi}
    K \Box_G \chi 
       + \pa_\mu K \pa^\mu \chi  
   - \frac{1}{2} \frac{\pa K}{\pa \chi} 
       \big(  (\pa \chi)^2 +  (\pa \psi)^2\big)       
    &=& \frac{\pa V}{\pa \chi}\,, 
  \label{scalarEOM5d}
\eea
and the Einstein equation is
\be
  R_{\m\n} =  
  2\Big[ 
     K \pa_\mu \chi  \pa_\nu \chi 
     + K \pa_\mu \psi  \pa_\nu \psi
     +  \pa_\mu \phi  \pa_\nu \phi 
     + \frac{2}{3} G_{\m\n} V
  \Big]\,.
  \label{Ein1}
\ee

We now use the metric Ansatz \reef{5dmetric} to express the five-dimensional equations of motion in terms of the metric $g_{ij}$ and $\rho$.
To rewrite the scalar equations of motion \reef{5dscEOMphi}--\reef{5dscEOMchi}, we  decompose the scalar Laplacian $\Box_G$ as (with primes denoting $\rho$-derivatives)
\be
  \Box_G \Phi =
  \rho \Box_g \Phi
  + 4 \rho^2 \Phi'' - 4\rho \Phi' + 2 \rho^2 \Phi' \big( \log (g)\big)'
  \,,
  \label{boxG} 
\ee
and also use the expressions
\be
   \pa_\mu K \pa^\mu \psi 
   = 4\rho^2 K' \psi' + \rho g^{ij} \pa_i K \pa_j \psi\,,\qquad
   ( \pa \psi )^2
   = 4\rho^2 ( \psi' )^2 + \rho g^{ij} \pa_i \psi \pa_j \psi\,.
\ee
In the asymptotic expansion, the terms with $\pa_i$-derivative or
 $\chi'$ will be subleading, so we drop them in the following. 
The scalar equations of motion  \reef{5dscEOMphi}--\reef{5dscEOMchi} are then written as
\bea
  \hspace{-2mm}
  \rho \Box_g \phi
  + 4 \rho^2 \phi'' - 4\rho \phi' + 2 \rho^2 \phi' \big( \log (g)\big)'
  - \frac{\pa V}{\pa \phi} &\!\!=\!\!& 0\,,~~~~~~~
    \label{phiEOM}
\\[2mm]  
  \hspace{-2mm}
  K \Big[ \rho \Box_g \psi
  \!+\! 4 \rho^2 \psi'' \!-\! 4\rho \psi' \!+\! 2 \rho^2 \psi' \big( \log (g)\big)' \Big]
  \!+\! 2\rho^2  \frac{\pa K}{\pa \psi} \psi'^2 
  \!-\! \frac{\pa V}{\pa \psi} 
  \!+\! \text{(subleading)}
  &\!\!=\!\!& 0\,,~~~~~~~{}
  \label{psiEOM}
\eea
The  $\chi$ EoM is similar to \reef{psiEOM}, but 
 for the purpose of determining the counterterms 
 in Section \ref{s:asymp}  we  need only the $\psi$ EoM \reef{psiEOM}.

The Ricci tensor decomposes as follows
\begin{equation}
\begin{split}
  - R_{ij}[G] &= 
  - R_{ij}[g]
  + \rho \big[
    2 g_{ij}'' - 2 (g' g^{-1} g')_{ij} + \Tr(g^{-1}g') g'_{ij}
  \big]
   - 2 g'_{ij} - \Tr(g^{-1}g') g_{ij}
   + \frac{4}{\rho} g_{ij}\,,\\
   - R_{\rho\rho}[G] &= 
   \frac{1}{2} \Tr(g^{-1}g'') 
   -\frac{1}{4} \Tr(g^{-1}g'g^{-1}g') +\frac{1}{\rho^2}\,.
   \label{RofG}
\end{split}
\end{equation}
(We do not need the $(i,\rho)$ components of the Einstein equations.) 
The last term on the RHS of each expression in \reef{RofG} can be written as a cosmological constant term
$\frac{4}{\rho} g_{ij} = - 2 \, \frac{2}{3\rho} V_0\, g_{ij}$ with $V_0 = -3$ the value of the scalar potential \reef{vs4} for $\phi=\chi=\psi=0$. The RHS of the Einstein equation \reef{Ein1} can then be written as
\be
  \begin{split}
  &
  \rho \big[
    2 g_{ij}'' - 2 (g' g^{-1} g')_{ij} + \Tr(g^{-1}g') g'_{ij}
  \big]-R_{ij}
   - 2 g'_{ij} - \Tr(g^{-1}g') g_{ij}\\[1mm]
   & \hspace{15mm}
   = -2 
   \bigg[ 
      \pa_i \phi  \pa_j \phi 
     + K \pa_i \psi  \pa_j \psi
     + K \pa_i \chi  \pa_j \chi
     + \frac{2}{3\rho} g_{ij}\big\{ V -  V_0 \big\}
  \bigg]\,,
  \end{split}
  \label{Einstein-ij}
\ee
(with $R_{ij}=R_{ij}[g]$) 
and
\be  
  \frac{1}{2} \Tr(g^{-1}g'') 
   -\frac{1}{4} \Tr(g^{-1}g'g^{-1}g')
   =  
   -2 
   \bigg[ 
       (\phi')^2  + K(\psi')^2 + K(\chi')^2 
     + \frac{1}{6\rho^2}\big\{ V -  V_0 \big\}
  \bigg]\,.~~~
    \label{Einstein-rr}
\ee
These decompositions of the Einstein equations \reef{Einstein-ij} and \reef{Einstein-rr} were previously given in (3.16) and (3.18) of \cite{Bianchi:2001kw}. However, we note  three differences:
\begin{enumerate}
\item on the LHS of the $ij$-Einstein equation  \reef{Einstein-ij}, 
 we have $-2 g'_{ij}$ whereas in (3.16) of \cite{Bianchi:2001kw} this term enters with a ``$+$". The minus sign is an important correction for fixing the relation between coefficients in the asymptotic expansion of the fields and the metric.
\item an overall factor of $1/2$ is missing from the LHS of (3.18) of \cite{Bianchi:2001kw} compared with  \reef{Einstein-rr}. 
Again, this correction is important for matching the asymptotics. 
\item the curvature conventions in the present paper differ from those in \cite{Bianchi:2001kw} by a minus sign; so to compare our equations with those in \cite{Bianchi:2001kw} one must take $R_{ijkl} \to - R_{ijkl}$.
\end{enumerate}

To summarize, we have rewritten the five-dimensional equations of motion \reef{5dscEOMphi}--\reef{Ein1} using the metric Ansatz \reef{5dmetric}. The results, \reef{psiEOM}, \reef{Einstein-ij}, and \reef{Einstein-rr}, will be used in the next section.

\subsubsection{Asymptotic expansion}
\label{s:asymp}

The expansion of the metric and scalar fields near the boundary of $AdS_5$ was given in \reef{fieldexp-MT}.   
We  now fix some of the coefficients in this expansion using the EoMs in the previous section: 
\begin{itemize}
\item Start with the Einstein equation \reef{Einstein-ij}. At leading order, $R_{ij}[g] = R_{ij}[g_0] \equiv R_{0\,ij}$. Also, we have $g_{ij}' = g_{2\,ij} + \ldots$ and
$V -  V_0 = - \frac{3}{2} \rho \psi_0^2 + \ldots$. Expanding to order ${\cal O}(\rho^0)$, we get from \reef{Einstein-ij} that
\be
   R_{0\,ij}
   - 2 g_{2\,ij} - \Tr(g_0^{-1}g_2) \,g_{0\,ij}
   = 2  g_{0\,ij}\psi_0^2 \,.
   \label{PREg2}
\ee 
Taking the trace with $g_0^{ij}$, we find
\be
  \Tr(g_0^{-1} g_2)
  = -\frac{1}{6} R_0 - \frac{4}{3} \psi_0^2  \,.
    \label{Trg2}
\ee
Plugging this back into \reef{PREg2}, we have a solution for $g_{2\,ij}$:
\be
  g_{2\,ij} = -\frac{1}{2} \Big( R_{0\,ij} - \frac{1}{6} R_0 \, g_{0\,ij}\Big)
     - \frac{1}{3} \psi_0^2 \,g_{0\,ij}\,.
  \label{g2}
\ee
This agrees with the result obtained in the GPPZ model, see Appendix A.2 of \cite{Bianchi:2001de}. 
Note that the K\"ahler metric $K$ and the quartic interaction in $V$ play no role in this result. It is
useful to record the trace 
\be
  \Tr(g_0^{-1} g_2 g_0^{-1} g_2) 
  = \frac{1}{4} \Big( R_{0}^{ij}R_{0\,ij} - \frac{2}{9} R_0^2 \Big)
     +\frac{1}{9} R_0 \psi_0^2 
     +\frac{4}{9} \psi_0^4\,.
  \label{Trg2g2}
\ee
\item The $\psi$ EoM \reef{psiEOM} does receive corrections from $K$. The first non-trivial term is ${\cal O}(\rho^{3/2})$ and it involves $\psi_0$, $g_0$ and $g_2$ as well as $\psi_2$; we solve it for $\psi_2$ to find
\be
  \psi_2 = - \frac{1}{4} \Box_0 \psi_0 - \frac{1}{4} \Tr(g_0^{-1} g_2)\,\psi_0 + \frac{c+2}{4} \psi_0^3\,.
  \label{psi2sol0}
\ee 
The term $\frac{1}{2} \psi_0^3$ comes from $K$. (It is absent in  the GPPZ model \cite{Bianchi:2001de}.) Using \reef{Trg2}, we can eliminate $\Tr(g_0^{-1} g_2)$ from \reef{psi2sol0} to get
\be
  \psi_2 = - \frac{1}{4} \Box_0 \psi_0 
   +\frac{1}{24} R_0 \,\psi_0 +
   \frac{1}{12}\Big (10 + 3 c\Big)\, \psi_0^3\,.
  \label{psi2}
\ee
\item The $\chi$ EoM determines the coefficients $\chi_4$, $\chi_2$, and $\tilde\chi_2$ at leading order.  However, we do not need those for the purpose of determining the counterterms.  The reason is that $\chi$ enters the action quadratically, so only the $\chi^2 \sim \rho^2 \chi_0^2 +\dots$ in the potential matters and it only  involves the leading  source term $\chi_0$.

\item The $\phi$ EoM \reef{phiEOM}  similarly determines $\phi_4$, $\phi_2$, and $\tilde\phi_2$ at leading order, but again we do not  
need them. The argument for terms quadratic in $\phi$ in the potential is as for $\chi$. The term $\phi\psi^2$
would affect counterterms, but it is not present in the potential.  (See the
discussion at the end of Section \ref{s:HRsetup}.)
\item 
Finally, the $\rho\rho$-component of the Einstein equations gives at leading orders---${\cal O}\big(\rho^0 (\log\rho)^2\big)$, ${\cal O}\big(\rho^0 \log\rho\big)$, and ${\cal O}\big(\rho^0\big)$---three conditions that allow us to solve for the three components of the ${\cal O}(\rho^2)$-part of the metric expansion:
\bea
  \label{trh2}
  \Tr(g_0^{-1}h_2) 
   \!\!&\!=\!&\!\! 
   - \frac{4}{3} \phi_0^2 - \frac{4}{3} \chi_0^2 \,,\notag\\[1mm]
  \label{trh1}
   \Tr(g_0^{-1}h_1)
    \!\!&\!=\!&\!\! 
    -\frac{8}{3} \phi_0 \tilde{\phi}_0 
     -\frac{8}{3} \chi_0 \tilde{\chi}_0 
    - 2 \psi_0 \psi_2
       \,,\\[1mm]  
  \nonumber
  \Tr(g_0^{-1}g_4) 
   \!\!&\!=\!&\!\! 
    \frac{1}{4} \Tr(g_0^{-1}g_2g_0^{-1}g_2) 
   -\frac{2}{3} \phi_0^2 - \frac{4}{3} \tilde{\phi}_0^2 
   -\frac{2}{3} \chi_0^2 - \frac{4}{3} \tilde{\chi}_0^2 
   -\frac{c+6}{12} \psi_0^4
   + \psi_0 \psi_2
    - 2 \psi_0 \tilde{\psi}_0 \,.\notag
       \label{trg4}
\eea

\end{itemize}
 
For the purpose of determining the divergent part of the on-shell action, we do not need to go further in the expansion of the fields and EoMs. We end this section by presenting the expansion of $\sqrt{g}$ to the order we need it:
\bea
  \nonumber
  \sqrt{g} &=& \sqrt{g_0} \bigg[
    1 +  \rho \frac{1}{2}  \Tr(g_0^{-1} g_2) 
    + \rho^2  
    \Big(
      \frac{1}{2} \Tr(g_0^{-1} g_4) + \frac{1}{2}\log\rho\, \Tr(g_0^{-1} h_1)
      + \frac{1}{2}(\log\rho)^2  \Tr(g_0^{-1} h_2) \\
      &&\hspace{5.2cm}
      + \frac{1}{8} \big(\Tr(g_0^{-1} g_2)\big)^2 
      - \frac{1}{4} \Tr(g_0^{-1} g_2g_0^{-1} g_2) 
    \Big)
    + \dots
  \bigg]\,.
  \label{sqrtg}
\eea
We are now ready to use these results to evaluate the on-shell action.

\subsubsection{On-shell action and counterterms}
\label{s:onshellaction}

As noted around \reef{trR}, the trace of the Einstein equation allows us to rewrite the action \reef{5Daction2-MT} as 
$S_\text{5D}=  \int_M d^5x\,\sqrt{G} \,
  \big\{ 
     - \frac{1}{3} V(\phi,\psi,\chi)
  \big\}$. 
Next, we use $\sqrt{G} = \sqrt{g} \frac{1}{2\rho^3}$ as well as the asymptotic expansions \reef{fieldexp-MT} of the fields. The result of the small-$\rho$ expansion must then be integrated over $\rho$ down to the near-boundary surface at  $\rho=\eps$: 
$\int d^5 x  \to \int d^4 x \, \int_\eps d\rho$. The result of the expansion, before using the constraints from the EoMs, is
\be
  S_\text{5D}
  ~=~
  \int_{\pa M_\eps} d^4x \, \sqrt{g_0}
  \bigg[
  \frac{1}{2\eps^2}
  + \frac{1}{2\eps} 
     \Big(
       \Tr (g_0^{-1}g_2) + \psi_0^2
     \Big)
     + {\cal O}\big((\log\eps)^3\big)
   \bigg]\,.
   \label{S5Dexpd0}
\ee
We do need to keep track of the $\log$-divergent terms, but for simplicity we do not display them until \reef{Sexpd1}. The 5D action $S = S_\text{5D} + S_\text{GH}$ also includes the Gibbons-Hawking boundary action $S_\text{GH}$. It is
\be
  S_\text{GH} = -\frac{1}{2} \int_{\pa M} \sqrt{\gamma}\, \mathcal{K}
  =  - \frac{1}{2} \int_{\pa M} \frac{1}{\rho^2} \sqrt{g} 
  \Big(4 - \rho \,\pa_\rho \log g \Big)
  \,.
\ee
In the second equality, we used  
$\mathcal{K}= - 2\rho \pa_\rho \log \sqrt{\gamma}$ and
 $\gamma_{ij} = \frac{1}{\rho} g_{ij}$.  The asymptotic expansion of the metric gives
$\pa_\rho \log g = \Tr(g_0^{-1} g_2) + \dots$, and after using \reef{sqrtg} the divergent contributions are 
\be
  S_\text{GH} = - \int_{\pa M_\eps}
  d^4 x \,\sqrt{g_0} \,\frac{1}{\eps^2} \Big( 2 + \frac{1}{2} \eps \, \Tr(g_0^{-1} g_2) + {\cal O}\big(\eps^0\big)\Big)\,.
  \label{SGB1}
\ee
Adding the actions \reef{S5Dexpd0} and \reef{SGB1} one is left with
\bea
  \nonumber
  S
  \!\!\!&=&\!\!\!
  S_\text{5D}+ S_\text{GH}\\[1mm]
  \nonumber
  \!\!\!&=&\!\!\!
  \int_{\pa M_\eps} d^4x \, \sqrt{g_0}
  \bigg[
   - \frac{3}{2\eps^2}
  + \frac{1}{2\eps}  \psi_0^2
     -(\log\eps)^3\,
      \frac{1}{6}  
      \Big(
          \Tr(g_0^{-1} h_2) 
          + \frac{4}{3} \phi_0^2 
          + \frac{4}{3} \chi_0^2
       \Big)
       \\[1mm]
       &&
       \hspace{2.4cm}
       - (\log\eps)^2\,
     \frac{1}{4}  
      \Big(
          \Tr(g_0^{-1} h_1) 
          + \frac{8}{3} \phi_0 \,\tilde{\phi}_0 
          + \frac{8}{3} \chi_0 \,\tilde{\chi}_0 
          + 2 \psi_0 \,\psi_2 
       \Big)
       \\[1mm]
         \nonumber
       &&
       \hspace{2.4cm}
       - (\log\eps)\,
       \frac{1}{2}  
      \Big\{
           \Tr(g_0^{-1} g_4) 
          + \frac{1}{4} \big(\Tr(g_0^{-1} g_2)\big)^2 
          -  \frac{1}{2}\Tr(g_0^{-1} g_2 g_0^{-1} g_2)
        \\[1mm]
         \nonumber
       &&
       \hspace{3.4cm}
          -2 \Tr(g_0^{-1} h_2) 
          + \frac{4}{3} \tilde{\phi}_0^2
          + \frac{4}{3} \tilde{\chi}_0^2
          + \frac{1}{2} \psi_0^2\,\Tr(g_0^{-1} g_2)
          + 2 \psi_0 \tilde{\psi}_0
          - \frac{c}{6} \psi^4
       \Big\}
   \bigg] \,,
   \label{Sexpd1}
\eea
plus finite terms. Next we impose the EoM constraints we found in Section~\ref{s:asymp} for the coefficients of the asymptotic expansions of the fields.  It follows from \reef{trh1} that the coefficients of the $(\log\eps)^3$- and $(\log\eps)^2$-terms vanish. The other terms also simplify significantly and after omitting finite terms, we have
\bea
  \nonumber
  S
  \!\!\!&=&\!\!\!
  S_\text{5D}+ S_\text{GH}\\[1mm]
  \!\!\!&=&\!\!\!
  \int_{\pa M_\eps} d^4x \, \sqrt{g_0}
  \bigg[
   - \frac{3}{2\eps^2}
  + \frac{1}{2\eps}  \psi_0^2
        \\[1mm]
         \nonumber
       &&
       \hspace{2.4cm}
       - (\log\eps)\,
       \frac{1}{2}  
      \bigg(
         \, \frac{1}{32}
      \Big[ R_0^{ij} R_{0\,ij} - \frac{1}{3} R_0^2 \Big]
      -\phi_0^2 -\chi_0^2
      + \frac{1}{8} \psi_0 \Box_0 \psi_0
      -\frac{1}{48} R_0\,\psi_0^2 
      \bigg)
   \bigg] \,.
   \label{Sexpd2}
\eea
The next step in the procedure of holographic renormalization is to rewrite the divergences \reef{Sexpd1} in terms of the  fields at the cutoff, not just the asymptotic components. The desired counterterm action is then \emph{minus} the result of this rewrite. We find
 \be
 \begin{split}
 S_\text{ct}
 &=
  \int_{\pa M_\eps} d^4x \, \sqrt{\gamma}
  \bigg[
  \frac{3}{2}
  +  \frac{1}{8}R[\gamma] 
  +\frac{1}{2} \psi^2 
   +\Big( 1 + \frac{1}{\log \eps}\Big) \big( \phi^2 + \chi^2 \big)
     \\[1mm]
   & \hspace{3.4cm}
   -\log\eps \, 
    \bigg\{
      \frac{1}{32}
      \Big[ R[\gamma]^{ij} R[\gamma]_{ij} - \frac{1}{3} R[\gamma]^2 \Big]
      + \frac{1}{4} \psi \Box_\gamma \psi
     \\[1mm]
   & \hspace{5cm}
      -\frac{1}{24} R[\gamma]\,\psi^2
      -\Big( \frac{5}{12} + \frac{1}{8}c \Big) \psi^4\bigg\}
      +   \text{finite}
  \bigg]\,.
  \end{split}
  \label{CTeom}
\ee 
Most of these terms are standard in similar models, for example the Coulomb branch flow and 
the GPPZ flow studied in \cite{Bianchi:2001de}. The effect of the quartic term $\psi^4$ in the scalar potential enters only in the term with coefficient $c$ in \reef{CTeom}; the only effect of the target space metric on the counterterm action is in the $c$-independent coefficient $\frac{5}{12}$ of $\psi^4$.
In our model, $c=-2$, so the coefficient of the $\psi^4\log{\eps}$ counterterm is $1/6$.  This is the result for the infinite counterterm action given in \reef{CTeom-MT}.

\subsection{Finite counterterms from supersymmetry}
\label{s:finiteCT}

As we discussed in Section \ref{HOLOREN} in addition to the infinite counter terms one calculates using holographic renormalization, there may also be finite counterterms required by supersymmetry. Here we provide the detailed derivation of such a term for our model.

\subsubsection{No superpotential for our bulk theory}
\label{app:noW}
The Bogomolnyi machinery requires that the scalar potential is quadratically related to a superpotential $W$ which is a real function for BPS RG flows in five-dimensional supergravity.  In our gravity theory there are three scalars $\eta, z,$ and $\tz$ with target space metrics $\hat{g}_{\h\h}=\Kt(\eta) = 6/\eta^2$ and $\hat{g}_{z\tz}= K(z,\tz)= 1/(1-z\tz)^2$.  The action and scalar potential are given in \reef{5Daction1} and \reef{vs4}. If a superpotential  $W(\eta,z,\tz)$ exists, it should be related to $V$ by
\be \lab{vwgen}
V = \frac12 \Kt^{-1} (\pa_\eta W)^2 + K^{-1}\pa_z W\pa_{\tz} W  - \frac43 W^2\,.
\ee
The Bogomolnyi manipulations will then inform us that the scalar fields of BPS solutions satisfy the first order flow equations: ( $'$ indicates the derivative with respect to the radial coordinate $r$):
\be   \lab{flow1}
 \eta' \equiv H = \Kt^{-1} \pa_\h W\,,   
 ~~\qquad z' \equiv Z = K^{-1} \pa_{\tz}W\,,
 ~~ \qquad  \tz' \equiv \tilde Z =
K^{-1} \pa_z W \,.
\ee
These are gradient flow equations in a 3-dimensional target space with the indicated inverse metric components. Since the ordinary derivatives commute, the flow equations \reef{flow1} are mutually consistent only when the following three integrability conditions hold
\be
 \pa_z (K Z) = \pa_{\tz} (K \tilde Z)\;, \qquad   
 \pa_{\tz} (\Kt H) = \pa_\eta(KZ)\;,\qquad
 \pa_z(\Kt H) = \pa_\eta (K \tilde Z) \, .
 \label{3intcond}
 \ee
Using the BPS equations from Section \ref{sec:supergravity} the second integrability condition requires the vanishing of
\bea 
\lab{secintcond} 
 \pa_\eta(KZ) - \pa_{\tz} (\Kt H)
  = \frac{3 \eta^3 (z^2-\tz^2) \big[ (2+\eta^6)z +(2-\eta^6)\tz \big]}
     {(1-z\tz )^2 \big[1-\eta^6 +(1+\eta^6 )z^2\big]^{1/2}\big[1-\eta^6 +(1+\eta^6 )\tz^2\big]^{3/2}}\,.
\eea
It is clear that this integrability condition fails and thus our \emph{full} system does \emph{not} possess  a superpotential $W$. However, there is a way out of this difficulty.  One can show that all integrability conditions are obeyed if one makes either of the two restrictions $\tz = \pm z$.  Thus we find two truncations of our system for which the Bogomolnyi analysis is valid.  Let us do the analysis.

\subsubsection{Bogomolnyi analysis with the constraint $\tilde z=-z$}
\label{s:bogo}

We carry out the Bogomolnyi analysis for flat Euclidean signature domain walls.  The metric and its scalar curvature are 
\be \lab{flatmetric}
ds^2 = dr^2 + e^{2A(r)}\d_{ij} dx^idx^j\,,
\qquad\qquad 
R= -4(2A'' +  5A'^2)\,.
\ee
The constraint implies that $z$ is pure imaginary, so we set $z=-\tz = i\psi(r)/\sqrt2$. We work with the fields $\psi(r)$ and $\eta(r)$ and use the canonical $\phi = \tfrac{1}{\sqrt6}\ln\eta$ when appropriate. It will be justified below that one  can  make this Ansatz 
for the system directly in the action \reef {5Daction1}. Thus we begin with the reduced action\footnote{The boundary term in the partial integration of $R$ is cancelled by the Gibbons-Hawking action.} 
\be\lab{consaction}
S = \int dr\,d^4x \, e^{4A(r)} \left[ -3 A'^2+ +\frac12 \Kt \eta'^2 + \frac12 K \psi'^2 + V\right] \,,
\ee
with  $G(\eta) = 6/\h^2$,  ~$K(\psi)=4/(2-\psi^2)^2$, and 
\be
V(\h,\psi) =   -\left[  \eta^{-4} + 2 \eta^2 \frac{2+\psi^2}{2-\psi^2}  -\eta^8 \frac{2\psi^2}{(2-\psi^2)^2}\right]  
= -3 -\frac12(4\phi^2 + 3 \psi^2) - \frac12\psi^4 + \ldots \;.
\lab{vz=-tz}
\ee
The simple but central technical point of the  discussion is that the superpotential 
\be \lab{supotnew}
W =\eta^{-2} + \frac12\frac{2+\psi^2}{2-\psi^2}\eta^4
\ee 
is related to $V$ of \reef{vz=-tz} by the BPS relation\footnote{One can observe that the same potential and superpotential occurred in the 5D $\cn=2$ supergravity theory whose BPS solutions were studied in \cite{Pilch:2000ue}.
The correspondence with our fields is $\r=\eta,~~ \cosh(2\chi) =(2+\psi^2)/(2-\psi^2). $ } 
 \be \lab{bpsnew}
 V =  \frac12\bigg(\Kt^{-1}(\pa_\h W)^2 + K^{-1} (\pa_\psi W)^2\bigg) - \frac43 W^2\,.
 \ee
We insert the relation \reef{bpsnew} in the action \reef{consaction},  complete  squares,  and partially integrate to find the desired Bogomolnyi form
\bea
\nonumber
S &=& \int^{r_0} \! dr \,d^4x \bigg( e^{4A} 
\Big[ -3\,\Big(A' -\frac23 W\Big)^2 
+\frac12 \Kt\left(\h' +\Kt^{-1}\pa_{\h} W\right)^2 \\
&&
\hspace{3.2cm}
+ \frac12 K\left(\psi'+ K^{-1}\pa_{\psi}W\right)^2 \Big] 
- \frac{d}{dr}\big(e^{4A}\,W\big) \bigg)\,.
 \lab{SBogo}
\eea
The quadratic factors above are the BPS equations for flat-sliced domain walls.\footnote{We have checked this by an explicit calculation of the supersymmetry variation of the $\mathcal{N}=8$ five-dimensional supergravity.}   
When the BPS equations are satisfied, the on-shell action reduces to the boundary term. To preserve
supersymmetry this must be cancelled by adding the counterterm 
\be \lab{ssusy}
   S_W
     =
     \int d^4 x  \, e^{4A} \,W \,,
\ee
in which fields are evaluated at the cutoff $r_0$.
If this is not done there would be a residual cosmological constant in a supersymmetric and Lorentz invariant state of the boundary theory.\footnote{The Bogomolnyi argument is essentially the same for Lorentz and Euclidean
signature.} In turn these flat-sliced BPS equations also imply second order equations (obtained by applying $\pa/\pa r$ and using \reef{bpsnew}) that are the limit as $\tilde{z} \to - z$ of the equations of motion of the full theory with flat slicing. This argument makes clear that theory with the  constraint $\tilde{z} = - z$ imposed is a consistent truncation.  We expand $S_W$ in a power series in the canonical fields
 \be \lab{swexp}
   S_W
     =
     \int d^4 x \,\sqrt{\gamma}\,
     \Big(
        \frac{3}{2}
        + \phi^2 + \frac{1}{2}\psi^2
        + \sqrt{\frac{2}{3}} \phi\, \psi^2
        + \frac{1}{4} \psi^4 + \dots
     \Big)\,,
\ee
and observe (using \reef{fieldexp-MT} with $\r= e^{-2r}$) that the first four terms are infinite as $r_0 \to \infty$, the $\psi^4$ term is finite, while omitted terms vanish.  Consistent truncation implies universality in the following precise sense:  i) the infinite part of $S_{W}$ must agree with $S_{\rm ct} $ of \reef{CTeom} for field configurations of flat-sliced BPS domain walls, and ii) the finite term $\tfrac{1}{4}\psi^4$ must be added to $S_{\rm ct}$ to obtain the complete
counterterm needed for all supersymmetric solutions of the equations of motion.

To prove the first assertion, we write the difference $S_\text{ct}-S_W$ with the near-boundary expansions of \reef{fieldexp-MT} included and with terms in \reef{CTeom} that do not contribute for flat-sliced BPS domain walls excluded.  This difference is 
\be \lab{ctdiff}
  S_\text{ct}-S_W
  =   \int d^4 x \, \sqrt{\gamma}\,
       \bigg(
         \phi_0^2 +  \frac{1}{6}\psi_0^4 -  \sqrt{\frac{2}{3}} \phi_0\, \psi_0^2
       \bigg)
         \log \eps 
       + \text{finite}\,.
\ee
The asymptotic expansion of the BPS equations is quite simple:
\be \lab{bpsflat}
\phi' = -\pa_\phi W = -\Big(2\phi + \sqrt{\frac23} \psi^2\Big)  \;,
\qquad\quad 
\psi'= -\pa_\psi W = -\Big(\frac12 \psi + 2 \sqrt{\frac23} \phi\,\psi \Big)\,.
\ee
Near the boundary we can neglect the nonlinear term in the $\psi'$ equation but not in the $\phi'$ equation.  Therefore the leading behavior of $\psi(r)$ is
$\psi = \psi_0 \,e^{-r}$.  Including this in the $\phi'$ equation we find the solution $\phi = \Big[-\sqrt{\frac23} \psi_0^2 \,r + \tilde\phi_0\Big]e^{-2r}.$   Since 
$r_0 = -\tfrac{1}{2}\log\e$, this means that the $\phi_0 =\psi_0^2/\sqrt6 .$  When these results are inserted in \reef{ctdiff}, 
the infinite term $\log\eps$ vanishes as ``predicted" above.

The physics of the analysis above is quite simple. The quantity $\psi_0$ is
the source for a fermion bilinear and thus a fermion mass. The quantity $\phi_0$ is the source for a scalar bilinear and thus a (mass)$^2$.  
Supersymmetry fixes the quadratic relation between these sources which we found by solving the BPS equations.

Operationally, the most important result of this section  is the last term of \reef{swexp}.  It is the finite counterterm
\be \lab{sfin}
S_\text{finite} = \int d^4x\, \sqrt{\g} ~\frac{1}{4} \psi^4\,,
\ee
which must be added to the infinite counterterms of \reef{CTeom} to obtain a renormalized on-shell action which incorporates the requirement of global supersymmetry. We will do this in Section \ref{s:freeE}.

\subsubsection{Bogomolnyi analysis with the constraint $\tilde z=z$}
\label{s:bogo2}

The story of the $\tilde z = +z = \chi/\sqrt2$  truncation is very similar to the previous one, so we will be brief. The scalar potential $V(\eta, z,\tz)$ of the full theory now reduces to
\be \lab{vztz}
  V~=~- \left( \eta^{-4} + 2 \eta^2\frac{2+\chi^2}{2-\chi^2} \right)\\
 ~\approx~-3 -\frac12\big(4\phi^2+4\chi^2\big) + \ldots \,,
\ee
so we now have a consistent truncation consisting of two scalars, each dual to an operator of scaling dimension two.  The superpotential that gives rise to $V$ via
 \be \lab{bpsnewer}
 V =  \frac12\bigg(\hat{K}^{-1}(\pa_\h W)^2 + K^{-1} (\pa_\chi W)^2\bigg) - \frac43 W^2\,,
 \ee
now with $K= 4/(2-\chi^2)^2$, is
\bea
  W~=~ \frac{1}{2}\,  \eta^{4} + \frac{2+\chi^2}{2-\chi^2}\eta^{-2}
  ~\approx~\frac32 + \phi^2 +\chi^2 +\ldots \,.
  \label{Wchi}
\eea
Omitted terms in the expansions in 
\reef{vztz} and \reef{Wchi} are not needed for renormalization. The Bogomolnyi manipulations are again easily performed. 
The BPS equations are
\be
  A'  = \frac{2}{3} W\,,
  ~~~~~~
  \chi' = - K^{-1}\pa_\chi W\,,
  ~~~~~~
  \phi' =- \pa_\phi W\,.
  \label{BPSeqflat2}
\ee
The residual surface term is again cancelled by
\be
S_{\rm SUSY} = \int d^4 x  \, e^{4A} \,W =  \int d^4 x  \, e^{4A}\Big(
        \frac{3}{2}
        + \phi^2 + \chi^2
         + \dots \Big)\,.
\ee
The infinite-counterterm action of \reef{CTeom} gives
\be
 S_\text{ct} = \int d^4 x 
   \, \sqrt{\gamma}
  \bigg[
  \frac{3}{2}
   +\Big( 1 + \frac{1}{\log \eps}\Big) \, 
        \big( \phi^2 + \chi^2 \big) \bigg]\,,
   \label{Sctchi}
\ee
so in this case the difference $S_\text{ct}-S_{\rm SUSY}$ amounts to
\be
  S_\text{ct}-S_{\rm SUSY}  \sim
  \frac{1}{\log \eps}  \, 
        \big( \phi^2 + \chi^2 \big) 
    \sim 
    \big( \phi_0^2 + \chi_0^2 \big)\,.
\ee
Solving the BPS equations \reef{BPSeqflat2} for $\phi$ and $\chi$ at leading order gives $\phi_0 = \chi_0 = 0$. So again we find that the Bogomolnyi machinery gives a boundary term that produces the infinite counterterms  correctly for the flat sliced domain walls. The physics is  quite simple.  Supersymmetry dictates that there can be no sources for bilinear scalar operators unless a fermion bilinear is also sourced.

The upshot of the analysis presented above, and of that in Section~\ref{HOLOREN}, is that although our full model does not admit a superpotential, holographic renormalization only requires the use of an approximate potential in which we only keep terms that are divergent or finite near the boundary. The approximate potential has a superpotential $W_{a \cup b}$ given in \reef{aunionb} and the corresponding counterterm 
$S_\text{susy} = \int d^4x\, \sqrt{\gamma} \,W_{a \cup b}$ contains the infinite and finite counterterms compatible with supersymmetry. Up to terms that vanish on the boundary, we have 
$S_\text{susy} =  S_\text{ct} +S_\text{finite}$ with $S_\text{finite}$ given in \reef{sfin}. We use 
$S_\text{ren}= S_\text{5D}+S_\text{GH} + S_\text{susy}$ to compute the free energy. 

\subsection{The free energy}
\label{s:freeE}

In the holographic description, the on-shell action encodes the free energy $F$ of the field theory. We calculate the derivative of $F$ with respect to the source-term parameter $\mu$:
\be
\label{dFpart0}
  \frac{dF}{d\mu}
  ~=~
  \frac{dS_\text{ren}}{d\mu}
  ~=~
  \frac{d}{d\mu}
  \int d^4 x \, \sqrt{\gamma} \,\mathcal{L}_\text{ren}\,
  ~=~
  \int d^4 x \, \sum_{\text{fields}\,\Phi} \frac{\d( \sqrt{\gamma} \,\mathcal{L}_\text{ren})}{\d \Phi} \,\frac{d\Phi}{d\m}
  \,.
\ee
The variation of the action with respect to the fields give the one-point functions via
\be
  \label{dFpart1}
  \begin{split}
  \frac{\d( \sqrt{\gamma} \mathcal{L}_\text{ren})}{\d \psi}
  = \eps^{3/2} \sqrt{\gamma}\,\<O_\psi\>  + \ldots\,,
  \hspace{1cm}&
  \frac{\d( \sqrt{\gamma} \mathcal{L}_\text{ren})}{\d \phi}
  = \frac{\eps}{\log\eps} \sqrt{\gamma}\,\<O_\phi\>  + \ldots\,,\\[1mm]
  \frac{\d( \sqrt{\gamma} \mathcal{L}_\text{ren})}{\d \chi}
  = \frac{\eps}{\log\eps} \sqrt{\gamma}\,\<O_\chi\>  + \ldots\,,
  \hspace{1cm}&  
  \frac{\d( \sqrt{\gamma} \mathcal{L}_\text{ren})}{\d \gamma^{ij}} = \frac{1}{2} \eps\, \sqrt{\gamma}\, \<T_{ij}\>+ \ldots\,.
  \end{split}
\ee
The expression for $\<T_{ij}\>$ can be found in (3.13) of \cite{Bianchi:2001de}.
Subleading terms in the small $\eps$-expansion are indicated with ``$+\dots$".

Next, consider the field derivatives $\frac{d\Phi}{d\m}$ in \reef{dFpart0}. 
For our scalar fields, we find
\be
  \label{dFpart2}
  \frac{d\psi}{d\m} ~=~ \eps^{1/2}\, \frac{d\psi_0}{d\m} + {\cal O}(\eps^{3/2})\,,
  ~~~~~~
  \frac{d\phi}{d\m} ~=~\eps (\log\eps)\, \frac{d\phi_0}{d\m} + {\cal O}(\eps^{2})\,,
\ee
and similarly for $d\chi/d\m$. We also need to vary the metric. 
Since $\gamma_{ij} = \frac{1}{\eps} g_{ij} = 
\frac{1}{\eps} (g_{0\,ij} + \eps g_{2\,ij} + \ldots)
= \gamma_{0\,ij} + \eps \gamma_{2\,ij} \ldots$ and $g_0$ is independent of $\mu$, we have
\be
  \label{dFpart3}
  \frac{d\gamma^{ij}}{d\mu} 
  = - \eps \,\gamma^{ik}\gamma^{jl} \frac{d\gamma_{2\,kl}}{d\mu} 
  = - \eps^2 \,g_0^{ik}g_0^{jl} \frac{d g_{2\,kl}}{d\mu} + \ldots \,.
\ee
Now, with the help of \reef{dFpart1}, \reef{dFpart2}, and \reef{dFpart3}, the expression \reef{dFpart0} becomes
\be
\label{dFpart4}
  \frac{dF}{d\mu}
  ~=~ 
  \frac{dS_\text{ren}}{d\mu}
  ~=~
  \int d^4 x\, \sqrt{\gamma} \eps^2
  \bigg( 
  \< O_\psi \> \frac{\pa \psi_0}{\pa\mu}
  + \< O_\phi \> \frac{\pa \phi_0}{\pa\mu}
  + \< O_\chi \> \frac{\pa \chi_0}{\pa\mu}
  -\frac{1}{2}\eps\, \< T_{ij} \> g_0^{ik}g_0^{jl} \frac{d g_{2\,kl}}{d\mu}
  \bigg)\,.
\ee
The contribution from the metric variation is suppressed by an extra power of $\eps$ compared to the other terms. Thus taking the limit $\eps \to 0$ we find
\be   
   \frac{dF}{d\mu}
   =
  \int d^4 x\, \sqrt{g_0}
  \bigg( 
  \< O_\psi \> \frac{\pa \psi_0}{\pa\mu}
  + \< O_\phi \> \frac{\pa \phi_0}{\pa\mu}
  + \< O_\chi \> \frac{\pa \chi_0}{\pa\mu}
  \bigg)\,.~
  \label{dFdmu}
\ee
This is the expression \reef{dFdmu-MT} used in Section \ref{HOLOREN}.

The parameter $\mu$ controls the source rate falloff of our fields as $\rho \to 0$.
Using $\chi = \sqrt{2} \,z_+$ and $\psi = - i \sqrt{2}\, z_-$ with $z_\pm = \frac{1}{2}(z\pm \tilde{z})$, one finds from the asymptotic expansion of our BPS solution \reef{UVsol} that
\be\label{psi00}
  \psi_0 = - i \sqrt{2} \m\,,~~~~~
  \phi_0 =- \sqrt{\frac{2}{3}} \mu^2\,,~~~~~
  \chi_0 = -\sqrt{2} \mu\,.
\ee
The subleading coefficients in the asymptotic expansion are
\be
  \begin{split}
    \label{psi0}
  &\psi_2 =  - i \sqrt{2} \m \frac{2}{3}(3-\mu^2)\,,~~~~~~
  \tilde\psi_0 =  - i \sqrt{2} \frac{1}{3}\big[ 2v (\mu^2-3) + \mu(4\mu^2-3)\big]\,,\\
  &\tilde{\phi}_0 =  \sqrt{\frac{2}{3}} \mu (\mu + v)\,,~~~~~~~~~~~\,
  \tilde{\chi}_0 =  \sqrt{2}  v\,.
  \end{split}
\ee
The boundary metric $g_{0\, ij}$ is that of a round four-sphere with radius $1/2$ and from the explicit form of the solution we extract the subleading contribution $g_{2\,\m\n}$:
\be
\label{theg0}
   g_0 = \frac{1}{4} g_\text{unit}\,,~~~~~~~
   g_2 = \frac{1}{6} (\m^2-3)\,  g_\text{unit}\,.
\ee
A simple consistency check is that the above results for 
$\Tr(g_0^{-1}g_2)$ and $\psi_2$ satisfy the conditions \reef{Trg2} and \reef{psi2}. To see this, use $R_0 = 4 \times R_\text{unit\,$S^4$} = 48$.

Using the results for the one-point functions summarized in Section \ref{HOLOREN} in \eqref{dFdmu} one obtains the result \reef{resdFdmu-MT} for $dF/d\mu$ that is then used to match the field theory free energy. 

The finite counterterm was essential in our analysis but suppose we did not want to rely on the Bogomolnyi method and the universality argument to fix this finite counterterm. To this end it is instructive to consider all possible candidate finite counterterm operators  (with $\sqrt{\gamma}$ implicit) 
\be
 \begin{array}{ccccc}
 &&& ~~~ \text{contribution to }\frac{dF}{d\mu}: \\
 R^{ij}[\g]R_{ij}[\g]\,,
 &
 (R[\g])^2\,, 
& \psi \Box_\gamma \psi\,, 
   &\text{0}
\\[1mm]
 \nonumber
 R[\g]\, \psi^2\,,
 &
 (\log\eps)^{-1} R \phi\,,
 &
 (\log\eps)^{-2}  \chi^2\,
  &\text{${\cal O}(\mu)$}
\\[1mm]
 (\log\eps)^{-2}  \phi^2\,,
 &
 (\log\eps)^{-1} \psi^2 \phi  \,, 
 &
 \psi^4
 &\text{${\cal O}(\mu^3)$}
 \end{array}
 \label{possibleCTs}
\ee
On the right, we indicate their contributions to $dF/d\mu$. The first two terms in the first line do not contribute at all because they are independent of the scalar fields and $\psi \Box_\gamma \psi$ does not contribute because it vanishes for our solution. The rest of the possible finite counterterms can only change the coefficients of the terms proportional to $\mu$ and $\mu^3$ in $dF/d\mu$ in \reef{resdFdmu-MT}; 
they cannot contribute any dependence on the ``vev'' parameter $v(\mu)$, in particular they cannot
affect the last term $v(\mu)$ in $dF/d\mu$. This means that if we take four more $\m$-derivatives, we get a result completely {\em independent of finite counterterms}. Thus, $d^5F/d\mu^5$ is independent of ambiguities of finite counterterms, so even without fixing those, we can compare  $d^5F/d\mu^5$ to the field theory result and obtain a perfect match.

\section{Analytic solutions with flat slicing}
\label{app:analytic}

We were not able to solve analytically the general system of BPS equations with $S^4$ slicing in Section \ref{subsec:BPSeqns} and we had to resort to numerics to extract the physics. If one studies the system of BPS equation in $\mathbb{R}^4$, however, one finds that it is consistent with equations of motion only if $z=\pm \tilde{z}$. In this case there are analytic solutions to the BPS equations which we present below.

\subsection{The solution for $z=\tilde{z}$}

The BPS equations for $z=\tilde {z}$ with $\mathbb{R}^4$ slicing can be derived either directly from the supersymmetry variation of the five-dimensional $\mathcal{N}=8$ supergravity theory or via the Bogomolnyi trick as in \eqref{BPSeqflat2}. The explicit result is (we set $L=1$)
\begin{equation}\label{BPSeqnspsi0}
\begin{split}
z'&= -\dfrac{2z}{\eta^2}\;, \\
\eta' &= \dfrac{\eta^6(1-z^2)-z^2-1}{3\eta(z^2-1)}\;,\\
A' &=  \frac{\eta^4}{3} + \frac{2}{3\eta^2}\left(\frac{1+z^2}{1-z^2}\right)\;.
\end{split}
\end{equation}
One can solve this system of equations analytically by eliminating $z(r)$ from the first equation and then integrating explicitly the other two to find
\begin{equation}\label{etaAzsol}
\eta^6 = \dfrac{1-z^2}{1+z^2 +C_1 z}\;, \qquad\qquad A = \frac{1}{6}\log\left[\frac{(1-z^2)^2(1+z^2 +C_1 z)}{z^3}\right]+C_2\;,
\end{equation}
where $C_1$ and $C_2$ are integration constants. Using \eqref{etaAzsol} in the first equation of \eqref{BPSeqnspsi0} one can solve for $z(r)$ in quadratures.\footnote{An analytic solution in terms of special functions exists for $C_1=0$.} It is clear that the solution develops a singularity at $z=\pm1$. The nature of the singularity is controlled by the constant $C_1$ and can be studied by using the criterion of \cite{Gubser:2000nd}. This singularity is of the same kind as the ones observed in the Coulomb branch RG flows in \cite{Freedman:1999gk}. In fact our solution is a generalization of the Coulomb branch flows in \cite{Freedman:1999gk}. The difference between our solution and those of \cite{Freedman:1999gk} is that we have two scalars in the $\bf{20}'$ turned on, as opposed to the single scalar used in \cite{Freedman:1999gk}, and the flow preserves $\mathcal{N}=2$ supersymmetry whereas the solutions in \cite{Freedman:1999gk} preserve $\mathcal{N}=4$.

\subsection{The solution for $z=-\tilde{z}$}

For $z=-\tilde{z}=i\psi/\sqrt{2}$ our truncation reduces to the one studied in \cite{Pilch:2000ue}. We can therefore derive the Pilch-Warner  solution \cite{Pilch:2000ue} dual to the $\mathcal{N}=2^*$ SYM on $\mathbb{R}^4$.  Again the BPS equation  for $z=-\tilde {z}$ with $\mathbb{R}^4$ slicing can be derived either directly from the supersymmetry variation of the five-dimensional $\mathcal{N}=8$ supergravity theory or via the Bogomolnyi trick. The result is
\begin{equation}\label{BPSeqnschi0}
\begin{split}
z'&= -z\eta^4\;, \\
\eta' &= \dfrac{\eta^6(z^2-1)+z^2+1}{3\eta(z^2+1)}\;,\\
A' &=  \frac{2}{3\eta^2} + \frac{\eta^4}{3}\left(\frac{1-z^2}{1+z^2}\right)\;.
\end{split}
\end{equation}
One can again eliminate $z(r)$ from the first equation and then integrate explicitly the other two to find the solution of Pilch-Warner \cite{Pilch:2000ue}
\begin{equation}
\eta^6 = \dfrac{1-z^4-4z^2\log z+C_1z^2}{(1+z^2)^2}\;, \qquad A = \log\left[\dfrac{1+z^2}{z}\eta^2\right] + C_2\;,
\end{equation}
where $C_{1,2}$ are integration constants. The solution is singular for $z^2=-1$ (or alternatively $\psi^2=2$) but the singularity is physical and well-understood \cite{Buchel:2000cn,Evans:2000ct}.



\begin{thebibliography}{99}

  
\bibitem{Witten:1988ze} 
  E.~Witten,
  ``Topological Quantum Field Theory,''
  {\em Commun.\ Math.\ Phys.}\  {\bf 117}, 353 (1988).

\bibitem{Pestun:2007rz} 
  V.~Pestun,
  ``Localization of gauge theory on a four-sphere and supersymmetric Wilson loops,''
  {\em Commun.\ Math.\ Phys.}\  {\bf 313}, 71 (2012)
  {\tt arXiv:0712.2824 [hep-th]}.

\bibitem{Kapustin:2009kz} 
  A.~Kapustin, B.~Willett and I.~Yaakov,
  ``Exact results for Wilson loops in superconformal Chern-Simons theories with matter,''
  {\em JHEP} {\bf 1003}, 089 (2010)
  {\tt arXiv:0909.4559 [hep-th]}.

\bibitem{Jafferis:2010un} 
  D.~L.~Jafferis,
  ``The Exact superconformal $R$-symmetry extremizes $Z$,''
  {\em JHEP} {\bf 1205}, 159 (2012)
  {\tt arXiv:1012.3210 [hep-th]}.

\bibitem{Benini:2012ui} 
  F.~Benini and S.~Cremonesi,
  ``Partition functions of ${\cal N}=(2,2)$ gauge theories on $S^2$ and vortices,''
  {\tt arXiv:1206.2356 [hep-th]}.

\bibitem{Doroud:2012xw} 
  N.~Doroud, J.~Gomis, B.~Le Floch and S.~Lee,
  ``Exact results in $D=2$ supersymmetric gauge theories,''
  {\em JHEP} {\bf 1305}, 093 (2013)
  {\tt arXiv:1206.2606 [hep-th]}.

\bibitem{Kallen:2012cs} 
  J.~Kallen and M.~Zabzine,
  ``Twisted supersymmetric 5D Yang-Mills theory and contact geometry,''
  {\em JHEP} {\bf 1205}, 125 (2012)
  {\tt arXiv:1202.1956 [hep-th]}.

\bibitem{Kallen:2012va} 
  J.~Kallen, J.~Qiu and M.~Zabzine,
  ``The perturbative partition function of supersymmetric 5D Yang-Mills theory with matter on the five-sphere,''
  {\em JHEP} {\bf 1208}, 157 (2012)
  {\tt arXiv:1206.6008 [hep-th]}.

\bibitem{Hosomichi:2012ek} 
  K.~Hosomichi, R.~-K.~Seong and S.~Terashima,
  ``Supersymmetric gauge theories on the five-sphere,''
  {\em Nucl.\ Phys.}\  {\bf B865}, 376 (2012)
  {\tt arXiv:1203.0371 [hep-th]}.

\bibitem{Hama:2010av} 
  N.~Hama, K.~Hosomichi and S.~Lee,
  ``Notes on SUSY gauge theories on three-sphere,''
  {\em JHEP} {\bf 1103}, 127 (2011)
  {\tt arXiv:1012.3512 [hep-th]}.

\bibitem{Klebanov:1996un} 
  I.~R.~Klebanov and A.~A.~Tseytlin,
  ``Entropy of near extremal black $p$-branes,''
  {\em Nucl.\ Phys.}\  {\bf B475}, 164 (1996)
  {\tt hep-th/9604089}.

\bibitem{Drukker:2010nc} 
  N.~Drukker, M.~Marino and P.~Putrov,
  ``From weak to strong coupling in ABJM theory,''
  {\em Commun.\ Math.\ Phys.}\  {\bf 306}, 511 (2011)
  {\tt arXiv:1007.3837 [hep-th]}.

\bibitem{Festuccia:2011ws} 
  G.~Festuccia and N.~Seiberg,
  ``Rigid supersymmetric theories in curved superspace,''
  {\em JHEP} {\bf 1106}, 114 (2011)
  {\tt arXiv:1105.0689 [hep-th]}.

\bibitem{Girardello:1998pd} 
  L.~Girardello, M.~Petrini, M.~Porrati and A.~Zaffaroni,
  ``Novel local CFT and exact results on perturbations of ${\cal N}=4$ super-Yang Mills from AdS dynamics,''
  {\em JHEP} {\bf 9812}, 022 (1998)
  {\tt hep-th/9810126}.

\bibitem{Freedman:1999gp} 
  D.~Z.~Freedman, S.~S.~Gubser, K.~Pilch and N.~P.~Warner,
  ``Renormalization group flows from holography supersymmetry and a $c$ theorem,''
  {\em Adv.\ Theor.\ Math.\ Phys.}\  {\bf 3}, 363 (1999)
  {\tt hep-th/9904017}.

\bibitem{Freedman:1999gk} 
  D.~Z.~Freedman, S.~S.~Gubser, K.~Pilch and N.~P.~Warner,
  ``Continuous distributions of D3-branes and gauged supergravity,''
  {\em JHEP} {\bf 0007}, 038 (2000)
  {\tt hep-th/9906194}.
  
\bibitem{Martelli:2011fu} 
  D.~Martelli, A.~Passias and J.~Sparks,
  ``The gravity dual of supersymmetric gauge theories on a squashed three-sphere,''
  {\em Nucl.\ Phys.}\  {\bf B864}, 840 (2012)
  {\tt arXiv:1110.6400 [hep-th]}.
  
\bibitem{Martelli:2011fw} 
  D.~Martelli and J.~Sparks,
  ``The gravity dual of supersymmetric gauge theories on a biaxially squashed three-sphere,''
  {\em Nucl.\ Phys.}\  {\bf B866}, 72 (2013)
  {\tt arXiv:1111.6930 [hep-th]}.

\bibitem{Martelli:2012sz} 
  D.~Martelli, A.~Passias and J.~Sparks,
  ``The supersymmetric NUTs and bolts of holography,''
  {\tt arXiv:1212.4618 [hep-th]}.
  
\bibitem{Freedman:2013oja} 
  D.~Z.~Freedman and S.~S.~Pufu,
  ``The Holography of $F$-maximization,''
  {\tt arXiv:1302.7310 [hep-th]}. 

\bibitem{Aharony:2008ug} 
  O.~Aharony, O.~Bergman, D.~L.~Jafferis and J.~Maldacena,
  ``${\cal N}=6$ superconformal Chern-Simons-matter theories, M2-branes and their gravity duals,''
  {\em JHEP} {\bf 0810}, 091 (2008)
  {\tt arXiv:0806.1218 [hep-th]}.

\bibitem{Gunaydin:1984qu} 
  M.~Gunaydin, L.~J.~Romans and N.~P.~Warner,
  ``Gauged ${\cal N}=8$ Supergravity in Five-Dimensions,''
  {\em Phys.\ Lett.}\  {\bf B154}, 268 (1985).
  
\bibitem{Pernici:1985ju} 
  M.~Pernici, K.~Pilch and P.~van Nieuwenhuizen,
  ``Gauged ${\cal N}=8$ $D=5$ Supergravity,''
  {\em Nucl.\ Phys.}\  {\bf B259}, 460 (1985).


\bibitem{Gunaydin:1985cu} 
  M.~Gunaydin, L.~J.~Romans and N.~P.~Warner,
  ``Compact and noncompact gauged supergravity theories in five-dimensions,''
  {\em Nucl.\ Phys.}\  {\bf B272}, 598 (1986).

\bibitem{Kim:1985ez} 
  H.~J.~Kim, L.~J.~Romans and P.~van Nieuwenhuizen,
  ``The Mass spectrum of chiral ${\cal N}=2$ $D=10$ supergravity on $S^5$,''
  {\em Phys.\ Rev.}\  {\bf D32}, 389 (1985).

\bibitem{Henningson:1998gx} 
  M.~Henningson and K.~Skenderis,
  ``The Holographic Weyl anomaly,''
  {\em JHEP} {\bf 9807}, 023 (1998)
  {\tt hep-th/9806087}.

\bibitem{Skenderis:2002wp} 
  K.~Skenderis,
  ``Lecture notes on holographic renormalization,''
  {\em Class.\ Quant.\ Grav.}\  {\bf 19}, 5849 (2002)
  {\tt hep-th/0209067}.

\bibitem{Bianchi:2001kw} 
  M.~Bianchi, D.~Z.~Freedman and K.~Skenderis,
  ``Holographic renormalization,''
  {\em Nucl.\ Phys.}\  {\bf B631}, 159 (2002)
  {\tt hep-th/0112119}.

\bibitem{Bianchi:2001de} 
  M.~Bianchi, D.~Z.~Freedman and K.~Skenderis,
  ``How to go with an RG flow,''
  {\em JHEP} {\bf 0108}, 041 (2001)
  {\tt hep-th/0105276}.

\bibitem{Russo:2012kj} 
  J.~G.~Russo,
  ``A Note on perturbation series in supersymmetric gauge theories,''
  {\em JHEP} {\bf 1206}, 038 (2012)
  {\tt arXiv:1203.5061 [hep-th]}.

\bibitem{Russo:2012ay} 
  J.~G.~Russo and K.~Zarembo,
  ``Large $N$ limit of ${\cal N}=2$ $SU(N)$ gauge theories from localization,''
  {\em JHEP} {\bf 1210}, 082 (2012)
  {\tt arXiv:1207.3806 [hep-th]}.

\bibitem{Buchel:2013id} 
  A.~Buchel, J.~G.~Russo and K.~Zarembo,
  ``Rigorous test of non-conformal holography: Wilson loops in ${\cal N}=2^*$ theory,''
  {\em JHEP} {\bf 1303}, 062 (2013)
  {\tt arXiv:1301.1597 [hep-th]}.

\bibitem{Russo:2013qaa} 
  J.~G.~Russo and K.~Zarembo,
  ``Evidence for large-$N$ phase transitions in ${\cal N}=2^*$ theory,''
  {\em JHEP} {\bf 1304}, 065 (2013)
  {\tt arXiv:1302.6968 [hep-th]}.

\bibitem{Russo:2013kea} 
  J.~G.~Russo and K.~Zarembo,
  ``Massive ${\cal N}=2$ gauge theories at large $N$,''
  {\tt arXiv:1309.1004 [hep-th]}.

\bibitem{Pilch:2000ue} 
  K.~Pilch and N.~P.~Warner,
  ``${\cal N}=2$ supersymmetric RG flows and the IIB dilaton,''
  {\em Nucl.\ Phys.}\  {\bf B594}, 209 (2001)
  {\tt hep-th/0004063}.

\bibitem{Buchel:2013fpa} 
  A.~Buchel,
  ``Localization and holography in ${\cal N}=2$ gauge theories,''
  {\tt arXiv:1304.5652 [hep-th]}.
  
\bibitem{Lu:1998nu} 
  H.~Lu, C.~N.~Pope and J.~Rahmfeld,
  ``A Construction of Killing spinors on $S^n$,''
  {\em J.\ Math.\ Phys.}\  {\bf 40}, 4518 (1999)
  {\tt hep-th/9805151}.

\bibitem{AlvarezGaume:1996mv} 
  L.~Alvarez-Gaume and S.~F.~Hassan,
  ``Introduction to S duality in ${\cal N}=2$ supersymmetric gauge theories: A Pedagogical review of the work of Seiberg and Witten,''
  {\em Fortsch.\ Phys.}\  {\bf 45}, 159 (1997)
  {\tt hep-th/9701069}.

\bibitem{Okuda:2010ke} 
  T.~Okuda and V.~Pestun,
  ``On the instantons and the hypermultiplet mass of ${\cal N}=2^*$ super Yang-Mills on $S^{4}$,''
  {\em JHEP} {\bf 1203}, 017 (2012)
  {\tt arXiv:1004.1222 [hep-th]}.

\bibitem{Buchel:2000cn} 
  A.~Buchel, A.~W.~Peet and J.~Polchinski,
  ``Gauge dual and noncommutative extension of an ${\cal N}=2$ supergravity solution,''
  {\em Phys.\ Rev.}\  {\bf D63}, 044009 (2001)
  {\tt hep-th/0008076}.

\bibitem{Intriligator:1998ig} 
  K.~A.~Intriligator,
  ``Bonus symmetries of ${\cal N}=4$ superYang-Mills correlation functions via AdS duality,''
  {\em Nucl.\ Phys.}\  {\bf B551}, 575 (1999)
  {\tt hep-th/9811047}.








\bibitem{D'Hoker:2002aw} 
  E.~D'Hoker and D.~Z.~Freedman,
  ``Supersymmetric gauge theories and the AdS / CFT correspondence,''
  {\tt hep-th/0201253}.

\bibitem{Freedman:2012zz} 
  D.~Z.~Freedman and A.~Van Proeyen,
  {\em Supergravity}.
  Cambridge University Press, 2012.

\bibitem{Breitenlohner:1982jf} 
  P.~Breitenlohner and D.~Z.~Freedman,
  ``Stability in gauged extended supergravity,''
  {\em Annals Phys.}\  {\bf 144}, 249 (1982).

\bibitem{Azeyanagi:2013fla} 
  T.~Azeyanagi, M.~Hanada, M.~Honda, Y.~Matsuo and S.~Shiba,
  ``A new look at instantons and large-$N$ limit,''
  {\tt arXiv:1307.0809 [hep-th]}.

\bibitem{Kachru:1998ys} 
  S.~Kachru and E.~Silverstein,
  ``4-D conformal theories and strings on orbifolds,''
  {\em Phys.\ Rev.\ Lett.}\  {\bf 80}, 4855 (1998)
  {\tt hep-th/9802183}.

\bibitem{Klebanov:1998hh} 
  I.~R.~Klebanov and E.~Witten,
  ``Superconformal field theory on three-branes at a Calabi-Yau singularity,''
  {\em Nucl.\ Phys.}\  {\bf B536}, 199 (1998)
  {\tt hep-th/9807080}.


\bibitem{Gubser:1998vd} 
  S.~S.~Gubser,
  ``Einstein manifolds and conformal field theories,''
  {\em Phys.\ Rev.}\  {\bf D59}, 025006 (1999)
  {\tt hep-th/9807164}.

\bibitem{Evans:2000ct} 
  N.~J.~Evans, C.~V.~Johnson and M.~Petrini,
  ``The Enhancon and ${\cal N}=2$ gauge theory: Gravity RG flows,''
  {\em JHEP} {\bf 0010}, 022 (2000)
  {\tt hep-th/0008081}.


\bibitem{Pilch:2003jg} 
  K.~Pilch and N.~P.~Warner,
  ``Generalizing the ${\cal N}=2$ supersymmetric RG flow solution of IIB supergravity,''
  {\em Nucl.\ Phys.}\  {\bf B675}, 99 (2003)
  {\tt hep-th/0306098}.

\bibitem{Gauntlett:2001ps} 
  J.~P.~Gauntlett, N.~Kim, D.~Martelli and D.~Waldram,
  ``Wrapped five-branes and ${\cal N}=2$ superYang-Mills theory,''
  {\em Phys.\ Rev.}\  {\bf D64}, 106008 (2001)
  {\tt hep-th/0106117}.

\bibitem{Bigazzi:2001aj} 
  F.~Bigazzi, A.~L.~Cotrone and A.~Zaffaroni,
  ``${\cal N}=2$ gauge theories from wrapped five-branes,''
  {\em Phys.\ Lett.}\  {\bf B519}, 269 (2001)
  {\tt hep-th/0106160}.

\bibitem{Girardello:1999bd} 
  L.~Girardello, M.~Petrini, M.~Porrati and A.~Zaffaroni,
  ``The Supergravity dual of ${\cal N}=1$ superYang-Mills theory,''
  {\em Nucl.\ Phys.}\  {\bf B569}, 451 (2000)
  {\tt hep-th/9909047}.
  
\bibitem{Polchinski:2000uf} 
  J.~Polchinski and M.~J.~Strassler,
  ``The String dual of a confining four-dimensional gauge theory,''
  {\tt hep-th/0003136}.
  
\bibitem{Dall'Agata:2001vb} 
  G.~Dall'Agata, C.~Herrmann and M.~Zagermann,
  ``General matter coupled ${\cal N}=4$ gauged supergravity in five-dimensions,''
  {\em Nucl.\ Phys.}\  {\bf B612}, 123 (2001)
  {\tt hep-th/0103106}.

\bibitem{Schon:2006kz} 
  J.~Schon and M.~Weidner,
  ``Gauged ${\cal N}=4$ supergravities,''
  {\em JHEP} {\bf 0605}, 034 (2006)
  {\tt hep-th/0602024}.
  
  
\bibitem{Gubser:2000nd} 
  S.~S.~Gubser,
  ``Curvature singularities: The Good, the bad, and the naked,''
  {\em Adv.\ Theor.\ Math.\ Phys.}\  {\bf 4}, 679 (2000)
  {\tt hep-th/0002160}.
  
    
\end{thebibliography}
\end{document}